\PassOptionsToPackage{table}{xcolor}

\documentclass[a4paper,fleqn]{cas-dc}

\usepackage[authoryear,sort]{natbib}

\usepackage{array}
\usepackage[caption=false,font=normalsize,labelfont=sf,textfont=sf]{subfig}
\usepackage{textcomp}
\usepackage{url}
\usepackage{verbatim}
\usepackage{booktabs}
\usepackage{tabularx}
\usepackage{cleveref} %
\usepackage{colortbl}
\usepackage{enumitem}
\usepackage{longtable}
\usepackage{tikz}
\usetikzlibrary{arrows.meta, positioning, calc}
\usepackage[most]{tcolorbox}
\usepackage{enumitem}
\usepackage{array}

\newif\ifshowchanges \showchangesfalse

\ExplSyntaxOn
\cs_set:Npn \__first_footerline: { \group_begin: \small \sffamily \__short_authors: :~ { \rmfamily \itshape Preprint } \group_end: }
\ExplSyntaxOff
\begin{document}
\let\WriteBookmarks\relax
\def\floatpagepagefraction{1}
\def\textpagefraction{.001}

\shorttitle{GUEST: Preliminary Guidelines for Using and Evaluating GenAI Tools to Support Systematic Literature Reviews}

\shortauthors{Kitchenham et al.}

\title [mode = title]{GUEST: Preliminary Guidelines for Using and Evaluating GenAI Tools to Support Systematic Literature Reviews}

\author[1]{Barbara Kitchenham}[orcid=0000-0002-6134-8460]
\fnmark[1]
\credit{Conceptualization, Methodology, Writing~–~original draft, Writing~–~review \& editing, Visualization, Project administration (initial submission)}
\affiliation[1]{
            city={Maidstone},
            country={United Kingdom}}

\author[2]{Sebastián Pizard}[orcid=0000-0002-5535-8646]
\fnmark[1]
\credit{Conceptualization, Investigation, Writing~–~review \& editing, Visualization}
\affiliation[2]{organization={Universidad de la República},
            city={Montevideo},
            country={Uruguay}}

\author[3]{Lech Madeyski}[orcid=0000-0003-3907-3357]
\cormark[1]
\fnmark[1]
\ead{lech.madeyski@pwr.edu.pl}
\credit{Conceptualization, Investigation,  Writing~–~review \& editing, Visualization, Project administration (revision process)}
\affiliation[3]{organization={Wrocław University of Science and Technology},
            city={Wroclaw},
            country={Poland}}

\author[4]{Ronnie de Souza Santos}[orcid=0000-0003-3235-6530]
\credit{Conceptualization, Investigation, Writing~–~review \& editing}
\affiliation[4]{organization={University of Calgary},
            city={Calgary},
            country={Canada}}

\author[5]{Martin Shepperd}[orcid=0000-0003-1874-6145]
\credit{Conceptualization, Writing~–~review \& editing}
\affiliation[5]{organization={Brunel University of London},
            city={London},
            country={United Kingdom}}

\author[6]{David Budgen}[orcid=0000-0001-7143-0241]
\credit{Validation, Writing~–~review \& editing}
\affiliation[6]{organization={Durham University},
            city={Durham},
            country={United Kingdom}}

\cortext[1]{Corresponding author}

\fntext[1]{These authors contributed equally to this work and are joint lead authors.}

\begin{abstract}
\relax\textbf{Context:} 
Generative AI (GenAI) and Large Language Models (LLMs) are increasingly used for academic tasks in software engineering and beyond, including systematic literature reviews (SLRs). However, whilst their capabilities are rapidly developing, there is no guarantee they can meet the rigour, reliability, and transparency that SLRs require.\\
\textbf{Objectives:} 
To support researchers intending to conduct SLRs using GenAI, or those conducting empirical studies to evaluate how well GenAI supports SLR tasks.\\
\textbf{Methods:} 
First, we conducted a rapid review to identify studies that propose guidelines for evaluating and using GenAI and LLMs to support SLRs. Second, we drew on thought experiments, relevant guidance from the literature, and our own experience from conducting SLRs and evaluating tools, to develop recommendations for how to use and assess GenAI in the context of SLRs. \\
\textbf{Results:} 
We classify methodological issues for the roles of evaluation and reviewing into four categories spanning SLR practice adjustments, general secondary-study standards, GenAI evaluation practices and core empirical rigour. Building on this, we introduce GUEST (GenAI Use and Evaluation in SLR Tasks) which is a comprehensive framework of 51 recommendations designed to guide researchers through planning, executing and transparently reporting GenAI-assisted literature reviews and validation studies.\\
\textbf{Conclusion:} 
Current empirical evidence indicates that GenAI cannot be relied upon to perform unsupervised systematic reviews so requires human oversight. Nevertheless, it can offer cost-effective assistance for repetitive screening tasks and secondary validation.  By standardizing how these tools are deployed and evaluated, the GUEST framework provides a necessary methodological baseline for trustworthy GenAI-assisted research in software engineering.  
\end{abstract}


\maketitle
\setcounter{footnote}{1}%

\section{Introduction}
{A}{s} advocates of the use of systematic literature reviews (SLRs),\footnote{In this paper, we use the term systematic literature review to refer to all types of systematically conducted literature review, including qualitative systematic reviews, quantitative systematic reviews, mapping studies, rapid reviews, tertiary studies, and updates to such studies.} we are used to conducting SLRs ourselves and also refereeing SLRs that have been performed by other researchers. Recently, however, we have been asked to referee empirical studies that investigate the use of Generative AI (GenAI) tools, such as Large Language Models (LLMs), for SLR tasks. Unfortunately, there is no well-established methodology for evaluating the performance of GenAI tools in the context of systematic reviews, and we have noticed some significant problems with the methods being adopted. For example, a review of 29 papers that evaluate LLMs for study selection in SLRs found that, among those using published SLRs as benchmarks, only one reported using SLRs that could not have been in the LLM's training data \citep{Madeyski26LLM4SCREENLIT}.

We believe that methodology problems must be addressed quickly before they become adopted as de facto standards by other researchers or even by GenAI tools. We note that bad practices can be extremely difficult to eradicate once they become established. An example of this in software engineering (SE) is the use of the Mean Magnitude Relative Error (MMRE) in cost estimation research. After the publication of \citeauthor{conte-1986}'s book on software metrics and models \citeyearpar{conte-1986}, MMRE was widely adopted as a means of measuring the performance of cost and fault estimation models. However, MMRE is a biased metric. MMRE bias was first subjected to detailed analysis in the early 2000s \citep{Kitchenham01,foss-2003}, but the problem of its continued use was later noted both in 2005 \citep{myrtveit-2005} and in 2012 \citep{myrtveit-2012}. Another example is incorrect analysis and meta-analysis of crossover experiments, where many researchers failed to realize that, when each participant in an experiment uses all experimental methods, correct statistical analysis is based on the change in the participants' outcomes, not a simple analysis of all observations \citep{kitchenham-2020}. 

The goal of this paper is to provide process recommendations to assist researchers who want either to evaluate or to use GenAI tools for conducting SLRs. We discuss a number of methodological issues concerning the use of GenAI tools and identify process recommendations for planning, conducting, and reporting that we hope will be of value to researchers using and evaluating GenAI tools to support SLRs. Our recommendations are primarily intended to support software engineering (SE) research, which is the focus of our own research, but may be of interest to the wider research community. For brevity, we refer to this preliminary set of recommendations as GUEST (GenAI Use and Evaluation in SLR Tasks). Like a guest, a GenAI tool should contribute to an SLR under the oversight of researchers, who remain responsible for the SLR.
These recommendations are preliminary because we expect them to evolve as both the technology and the empirical evidence mature. The box below gives an overview of GUEST, and \Cref{sec:Recommendations} presents the full set of recommendations.

\begin{tcolorbox}[colback=blue!3, colframe=blue!60, title=\textbf{GUEST at a Glance}, fonttitle=\sffamily, boxrule=0.6pt, arc=2mm, breakable, left=4pt, right=4pt]
\small
GUEST has 51 recommendations for two roles undertaken by researchers.
\emph{Reviewers} conduct an SLR with the help of GenAI tools.
\emph{Evaluators} assess the performance of GenAI tools in SLRs, independently of the tool developers.
The recommendations cover planning, conduct and reporting, and four SLR processes: literature screening, data extraction, qualitative synthesis, and risk-of-bias and strength-of-evidence assessment.
\emph{SLR-specific} recommendations (kind~(i) in \Cref{tab:RecCategories}) change SLR practice when GenAI is used.
\emph{General} recommendations (kinds (ii)--(iv)) apply good practice for secondary studies or evaluations.

\smallskip
\noindent\begin{tabularx}{\linewidth}{@{}>{\raggedright\arraybackslash}Xrrr@{}}
\toprule
Number of recommendations & SLR-specific & General & Total \\
\midrule
Planning, conduct and reporting (\Cref{tab:GUESTPlanConRep}) & 6 & 25 & 31 \\
SLR processes (\Cref{tab:GUESTProcess}) & 17 & 3 & 20 \\
\midrule
Total & 23 & 28 & 51 \\
\bottomrule
\end{tabularx}

\smallskip
Key messages from the SLR-specific recommendations, with their IDs from \Cref{tab:GUESTPlanConRep,tab:GUESTProcess}:
\begin{enumerate}[leftmargin=*, itemsep=1pt, topsep=3pt, label=\textbf{\arabic*.}]
    \item \textbf{Use GenAI only to check human work in synthesis and assessment} (RevA1; reviewers). This applies to qualitative synthesis and to risk-of-bias and strength-of-evidence assessment.
    There are no agreed risk-of-bias instruments for the many study types used in SE.
    Without such instruments, GenAI performance in risk-of-bias assessment cannot be measured consistently.
    \item \textbf{Define the sub-tasks that GenAI performs} (Ext5--Ext7, A1--A3; both roles).
    This applies to complex data extraction and to risk-of-bias and strength-of-evidence assessment.
    Also state the input that each sub-task needs and how each sub-task's output is compared with the work of human researchers.
    In our view, the variety of SE study types makes full automation of data extraction for synthesis unlikely.
    \item \textbf{Use screening metrics appropriate for SLRs} (Scr1--Scr3, from \citealp{Madeyski26LLM4SCREENLIT}; both roles).
    Most candidate studies are irrelevant, so accuracy and $F_1$ scores mislead (\Cref{tab:FailureModes}).
    Report full confusion matrices and the Matthews correlation coefficient (MCC), which show whether results are better than chance.
    Add the weighted MCC (WMCC), which gives more weight to a missed relevant study than to a wrongly included one.
    A wrongly included study can be excluded later, but a missed study may be lost forever.
    Accuracy, $F_1$, and MCC give both mistakes the same weight.    
    \item \textbf{Evaluate GenAI tools prospectively, inside new SLRs} (PlanR3: reviewers; Con3: both roles).
    Reviewers should treat their use of GenAI in a new SLR as a case study of the tool~\citep{thomas-2026-RAISE1,Madeyski26LLM4SCREENLIT}.
    A new SLR reduces the risk of data contamination, which arises when a published SLR may be part of the tool's training data.
    Report all the human effort that the tool needs.
    In our experience, SE researchers often both evaluate tools and conduct SLRs, so they are well placed to run such case studies.
\end{enumerate}

In addition to the SLR-specific recommendations, we include the 28 general recommendations, not because they are new, but to allow GUEST to be used on its own.
For example, the general recommendations ask researchers to justify the choice of tool and of study design, to compare tool output with the results of at least two human researchers, and to report tool versions, prompts, settings and costs in full (Plan1, Plan2, Plan6, Rep1--Rep4).
\Cref{tab:RecCategories} links each kind of general recommendation to existing guidance: (ii) for secondary studies, (iii) for evaluations of GenAI tools, and (iv) for evaluation studies in general. 
\end{tcolorbox}

\Cref{sec:GuidelineBackground} describes the nature of SLRs, the limitations of GenAI tools that matter for SLRs, and prior work on SLR automation.
The section also explains how GUEST differs from existing guidance.
\Cref{sec:DevelopingLLMEvaluationANdUseGuidlines} describes how we developed GUEST: a rapid review of guidelines on using GenAI tools, followed by thought experiments that drew on the review results and on our experience of conducting SLRs and evaluating tools.
\Cref{sec:SummaryOfThougthExperimentResults} reports the outcomes of the thought experiments, presents the GUEST recommendations, and links the recommendations to failure modes reported in empirical studies.
\Cref{sec:Validation} reports a first validation in which we compare GUEST with other recently published guidelines.
\Cref{sec:ThreatsToValidity,sec:Discussion,sec:OpenQuestions,sec:Conclusions} cover threats to validity, the implications of GUEST, open research questions and conclusions.

\section{Background and Related Work}\label{sec:GuidelineBackground}
This section discusses factors that affect the way researchers interact with and evaluate GenAI tools, including LLMs, in the context of conducting SLRs. In addition, we report related work on automating the SLR process using GenAI tools.

\subsection{The Nature of SLRs}\label{sec:TheNatureOfSRs}
We undertake SLRs to reduce the risk of validity problems associated with expert reviews, such as:
\begin{itemize}[noitemsep]
    \item Personal bias.
    \item Incomplete personal knowledge of the relevant literature.
    \item Inclusion of poor-quality studies.
    \item Transcription errors during data collection (as a result of fatigue, misreading, or misunderstanding).
    \item Analysis and synthesis errors (as a result of personal bias, fatigue, or personal knowledge gaps).
\end{itemize}
Because of such problems, expert literature reviews can be unreliable, so there is no way to determine whether specific results and recommendations can be trusted. 

Such validity risks were undermining scientific progress in many disciplines, particularly medicine and health care \citep{mulrow-1987}. Hence, to address these problems, medical researchers developed guidelines for conducting SLRs \citep[see][for the recent guidelines]{cumpston-2019}, which were subsequently accepted by other empirical disciplines, including Software Engineering (SE). An important feature of these guidelines is that they do not only define a set of SLR activities, but also identify associated SLR process control activities intended to minimize errors. We accept the overheads of process control (e.g., broad searches often adopting a variety of different strategies and multiple researchers doing the same tasks) because we want to reduce the risks of invalid results that are inherent in expert reviews. These process controls cannot guarantee SLR reliability. However, they should increase the likelihood of identifying and correcting human errors.

We want our SLRs to produce evidence that is Valid, Traceable, Qualified, and Verifiable. \emph{Validity} requires that authors find all the relevant literature and draw trustworthy conclusions. It also demands that authors use appropriate methods for all of their SLR process and SLR process control tasks. \emph{Traceability} requires authors to make clear the links among their research questions, search and selection process, data collection process, and data aggregation process and how each contributes to the final conclusions and recommendations. In the context of evidence from SLRs, \emph{Qualified} means that authors assess the strength of the evidence supporting their recommendations and conclusions, which requires assessing both the methodological validity of the individual primary studies and the combined strength of the evidence from the set of available primary studies. Finally, to be \emph{Verifiable} (also referred to as \emph{Reproducible} or \emph{Auditable}), authors need to take care to fully document both their study protocols and the conduct of their SLRs, using current reporting standards \citep{kitchenham-2023}.   

Rapid Reviews (RR) and Mapping studies (MS) are sometimes considered to be forms of SLR. But it is important to note that they have different constraints:
\begin{itemize}
    \item RRs trade-off some aspects of the SLR process to obtain results in a short time scale and/or with limited resources.
    \item Mapping studies (other than tertiary studies) seldom require an assessment of the weaknesses of individual primary study methodology or the strength of evidence but instead typically aim to classify primary studies related to a topic area and sometimes to develop a taxonomy of primary study types.
\end{itemize}

Nonetheless, such studies can still claim to be systematic if researchers apply process control to the subset of SLR processes that they do use and to deliver reports that conform with SLR reporting standards.

\paragraph{Specific SLR tasks.} An SLR is a structured research method that answers defined questions by systematically locating, selecting, and analysing all relevant scientific literature. This involves a number of very different but related tasks, each of which requires a valid starting point and needs to deliver a valid input to the subsequent activity. Typically, this is done by inspecting the output of each task, not only to assess whether it is valid, but also to ensure it was produced appropriately (that is, that the process used to obtain the output does not invalidate the chain of evidence from the research question to the final conclusion). To support this, well-established good practices exist for each task (for example, study selection generally involves two or more reviewers independently performing the same activity), along with validation procedures (for example, in independent selection, calculating inter-reviewer agreement metrics). The different SLR tasks are discussed in more detail in Appendix~\ref{sec:SLRProcessDetails}.
Appendices~\ref{sec:DetailsOfRR}, \ref{sec:ThoughExperimentResults}, \ref{sec:SLRProcessDetails}, and~\ref{sec:FirstValidationDetails} follow \Cref{sec:Conclusions}.

\subsection{GenAI Tool Limitations Relevant to SLRs}\label{sec:TheNatureOfLLMs}

Generative AI refers to algorithmic approaches based on deep generative models trained on large datasets to learn the underlying data distribution, enabling them to generate new, realistic content (for example, text, images, or code) from simple prompts \citep{Banh2023}. They are able to produce plausible natural-language answers to textual questions on any topic. However, GenAI tools have basic limitations that make their use in the context of SLRs challenging:
\begin{enumerate}
    \item They create something of a moving target for evaluators and users, and this happens at two levels. The first level is the major model updates that occur (e.g., GPT-4 to GPT-4o to GPT-5.6), though fortunately, these releases are differentiated by formal release announcements, version numbers, and some documentation. Researchers can at least explicitly reference these version details. More problematic is the second, or infrastructure level. This includes: (i) safety filters and content policies, (ii) system prompts and guardrails, (iii) input preprocessing and output filtering, which can be highly relevant for say PDF processing capabilities, (iv) rate limiting and abuse detection, and finally (v) general UI/API wrapper changes. Unfortunately, Level 2 changes can happen daily or even hourly without any public notice, leading to reproducibility problems.
    \item They can occasionally produce invalid or nonsensical output. This is called hallucinating and is the focus of much current research. \citet{Huang25} provide a deeper discussion and taxonomy. The causes have been discussed from different perspectives: \citet{Kalai25} argue that hallucinations arise because standard training and evaluation reward confident guessing over acknowledging uncertainty, while He and Thinking Machine Labs \citeyearpar{he-2025} attribute the related non-determinism of LLM endpoints to load- and batch-size variation. Mitigation is an active research area (for example, retrieval-augmented generation and fact-checking; \citealp{Greengard-2025}), but none of these approaches yet eliminates the problem, so hallucination remains a limitation that users must guard against.
    \item They can produce inconsistent outputs from the same input. They respond to questions based on the frequency of textual patterns with an added element of randomization to simulate variations in human responses. In contrast to hallucinations, inconsistencies are not invalid. They represent the variations in responses we find in spoken and written communication. However, it is the responsibility of the GenAI tool user to distinguish between responses that are harmless linguistic variations and those that are invalid hallucinations.
    \item They are designed to identify frequent conversational threads and move such threads into their long-term memory. This means that frequent use of the same test material may lead to the test questions and answers being stored in long-term memory, so the GenAI tool can obtain the answers by accessing the long-term memory without actually processing the specific questions. This phenomenon is referred to as data leakage or data contamination. Data contamination represents a serious threat to the use of publicly available datasets to test GenAI tools \citep{woelfle-2022}. In addition, in the context of aggregating evidence from primary studies, the risk of data contamination means that without extremely careful prompting, LLMs may give priority to the most commonly reported outcomes among a set of papers, while ignoring minority opinions and the possibility of publication bias.
    \item Some, but not all, GenAI tools allow researchers access to parameters that influence their behaviour. For example, to reduce the risk of inconsistent outputs, researchers can adjust the \texttt{temperature} parameter in LLMs that directly affects the variability and randomness of generated responses. A lower LLM temperature value (close to 0) produces more deterministic and focused outputs for tasks requiring greater levels of reproducibility. In addition, there are different parameters (i.e., \texttt{top\_p}, max tokens, frequency penalty, or presence penalty) that can also be adjusted \citep{Brown20}. However, parameters can interact in unpredictable ways, making the effects of changing multiple settings hard to predict.
    \item Currently, GenAI tools are accessed via chatbot \emph{prompts}. A recent paper discussing the prompting interface has identified some problems with prompting in the context of SE research \citep{morris-2024}. In particular, Morris identified the problem of researchers not reporting how many prompts they used to obtain the results they reported, and failing to assess and report the robustness of their prompts across different GenAI models and different generations of the same model.
    \item Most LLMs are optimized to produce fluent, plausible responses to prompts \citep{Banh2023} rather than to explain how they arrived at an answer or to justify why it is valid --- a limitation that matters for SLRs, where conclusions must be traceable and justified. Furthermore, the major commercial chatbots tend to keep internal reasoning private. The default approach is to hide chain-of-thought or other internal reasoning steps (due to safety and misuse prevention, as well as to keep intellectual property away from competition) and offer only curated, higher-level explanations that aim to foster user trust without compromising safety or competitive advantage. Although there is currently more understanding of the importance of providing justifications and explanations of outputs, it remains the responsibility of the user to put in place a prompt process that will provide any necessary explanatory information. 
    \item They are likely to exhibit bias. They are trained on enormous volumes of Internet-based data and, therefore, ``inherit stereotypes, misrepresentations, derogatory and exclusionary language'' \citep{Gallegos-2024}. In the context of SLR tasks, this bias could lead a tool to downplay or overlook relevant studies --- for example, studies not written in English, or whose authors are women or have non-European-sounding names. A recent preprint reports a preference for male-authored references in LLM-driven reference selection \citep{He2025refbias}. However, we are not aware of peer-reviewed, SE-specific evidence that quantifies this risk for SLR screening, so we flag it as a concern that warrants empirical investigation rather than as an established finding.
\end{enumerate}

Thus, like humans, GenAI tools can make mistakes (although the causes of error are different), and they can behave without regard for scientific ethics or scientific method. In addition, they are designed to be non-deterministic and to conceal their working methods. This means that while some systems are able to produce seductively plausible literature reviews, such reviews are, at best, equivalent to expert opinion reviews, not \emph{systematic literature reviews}. 

\subsection{Related Work on SLR Automation}

SLRs have been adopted by many different academic disciplines and, particularly in healthcare, form an important source of information for practitioners. In SE, we find wide adoption of SLRs, although primarily by academics. For example, \cite{kamei2021} identified 446 SLRs published only in the top SE journals and conferences before 2019, and many more have been published subsequently. However, planning and conducting SLRs are complex activities that require substantial resources and effort. This has long motivated work on their automation. 

The International Collaboration for the Automation of Systematic Reviews (ICASR)\footnote{https://icasr.github.io/} is an international collaboration among several groups working on systematic review automation since 2015. In their first paper \citep{Beller2018}, they argued that automation for SLRs should support and improve tasks across the full evidence-synthesis workflow, be developed through continuous, user-informed refinement (starting with key bottlenecks), and maintain rigorous review standards (for example, PRISMA and MECIR). They also emphasized building flexible, interoperable, modular tools; fostering multidisciplinary collaboration; openly sharing methods, tools, and resources; and evaluating automation rigorously and reproducibly, including both technical performance and usability in real review workflows. They meet annually to share their progress on SLR automation.

While early work on SLR automation showed promise, it was only in the last few years, with the emergence of GenAI, that broad automation began to seem realistically achievable. In an SLR on the topic, \citet{vandinter2021} found only 41 papers on SLR automation up to June 2020, and these studies were confined to just two domains: software engineering and medicine. Since around that time, there has been a rapid increase in both the number and the diversity of studies investigating how LLMs and other GenAI techniques can automate one or more stages of the SLR process \citep[see, e.g.,][]{Dennstadt2024, Guo2024b, Guo2024}. Several commercial tools tailored to SLRs have also emerged (see, e.g., Elicit\footnote{Elicit is an AI research assistant that searches, summarizes, and extracts data from academic papers to automate parts of literature reviews and SLRs. https://elicit.com/}
). However, guidance on how to use and evaluate these tools in SLRs is still limited, and without it, users may adopt them in ways that compromise the rigour SLRs require.

Against this backdrop, a group consisting mostly of health researchers is developing guidelines to support the use of GenAI in SLRs \citep{thomas-2026-RAISE1}. The Responsible use of AI in Evidence SynthEsis (RAISE 2026) guidelines are a community-based set of recommendations and companion guidance intended to ensure the responsible, transparent, and ethical use of AI in evidence synthesis, grounded in research integrity. They make clear that their guidelines apply to several different researcher roles. In particular, they point out the distinction between researchers using GenAI tools and LLMs to support the conduct of their SLRs and researchers studying the performance of GenAI and LLM tools when applied to SLR-related tasks. They also make clear that researchers who want to use GenAI tools to support their SLRs need to be able to justify the use of the GenAI tools and report how the tools were trained and their performance evaluated.

Additionally, given the risks of using GenAI in research within our community, a group of researchers has begun developing guidelines for experimenting with LLMs in software engineering \citep{wagner2025}. Although these preliminary guidelines apply more broadly to empirical studies that use LLMs, they are also relevant to research that involves planning and conducting SLRs.

Lastly, two further guideline efforts help to position our work. \citet{baltes-2025} provide community guidelines for empirical SE studies involving LLMs, organized around a taxonomy of seven study types. In that taxonomy, systematic reviews are one study type rather than the focus, so the guidance is not tailored to specific SLR tasks. In the evidence-synthesis community, \citet{farotimi-2025}, in an editorial in \emph{Research Synthesis Methods}, give guidance for manuscripts that test GenAI for systematic review and meta-analysis, framed around reporting, validation, and reproducibility requirements rooted in the medical and broader evidence-synthesis literature.

\begin{table*}[!t]
\centering
\footnotesize
\caption{Positioning of our guidelines relative to related GenAI evidence-synthesis guidance.}
\label{tab:related-guidelines}
\begin{tabular}{@{}>{\raggedright\arraybackslash}p{2.3cm}>{\raggedright\arraybackslash}p{3.2cm}>{\raggedright\arraybackslash}p{1.9cm}>{\raggedright\arraybackslash}p{4.8cm}>{\raggedright\arraybackslash}p{3.5cm}@{}}
\toprule
\textbf{Work} & \textbf{Domain} & \textbf{SLR-specific?} & \textbf{Roles / granularity} & \textbf{Grounding} \\
\midrule
This paper & Software engineering & Yes & Two researcher roles (\emph{using} vs \emph{evaluating}); per-SLR-task, SE-tailored methods & Rapid review, thought experiments, companion SE studies \\
\addlinespace[2pt]
RAISE 2026 \citep{thomas-2026-RAISE1,thomas-2025-RAISE2,Thomas-2026-RAISE3} & Evidence synthesis; health/Cochrane-rooted; SE not addressed & Yes & Eight ecosystem roles; principles plus task-level metrics (RAISE-2) and tool-use recommendations (RAISE-3) & Large multi-organization expert consensus \\
\addlinespace[2pt]
\citet{farotimi-2025} & SRMA; health/cross-discipline & Yes & Manuscript authors; reporting/validation by SRMA stage & Editorial; adopts RAISE and DEST \\
\addlinespace[2pt]
\citet{baltes-2025} & Empirical SE (general) & No (seven study types) & Researchers and reviewers; study-type level, not SLR-stage & Community guidelines plus checklist \\
\addlinespace[2pt]
\citet{wagner2025} & Empirical SE (general) & No & SE researchers; foundational, high-level & Position paper (precursor to Baltes et al.) \\
\bottomrule
\end{tabular}
\end{table*}

\Cref{tab:related-guidelines} summarizes how our guidelines relate to these efforts. RAISE is the closest related work: it is task-aware (RAISE-2 discusses evaluation metrics and RAISE-3 gives tool-use recommendations by synthesis stage) and addresses both the using and evaluating roles within a broader, eight-role evidence-synthesis ecosystem. RAISE originates in the health and broader evidence-synthesis tradition, where evidence often takes relatively standardized forms such as randomized controlled trials. Software engineering, by contrast, exhibits greater heterogeneity of SLR and study types, and RAISE does not specifically address this. The same applies to the Farotimi et al. editorial, which addresses authors in the evidence-synthesis community rather than software engineering. Baltes et al. and Wagner et al. are SE-focused but target empirical LLM studies in general rather than SLRs. Our contribution is distinctively SE-specific providing: (i)~a focus on the heterogeneity of SLR and study types in software engineering, especially the prevalence of qualitative SLRs; (ii)~an in-depth treatment of the two researcher roles (\emph{using} versus \emph{evaluating} GenAI for SLRs); (iii)~SE-tailored, task-level methods; and (iv)~empirical grounding through our companion studies on literature screening \citep{Madeyski26LLM4SCREENLIT} and qualitative synthesis \citep{Pizard26}.

\paragraph{Relationship to our companion studies.}
Two of these works are our own peer-reviewed studies, and we use them here as the empirical backbone for our screening and synthesis recommendations rather than re-deriving that evidence.
Our study of literature screening \citep{Madeyski26LLM4SCREENLIT} (published in \emph{Information and Software Technology}) provides the performance-metric analysis, the reanalyses of published SE screening evaluations, and the validation guidance that underpin our screening recommendations.
Our study of qualitative synthesis \citep{Pizard26} (published at WSESE'26) provides the trial-based failure modes and soundness criteria that underpin our synthesis recommendations.
This paper makes a distinct, higher-level contribution: a role-based framework that covers all the main SLR tasks, integrates GenAI tool validation into the SE SLR planning, conduct, and reporting process, and positions this guidance against related efforts (\Cref{tab:related-guidelines}).
To keep the present paper focused on that framework, we summarize the companion findings where they motivate a recommendation and refer readers to the companion studies for the task-level detail, which we do not reproduce here.

\subsection{Rationale for Our Guidelines}

As noted earlier, our goal is to provide recommendations to guide researchers when they evaluate or use GenAI to support SLRs. With this in mind, our guidelines contribute to, and differ from, prior work in the following ways:

\begin{itemize}
    \item We build on prior recommendations from the field where SLRs originated, but also draw on other related work and our own experience. Adapting research guidelines from other disciplines is common in SE, and it is also necessary. Software engineering differs from those disciplines in the nature of its primary studies and evidence. In addition, SE researchers are used to methodological guidelines that are written for their field and illustrated with examples from it. Well-known adaptations include the guidelines for conducting and reporting case study research \citep{runeson2009guidelines}, for conducting SLRs \citep{kitchenham2007guidelines}, systematic mapping studies \citep{petersen2015guidelines}, and for reporting secondary studies \citep{kitchenham-2023}. Following this tradition, our work complements and adapts guidance developed in the health and broader evidence-synthesis tradition, such as RAISE \citep{thomas-2026-RAISE1} and the \citet{farotimi-2025} editorial, by focusing specifically on software engineering, where SLRs and primary studies are more heterogeneous and qualitative SLRs are common. Our contribution is SE-specific: SE-tailored, task-level methods and a detailed account of how to integrate GenAI tool validation into the SE SLR workflow.
    \item We focus only on the two predominant roles in our discipline: (1) researchers who evaluate GenAI to support SLRs and (2) reviewers who conduct an SLR using GenAI. This both simplifies our proposal by focusing on the roles we believe will most benefit from the guidelines and allows us to go deeper, providing more deliberate analyses of the challenges these roles face and more detailed guidance to support them.
    \item We also draw on our computing expertise regarding the nature, limitations, and risks of GenAI and LLMs. We see this as essential, especially after observing that many studies evaluating or using these technologies in SLRs underestimate risks or conflate key concepts related to GenAI.
\end{itemize}

\section{Developing Our Recommendations: Method and Inputs}\label{sec:DevelopingLLMEvaluationANdUseGuidlines}

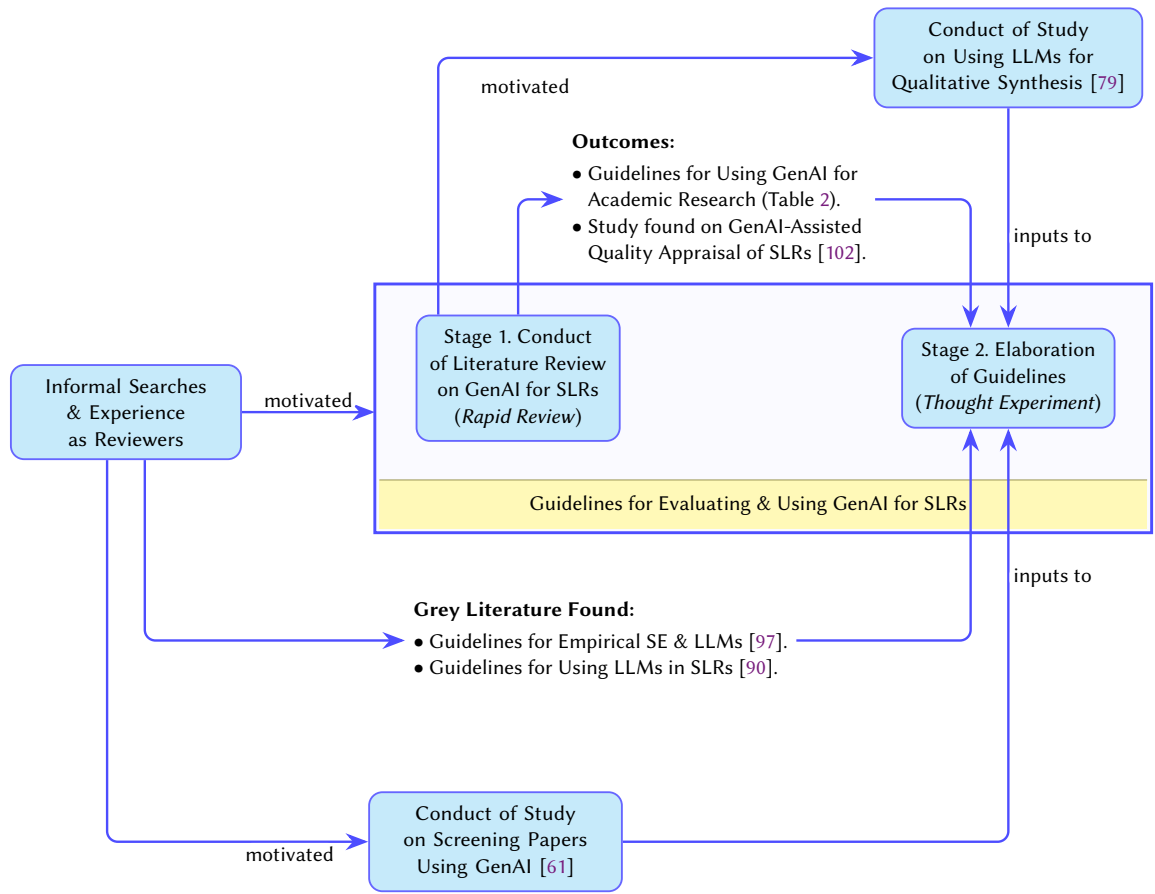
\begin{figure*}[!t]
\centering
{\sffamily\resizebox{0.95\textwidth}{!}{%
\begin{tikzpicture}[
  font=\small,
  box/.style={rounded corners=5pt, draw=blue!60, fill=cyan!20, line width=0.7pt, align=center, inner sep=5pt},
  frame/.style={draw=blue!70, line width=1.1pt},
  ar/.style={-{Stealth[length=3mm]}, draw=blue!70, line width=0.9pt, rounded corners=4pt},
  lbl/.style={font=\small, fill=white, inner sep=1.5pt},
  txt/.style={align=left, font=\small},
]
\draw[frame, fill=blue!2] (5.0,3.9) rectangle (15.3,7.2);
\fill[yellow!35] (5.06,3.96) rectangle (15.24,4.6);
\draw[yellow!55!black, line width=0.4pt] (5.06,4.6) -- (15.24,4.6);
\node[xshift=-12mm] at (10.65,4.27) {Guidelines for Evaluating \& Using GenAI for SLRs};
\node[box, text width=2.7cm] (informal) at (1.7,5.5) {Informal Searches \& Experience as Reviewers};
\node[box] (stage1) at (6.9,5.95) {Stage 1. Conduct\\ of Literature Review\\ on GenAI for SLRs\\ (\emph{Rapid Review})};
\node[box] (stage2) at (13.4,5.95) {Stage 2. Elaboration\\ of Guidelines\\ (\emph{Thought Experiments})};
\node[box] (valid) at (18.7,5.5) {A First Validation\\ \emph{(Comparison with}\\ \citealp{baltes-2025}; \\ \citealp{thomas-2026-RAISE1}\emph{)}};
\node[box] (synth) at (13.4,10.65) {Conduct of Study\\ on Using LLMs for\\ Qualitative Synthesis\\ \citep{Pizard26}};
\node[box] (screen) at (6.6,-0.2) {Conduct of Study\\ on Screening Papers\\ Using GenAI\\ \citep{Madeyski26LLM4SCREENLIT}};
\node[txt, anchor=north west] (outcomes) at (7.5,9.75)
  {\textbf{Outcomes:}\\[2pt]
   $\bullet$~Guidelines for Using GenAI for\\
   \phantom{$\bullet$}~~Academic Research (\Cref{tab:Guidelines}).\\
   $\bullet$~Study found on GenAI-Assisted\\
   \phantom{$\bullet$}~~Quality Appraisal of SLRs\\
   \phantom{$\bullet$}~~\citep{woelfle-2022}.};
\node[txt, anchor=north west] (grey) at (5.4,3.1)
  {\textbf{Grey Literature Found:}\\[2pt]
   $\bullet$~Guidelines for Empirical SE \& LLMs\\
   \phantom{$\bullet$}~~\citep{wagner2025}.\\
   $\bullet$~Guidelines for Using LLMs in SLRs\\
   \phantom{$\bullet$}~~\citep{thomas-2026-RAISE1}.};
\draw[ar] (informal.east) -- node[lbl,above,fill=none]{motivated} (5.0,5.5);
\draw[ar] ([xshift=3mm]stage1.north west) |- node[lbl,pos=0.60,below,yshift=-2mm]{motivated} (synth.west);
\draw[ar] ([xshift=2.5mm]informal.south) |- (grey.west);
\draw[ar] ([xshift=-2.5mm]informal.south) |- node[lbl,pos=0.85,below]{motivated} (screen.west);
\draw[ar] (stage1.north) |- (outcomes.west);
\draw[ar] (synth.south) -- node[lbl,pos=0.59,right]{inputs to} (stage2.north);
\draw[ar] (outcomes.east) -| ([xshift=-5mm]stage2.north);
\draw[ar] (screen.east) -| node[lbl,pos=0.82,right]{inputs to} (stage2.south);
\draw[ar] (grey.east) -| ([xshift=-5mm]stage2.south);
\draw[ar] (15.3,5.5) -- node[lbl,above,fill=none]{assessed by} (valid.west);
\end{tikzpicture}%
}}
\caption{Context and development process for the guidelines on evaluating and using GenAI for SLRs.}
\label{fig:Process}
\end{figure*}

Our goal is to propose guidelines for evaluating and using GenAI to support SLR planning and conduct. As mentioned earlier, these guidelines were motivated by our experiences as reviewers of papers reporting such studies. This motivation also led some of us to conduct two complementary studies that have informed the guidelines: one on using GenAI for paper screening \citep{Madeyski26LLM4SCREENLIT}\footnote{When we began to develop our recommendations, we used a preliminary draft of the paper.} and another on using LLMs for qualitative synthesis \citep{Pizard26}. \Cref{fig:Process} positions the guidelines within this broader body of work, showing the stages we carried out and the inputs used in their development.

\subsection{Background}
When we started to develop process recommendations, we were aware of two extremely important grey literature reports that have directly influenced our process recommendations:
\begin{enumerate}[noitemsep]
    \item Wagner et al., ``Towards Evaluation Guidelines for Empirical Studies involving LLMs'' \citep{wagner2025}\footnote{When we began our study, only the preliminary guidelines were available.}. %
    \item Thomas et al., ``Responsible use of AI in Evidence SynthEsis (RAISE 2026) 1: Recommendations for practice'' \citep{thomas-2026-RAISE1}\footnote{When we began our study we used an earlier version of this document \citep{Thomas-2024}. However, because the list of recommendations we used when developing our recommendations were unchanged in the different versions of the document, we reference the most recent version of the article throughout this article.}.
\end{enumerate}

\subsection{Stage 1: Literature Search}
Given the possibility of other useful papers, we conducted a \emph{Rapid Review (RR)} of the literature \citep{pizard-2025}, searching for any English-language papers that reported guidelines for evaluating the performance of GenAI tools and LLMs in the context of SLRs. The RR process, including the research questions, eligibility criteria, search strategy, the search strings, search results, data extraction, data synthesis and limitations, is discussed in Appendix~\ref{sec:DetailsOfRR}. Here we summarize the review protocol and its main results and conclusions.

In brief, the protocol was as follows.
\begin{itemize}
\item \emph{Research question.} Which papers report guidelines or recommendations for evaluating the performance of GenAI tools in the context of conducting SLRs?
\item \emph{Sources.} We searched a single digital library, SCOPUS, because it indexes a wide range of important journals and conferences across many domains, including the computer science and medical domains (where SLR guidance has often been developed first); it supports advanced search capabilities including Boolean search strings; and drawing candidate papers from one source reduces the risk of including multiple versions of the same study report in the candidate set, which could distort the synthesis. We did not restrict the search by discipline or by GenAI tool, and we checked sufficiency by verifying that the searches recovered a pre-defined set of known relevant papers (Tables~\ref{tab_knownpapers}).
\item \emph{Eligibility.} We included peer-reviewed papers (journal or conference), editorials and letters that presented guidelines or recommendations for evaluating or using GenAI tools in an SLR context and whose full text was available in English. We excluded papers available only on self-publication sites (such as arXiv), and papers that used SLR results to make practice recommendations (for example, clinical guidelines).
\item \emph{Search and validation.} We ran two SCOPUS searches: Search~1 (13/03/2025) targeted GenAI use in SLRs in general, and Search~2 (29/03/2025) broadened the terms to cover specific SLR tasks such as screening, data extraction, risk-of-bias assessment, and qualitative synthesis. The search strings (Tables~\ref{tab:search1} and~\ref{tab:search2}) were drafted by one author and refined by another, and their effectiveness was validated against the known papers. Search~1 recovered all 13 papers in our known set at that stage, and Search~2 recovered all 18 known papers before it was restricted to guideline terms, which indicates that a single-source SCOPUS search was sufficient for our goal of covering the important themes rather than retrieving every relevant paper.
\end{itemize}

We analysed guidelines obtained from six papers that reported general guidelines for using GenAI tools to support academic research, since any research involving an SLR must also comply with them. From these, we identified the six major guidelines reported in \Cref{tab:Guidelines} (from now on, these are referred to as G1 to G6). G6 specifically addresses peer review, so only the first five guidelines are of direct relevance to our process recommendations. These guidelines provided a legal and ethical framework within which to develop our process recommendations. The integrated guidelines are not completely independent. Human accountability is an overriding requirement, while guidelines G2 to G6 represent specific means of ensuring human accountability.

\begin{table*}[!t]
\caption{General Guidelines for Using GenAI tools for Academic Research}
\label{tab:Guidelines}
\footnotesize
\begin{tabular}{p{0.4cm}p{8cm}p{7.5cm}}
\toprule
\textbf{ID} & \textbf{Guideline} & \textbf{References} \\
\midrule

    G1 & Humans are accountable for AI tool use in terms of outcome validity and adherence to legal and scientific ethics. & \citealp{Veiga-2025,uribe-2025,kim-2023,perkins-2024guidelines} \\
    G2  & All AI-assisted outputs should be reviewed by humans.  & \citealp{Veiga-2025,inam-2024,kim-2023,lubowitz-2024,perkins-2024guidelines}\\
    G3  &  Do not assign authorship to AI tools. 
    & \citealp{Veiga-2025,uribe-2025,inam-2024,kim-2023,lubowitz-2024,perkins-2024guidelines} \\
    G4   &  All AI tool use, with the exception of editing and refining text, should be fully reported.
   & \citealp{Veiga-2025,uribe-2025,inam-2024,lubowitz-2024,kim-2023} \\
   G5 & AI tools should not create the full text of reports, nor create, alter or manipulate images, data,  citations, results, conclusions or recommendations. & \citealp{Veiga-2025,inam-2024,kim-2023,lubowitz-2024}\\
   G6 & AI tools should not be used for peer review. & \citealp{inam-2024,lubowitz-2024}\\
\bottomrule
\end{tabular}
\end{table*}

Subsequent to our RR, \emph{Communications of the ACM} published a paper by \citet{nahar-2025} describing how computer science conferences are reacting to the use of GenAI tools in scientific writing. Apart from G6, their recommendations mirror the guidelines in \Cref{tab:Guidelines}. Nahar et al. note that risks related to exposing sensitive information during peer review can be mitigated by configuring LLM settings to prevent models from learning or retaining user-provided data. This suggests that G6 is over restrictive in the context of GenAI tools, although editors would need to ensure that appropriate safeguards are in place\footnote{However, configuring model settings may not be a sufficient safeguard for Agentic AI tools: whether built on GenAI models or not, they act autonomously and may go beyond their instructions.}. 

After completing our RR, we concluded that:
\begin{enumerate}
    \item There were no readily available English language research papers indexed by SCOPUS (up to 29/03/2025) that addressed our goal of producing recommendations for using GenAI tools to support SLRs.
    \item The literature screening recommendations proposed by \citet{Madeyski26LLM4SCREENLIT} were at a lower level of abstraction than \citeauthor{thomas-2026-RAISE1}'s recommendations \citeyearpar{thomas-2026-RAISE1}, and recommendations from the two articles could not be easily integrated. 
    \item Although we would continue to develop general recommendations for evaluating and using GenAI tools for conducting SLRs, more detailed recommendations (e.g., similar to those produced by \citet{Madeyski26LLM4SCREENLIT} and later by \citet{Pizard26}) related to different SLR tasks (including risk of bias assessment, data extraction, qualitative synthesis, strength of evidence assessment) would also be of value. Detailed task-specific investigations of all SLR processes are beyond the scope of our current study, but we produced briefing notes describing the individual tasks included in each of these SLR processes to assist the development of preliminary task-related recommendations.
    \item The paper by \citet{woelfle-2022} provided very valuable insight into the problems associated with using GenAI tools to support quality assessment.
\end{enumerate}
Our RR deliberately targeted papers that report guidelines or recommendations, as stated in its research question.
It was not designed to aggregate the empirical literature on how GenAI tools actually perform on SLR tasks.
We address that empirical evidence separately.
\Cref{sec:FailureModes} synthesizes the failure modes reported in empirical studies --- including our companion studies on literature screening \citep{Madeyski26LLM4SCREENLIT} and qualitative synthesis \citep{Pizard26} --- and maps them to the SLR tasks they affect.
We have used this synthesis as an evidence backbone for our recommendations.
Many of these empirical studies were also among the known papers our searches recovered (Tables~\ref{tab_knownpapers} and~\ref{tab_knownpapersnot}).
Although the searches were run in March 2025, our recommendations do not rest on the guideline literature frozen at that date: we continued to track the field, verifying that successive versions of the RAISE recommendations left unchanged the list we had used \citep{thomas-2026-RAISE1}, positioning our work against guidance published after the searches --- \citet{baltes-2025} and the \citet{farotimi-2025} editorial (\Cref{tab:related-guidelines}) --- and cross-checking our integrated guidelines against the conference policies analysed by \citet{nahar-2025}.

These conclusions and \Cref{tab:Guidelines} are the final outcomes of our rapid review. The approach used to develop recommendations addressing the use of GenAI tools to support SLRs is presented in \Cref{sec:DevelopingRecommendations}.

\subsection{Stage 2: Researcher Roles and Inputs}\label{sec:DevelopingRecommendations}
The general guidelines in \Cref{tab:Guidelines} do not address the specific process issues related to using GenAI in the context of SLRs. However, they make it clear that our recommendations must recognize that academic research involving GenAI tools, including conducting SLRs, requires human oversight.
In addition, the results of our RR made it clear that producing an answer to our research question could not rely on published research and a different approach was needed in order to develop useful recommendations for evaluating GenAI tools in the context of SLRs.

We considered it important that our recommendations should support two different researcher roles in the context of SLR-related research:
\begin{enumerate}
\item \emph{Tool Evaluators} whose primary interest is assessing the performance of SLR tools. Such researchers will need to consider amending their standard research practices when studying GenAI tool capability. In this study, we restrict ourselves to researchers undertaking \emph{independent} assessments of GenAI tools. Evaluators working as part of an AI tools development team should refer to \citet{thomas-2025-RAISE2}.
\item \emph{Reviewers} whose primary interest is conducting high-quality SLRs. In order to use such tools, researchers will need to know how to conduct any necessary pre-deployment validation of GenAI tools, to assess the reliability of tool outputs and to maintain human oversight. We do not include any discussion of issues related to the selection of specific GenAI tools.  For tool selection advice refer to \citet{Thomas-2026-RAISE3}.
\end{enumerate}
In addition, useful recommendations should support both SLR article reviewers and subsequent readers.

As well as complying with the general recommendations in \Cref{tab:Guidelines}, we wanted our recommendations to be consistent with the Responsible use of AI in Evidence SynthEsis (RAISE 2026) guidelines \citep{thomas-2026-RAISE1} and the guidelines for empirical SE studies \citep{wagner2025}. In addition, we wanted to take advantage of the results reported in three other papers:
\begin{enumerate}
    \item \citet{Madeyski26LLM4SCREENLIT} discuss the selection and interpretation of performance metrics used to evaluate GenAI tools when screening titles, abstracts and keywords (TAK) of candidate primary studies for eligibility. This article informed our initial recommendations, and a later revision, which extended its recommendations to full-text screening and (in principle, pending empirical validation) to data extraction, was used to refine them.
    \item \citet{woelfle-2022} discuss in detail a study investigating the use of GenAI tools to evaluate the quality and rigour of SLR reports using three different evaluation instruments.
    \item \citet{Pizard26} discuss the issues involved when using GenAI tools to support qualitative synthesis.
\end{enumerate}
\noindent We also found another relevant SE study by \citet{baltes-2025}, which we used to validate our recommendations.

There are also two other RAISE-related studies that cover issues not explicitly addressed in our recommendations:
\begin{enumerate}
    \item \citet{thomas-2025-RAISE2} which discusses issues relating to building GenAI tools.
    \item \citet{Thomas-2026-RAISE3}, which discusses the selection of tools, summarizes the current capabilities of a number of existing tools.
\end{enumerate}
Although these papers are of value for researcher teams producing and evaluating GenAI tools for SLRs, they are outside the scope of our recommendations.

\subsection{Stage 2: Method Used to Develop Recommendations}
To develop role-related recommendations, we conducted \emph{Thought Experiments} on the process issues we would need to address when planning, conducting and reporting evaluations of GenAI tool performance. A thought experiment is an organized exercise of the mind in which a hypothetical situation is imaginatively devised and explored to examine, assess, refine, or illuminate ideas, theories, or propositions \citep{Stuart2018}. This method has been used previously in software engineering \citep[see, e.g.,][]{Melegati2024,Ko2019, Monperrus2014}, and several authors describe it as a useful complement to purely empirical research \citep[e.g. see][]{Genova202}. 

The advantages of a thought experiment based approach are:
\begin{enumerate}
    \item Considering the processes used for planning, conducting and reporting, should provide a common level of abstraction for recommendations.
    \item The two roles and the three process stages provide a framework for assessing the completeness of the recommendations.
    \item A thought experiment would allow us to utilize our own experience of SE experimentation and conducting SE SLRs to identify process recommendations that are proactive rather than retrospective. 
    \item A process-based presentation identifying the difference between the standard process for tools evaluation and SLR conduct would provide a practical context for researchers wanting to use the recommendations. It would also make clear the scope and limitations of the guidelines. If researchers find they require a process step not covered by the guidelines, they could use the issues raised in \Cref{sec:GuidelineBackground} or their own thought experiment to assess whether there are any GenAI tool or SLR requirements they need to be aware of in the context of the novel process step.
    \item SE researchers with experience of conducting SLRs should be familiar with the use of thought experiments, because it is the method used to develop SLR protocols.
\end{enumerate}

We focused our thought experiment on identifying how standard evaluation and SLR processes would need to be changed to incorporate GenAI tools. We decided to conduct a thought experiment for each researcher role because, although the roles overlap, there are significant differences. %

Our RR of existing guidelines identified that a general problem with developing recommendations from existing guidelines was that both the existing guidelines we found and other related articles treated the evaluation at different levels of abstraction. For example, our existing recommendations related to screening the literature \citep{Madeyski26LLM4SCREENLIT} and qualitative synthesis \citep{Pizard26} provided some detailed phase-related recommendations, while other guidelines \citep[i.e.,][]{wagner2025,thomas-2026-RAISE1} were generally at a higher level of abstraction. In order to support SLR researchers, we felt that we needed to consider the specific issues related to all the main SLR tasks. 

To do this, we developed briefing notes detailing the tasks required to conduct risk of bias assessment\footnote{Often referred to as quality assessment in the SE literature.} and strength of evidence assessment (also referred to as certainty of evidence). These are presented in Appendix~\ref{sec:SLRProcessDetails}. We also include briefing notes on literature screening (both abstract-based and full-text), based on \citet{Madeyski26LLM4SCREENLIT}; on data extraction, which draws on that study's principled extension of its metric recommendations to extraction tasks (pending empirical validation); and on qualitative synthesis, based on \citet{Pizard26}. This allowed us the opportunity to check that all appropriate issues were included in our recommendations. We exclude discussion of the initial literature search process because it usually depends on the capabilities of external search engines and data libraries, which are outside the control of individual researchers or research groups.

\subsection{Stage 2: Context and Scope of the Thought Experiments}\label{sec:Context}
This section discusses the starting point for our thought experiments by highlighting some of the similarities and differences between the two researcher roles. 

For the evaluator role, we adopted the viewpoint of researchers with experience in empirical research methods and GenAI tools, intending to undertake independent evaluations of GenAI tools' performance. We assumed that any rigorous evaluation of a GenAI tool requires an experimental framework consistent with  best practice that: 
\begin{itemize}[noitemsep]
    \item Allows the performance of a GenAI tool to be compared with current practice.
    \item Does not unfairly favour either the GenAI tool or current practice.
    \item Allows limitations of the tool and risks associated with tool use to be identified, as well as tool benefits and opportunities.
    \item Is reported at a level of detail that supports repeatability and meta-analysis.
\end{itemize}

Evaluators can have many different goals, including but by no means limited to:
\begin{itemize}[noitemsep]
    \item Comparative evaluation and benchmarking of different GenAI tools.
    \item Comparative evaluation of GenAI tools and other more deterministic SLR support tools.
    \item Evaluation of different prompting strategies.
    \item Evaluation of methods to predict the performance of GenAI tools in the context of a specific SLR.
\end{itemize}

In addition, evaluators need to be ready to assess new GenAI tools, intended to address SLR tasks that are beyond the capability of existing tools. In particular, tools intended to address qualitative analysis and synthesis, and tools intended to support (or assist) the evaluation of the reliability of primary studies and the strength of evidence supporting SLR recommendations.

For the reviewer role, we adopted the viewpoint of members of a review team looking to adopt GenAI tools. We expect members of a review team to have experience with the basic SLR process standards and reporting guidelines, and a general understanding of the nature of GenAI tools. We assumed that the review team would want to ensure that:
\begin{itemize}
    \item Any use of GenAI tools would conform with all relevant legal and ethical requirements for academic use of GenAI tools in terms of ensuring human oversight and full disclosure of use, as outlined in \Cref{tab:Guidelines}.
    \item The use of GenAI tools did not compromise SLR validity.
    \item Any use of GenAI tools to replace a human researcher during the conduct of an SLR was justified by a well-conducted validation of tool performance, which provides empirical support for tool adoption. Note, undertaking a validation intended to decide whether to use an AI tool to replace a human researcher is an example of an important overlap between the evaluator and reviewer roles. 
\end{itemize}

For both roles, our thought experiments were aimed at identifying any changes researchers would need to make to their study planning, conduct and reporting process when using or evaluating GenAI tools. However, we have assumed that any research study is being undertaken in the context of academic research where the general ethos supports free and open reporting of research studies. We have not considered the issues involved when software engineers are working in any industry context and may have a requirement to protect their company's commercial interests. For industry-based researchers, the basic planning and conduct recommendation may be of value, but some of the reporting requirements might be unnecessary.

In addition to considering the process changes needed to support GenAI tool evaluation and use of GenAI tools, we also wanted to explore issues related to the individual SLR processes. SLR processes have different purposes and characteristics, which any GenAI tools aimed at automating specific SLR processes need to properly address. For this reason, we constructed briefing notes that describe the different processes and their sub-tasks and how GenAI tools could support them. Our briefing notes are reported in Appendix~\ref{sec:SLRProcessDetails} and provide an important addition to our recommendations for evaluators and reviewers.

We did not produce any briefing notes related to preliminary SLR planning, the search for candidate primary studies, quantitative analysis, or production of SLR reports, because:
\begin{itemize}
\item Preliminary planning includes developing research questions, search strings and eligibility criteria. We discuss these topics in Appendix~\ref{sec:ToolTestingAndValidation}. \citet{Wang2023} reported an evaluation of automating the construction of Boolean queries. One of the  prompting methods Wang et al. tested was a designed guided prompt based on the method human researchers use to create a prompt. This approach is  similar to the approach we used to develop our technical briefing notes. However, in our opinion, automating these initial planning tasks may not require a formal validation, as long as such GenAI tool use is fully reported, and reviewers take responsibility for how any GenAI tool output is used.
\item Searching for candidate studies relies on the facilities provided by digital libraries, which are outside the control of SLR researchers. 
\item For quantitative analysis, the use of meta-analysis is based on existing statistical methods and GenAI tools are not needed.
\item When using GenAI tools to assist the production of SLR reports, as long as assistance is restricted to linguistic and grammatical improvements and all such help is overseen and assessed by human researchers, it is sufficient to report such usage without any additional tests or validation. 
\end{itemize}

\section{Results}\label{sec:SummaryOfThougthExperimentResults}
 
This section reports the outcomes of the thought experiments for each SLR task, presents the GUEST recommendations for evaluators and reviewers, and links the recommendations to failure modes reported in empirical studies. 

\subsection{Thought Experiment Outcomes by SLR Task}
In the context of evaluation studies aimed at benchmarking GenAI tool performance, it is critical to properly address the risk of data contamination associated with retrospective analyses of previously published SLRs. Recommendations we found in the literature emphasized that prospective studies are the most reliable means of addressing this risk. Fortunately, reviewers using GenAI tools to assist with performing an SLR are in an excellent position to undertake prospective studies of GenAI performance. Thus, SE researchers who usually have experience in both SE tool evaluation and using  SLRs as an integral part of their standard research process are in a particularly good position to contribute to evaluating GenAI tool support for SLRs. Furthermore, the need for prospective studies confirms the value of SE-oriented recommendations for GenAI tool performance studies for conducting GenAI-supported SLRs that consider both roles.

Our briefing notes in Appendix~\ref{sec:SLRProcessDetails} highlight major differences between the types of SLR tasks that would impact both reviewers and evaluators:
\begin{itemize}
    \item Screening candidate primary studies for inclusion in an SLR involves multiple repetitions of the same task, often involving processing thousands of individual research papers. It is the main scenario in which a review team might want to perform a full validation based on a random selection of candidate studies with the aim of deciding whether a human researcher could be replaced. The literature screening process is repetitive, time-consuming and error-prone, and represents a clear opportunity for GenAI tools to increase SLR efficiency even if validation samples are relatively large, as demonstrated by \citet{Cao2024}. The use of GenAI tools in the context of an ongoing SLR can deliver prospective assessments of GenAI tool performance while reducing the risk of data contamination. However, the risk is not eliminated if evaluators use multiple GenAI tools and multiple prompt strategies on the same data sample.
    \item Data extraction, including classification of primary studies for mapping studies, is also repetitive, but is also more varied, with major differences between identifying the numerical values required for quantitative primary studies and the textual data required for qualitative primary studies. Also, data extraction is based only on the available primary studies, and while the number of primary studies might be a relatively large number for mapping studies, it will probably be relatively small for quantitative and qualitative SLRs. Thus, apart from mapping studies, the opportunity for reducing human researcher effort is likely to be relatively small, while the effort required to construct GenAI tool prompts and related data collation processes could be considerable.
    \item Qualitative synthesis is problematic because there are many different methods of qualitative analysis and synthesis, and they are less well defined than quantitative methods. Using interpretive approaches, it may be difficult for different human researchers to agree on what constitutes a valid synthesis, making it difficult to establish a baseline for comparison with a GenAI-based synthesis. In general, outcomes are extremely varied, including, for example, textual descriptions of identified themes, models of the relationships between influencing factors, or lists of risks and benefits. Only simple lists lend themselves easily to the use of quantitative agreement metrics, without which judging the relative performance of GenAI tools and human researchers objectively is difficult.
    \item Risk of bias (RoB), which is often referred to as quality assessment in the SE literature, involves assessing the methodology used in each primary study to identify whether there are any serious weaknesses that could cast doubts on the study results. Primary studies with serious methodological weaknesses may be excluded from the SLR synthesis. RoB assessment is similar to the process researchers use when they are asked to review manuscripts submitted to conferences or journals. Although unnecessary for mapping studies, RoB is required when reviewers intend to synthesize primary study findings. Appendix~\ref{sec:RiskOfBiasAssessment} identifies that in the case of SE, the problems associated with automating RoB relate to identifying appropriate evaluation instruments for SE primary studies and to the expertise required to understand and use those criteria. In terms of researchers conducting SLRs, the individual sub-tasks discussed in~\ref {sec:RiskOfBiasAssessment} highlight the opportunity for GenAI tools to act as an additional research assistant, for example, by highlighting primary study text that addresses a specific criterion, or making suggestions about the value of a specific criterion.
    \item Strength of evidence assessment involves assessing whether the evidence supporting \emph{a specific SLR finding} is trustworthy. Noting that SLRs are retrospective studies, the strength of evidence assessment is similar to identifying and investigating possible confounding factors, which is required when assessing the limitations of a quantitative primary study or the search for alternative viewpoints in the context of a qualitative study. Like the RoB assessment, this process is not required for mapping studies. However, it is essential for assessing the reliability of findings derived by synthesizing results from multiple primary studies.
\end{itemize}

Overall, the briefing notes suggest that GenAI tools are well-suited to automating mapping study processes. With well-defined eligibility criteria and well-defined research questions, it would seem possible to fully automate literature screening and primary study classification for mapping studies. Furthermore, mapping studies do not usually require risk of bias (RoB) assessment or strength of evidence assessment, which are more complex processes. Even so, for a \emph{systematic} mapping, providing traceability between study goals and study findings and supporting reproducibility are essential, so GenAI tools would need to provide a final report that conforms with mapping study reporting requirements \citep{kitchenham-2023}. 

For formal SLRs that require RoB assessment, detailed data extraction, data syntheses and strength of evidence assessment, the prospect of full automation is poor. In the context of data extraction for quantitative and qualitative synthesis, the challenge posed by the variety of different study types typically found in SE SLRs makes full automation unlikely. RoB and the strength of evidence rely on knowledge not contained in any specific primary study text. In the case of RoB, this process requires both knowledge of the methodological weaknesses of different empirical methods and the experience to make inferences from what is not reported, as well as what is reported in individual primary studies. In the case of strength of evidence assessment, reviewers need to make an assessment of the reliability of individual findings obtained from a set of primary studies. This requires a rigorous assessment of the factors that could bias findings, even from a group of good-quality primary studies. For example, a group of laboratory experiments may all be of good methodological rigour, but may be based on student participants or relatively simple tasks, so experimental findings might not generalize to practitioners and industry-scale tasks. 

In terms of researchers conducting SLRs, the individual sub-tasks for complex tasks such as RoB assessment, qualitative synthesis and strength of evidence assessment offer the opportunity for GenAI tools to act as an additional research assistant, for example by highlighting primary study text snippets that address a specific issue and/or issues related to a specific evaluation criterion, or suggesting an answer to an evaluation question, or proposing a qualitative synthesis finding. In such a scenario, the purpose of the GenAI tool usage would not be to save research effort, but:
\begin{enumerate}[noitemsep]
    \item To provide independent validation of the task performance of the human researchers and, thus increase the reliability of the SLR findings.
    \item To provide a prospective assessment of GenAI tool performance on more difficult SLR processes.
\end{enumerate}

\subsection{GUEST Recommendations}\label{sec:Recommendations}
This section presents the detailed recommendations for evaluators and reviewers arising from integrating the results of our role-based, process-oriented thought experiments (see Appendix~\ref{sec:ThoughExperimentResults}) with the issues identified by our briefing notes (see Appendix~\ref{sec:SLRProcessDetails}). Our thought experiments make it clear that evaluators interested in studying the GenAI tools and reviewers wanting to use GenAI tools to support the conduct of a specific SLR share a number of process-related issues, particularly issues related to planning and reporting the use of GenAI tools. To avoid repetition, we organized our recommendations into three groups: those common to both roles and those specific to each role. We report the more general recommendations first, in terms of planning, conducting, and reporting recommendations, followed by the recommendations related to individual SLR processes.

Our recommendations are of four kinds, summarized in~\Cref{tab:RecCategories}, and we separate them so that readers can tell which ones are specific to SLRs.
The first kind covers SLR practice changes due to GenAI tool use.
Such changes arise in the tasks that belong to systematic reviews in particular: literature screening, data extraction, qualitative synthesis, and risk-of-bias and strength-of-evidence assessment.
These recommendations are the distinctive contribution of this paper, and most of them form the SLR-process recommendations below.
The second kind restates good practice that is already well established for secondary studies in general, with or without GenAI.
Examples are documenting exclusion decisions and guarding against lost evidence.
We treat these as known concerns for secondary studies and align them with established reporting guidance: SEGRESS \citep{kitchenham-2023} for software engineering, and PRISMA 2020 \citep{page-2021} for evidence synthesis in a broader context.
Some of these concerns also take an SLR-specific form.
Guarding against lost evidence is a general aim of any secondary study, but measuring and limiting it when a GenAI tool screens the literature is specific to our setting: it needs cost-sensitive metrics that weight a missed relevant study above a wrongly retrieved one, which our companion study \citep{Madeyski26LLM4SCREENLIT} develops in detail (see Scr1--Scr3 and ScrRev1--ScrRev3 below).
The third kind is generic to almost any empirical evaluation of a GenAI tool, not just to SLRs.
Examples are justifying the choice of tool, reporting the model version, its parameters and the prompts used, and using an open tool as a baseline.
We align these with the emerging SE guidance on empirical studies that use LLMs: \citet{baltes-2025}, \citet{wagner2025}, and RAISE \citep{thomas-2026-RAISE1}.
The fourth kind relates to good practice for any evaluation study, whether or not it involves a GenAI tool.
This includes ensuring that objectives, goals and research questions are formally stated, using appropriate evaluation methods, and reporting evaluation results with reproducibility in mind, for example by providing replication packages and open data \citep{Huotala25repl}.
There are many academic papers discussing such issues.
In~\Cref{tab:RecCategories} we cite only a few, including studies in which we have been directly involved. 
Most of our planning, conduct, and reporting recommendations are of the second, third or fourth kind.
We include them for completeness, so that the guidance can be used on its own, not as a claim to novelty.
The SLR-process recommendations, where our distinctive contribution lies, are mostly of the first kind.
In \Cref{tab:RecCategories} the four kinds are labelled (i) to (iv).
\Cref{tab:GUESTPlanConRep} and \Cref{tab:GUESTProcess} annotate each recommendation with its kind.

\begin{table*}[!t]
\caption{Classification of our recommendations: which are specific to SLRs, and which existing guidance we align the general ones with.}
\label{tab:RecCategories}
\footnotesize
\begin{tabular}{@{}>{\raggedright\arraybackslash}p{2.0cm}p{10.2cm}>{\raggedright\arraybackslash}p{4.35cm}@{}}
\toprule
\textbf{Category} & \textbf{What it covers (examples)} & \textbf{Related existing guidance} \\
\midrule
(i) SLR practice changes due to GenAI tool use & Tasks particular to SLRs: literature-screening metrics, data-extraction performance assessment, qualitative synthesis, and risk-of-bias and strength-of-evidence assessment. The distinctive contribution of this paper. & --- (this paper; see also our companion studies \citep{Madeyski26LLM4SCREENLIT,Pizard26}) \\
\addlinespace[2pt]
(ii) General secondary-study practice & Good practice for any secondary study, with or without GenAI: documenting exclusion decisions and guarding against lost evidence (general aim; SLR-specific control under screening metrics). & SEGRESS \citep{kitchenham-2023}; PRISMA 2020 \citep{page-2021} \\
\addlinespace[2pt]
(iii) General GenAI evaluation & Practice generic to almost any empirical evaluation of a GenAI tool: justifying tool choice, reporting model version, parameters and prompts, and using an open tool as a baseline. & \citet{baltes-2025}; \citet{wagner2025}; RAISE \citep{thomas-2026-RAISE1} \\
\addlinespace[2pt]
 (iv) General evaluation practice & Good practice for almost any evaluation study whether or not GenAI-based: defining goals and hypotheses, using an appropriate study design and data analysis methods, and supporting reproducibility, including replication packages and open data. & \citet{kitchenham-2002preliminaryguidlines,kitchenham-2019}; \citet{ralph-2021acm}; \citet{Huotala25repl} (replication packages) \\
\bottomrule
\end{tabular}
\end{table*}

\subsubsection{Planning, Conduct, and Reporting Recommendations}
These recommendations are based on an assessment of the different roles.
\Cref{tab:GUESTPlanConRep} lists them, grouped into planning, conduct, and reporting, together with the role each recommendation applies to and its kind.

\begin{table*}[!t]
\caption{GUEST planning, conduct and reporting recommendations. ID prefixes: Plan = planning, Con = conduct, Rep = reporting. E~or~R~marks a recommendation for evaluators or reviewers only.}
\label{tab:GUESTPlanConRep}
\footnotesize
\begin{tabularx}{\textwidth}{@{}lXll@{}}
\toprule
\textbf{Id} & \textbf{Recommendation} & \textbf{Role} & \textbf{Kind} \\
\midrule
\rowcolor{gray!20}\multicolumn{4}{l}{\emph{Planning}} \\ \midrule
Plan1 & Justify the choice of GenAI tools and critically appraise them. & Both & (iii) \\
Plan2 & Justify the choice of study design. & Both & (iv) \\
Plan3 & Define the methods to be used to evaluate GenAI tool performance. & Both & (iv) \\
Plan4 & Ensure that the evaluation processes are designed to detect tool errors and to minimize risks to validity from built-in tool biases and from data contamination caused by repeated analyses of the same dataset within the study. & Both & (iii) \\
Plan5 & Identify the data needed to assess tool performance and plan the data collection, necessary data labelling or annotation, storage processes and any related quality assurance procedures. & Both & (iv) \\
Plan6 & Base comparisons on results from at least two human researchers. & Both & (ii) \\
PlanE1 & Clearly define evaluation goals, research questions, and hypotheses. & Evaluators & (iv) \\
PlanE2 & Use an open GenAI tool as a baseline. & Evaluators & (iii) \\
PlanE3 & Identify any assumption made at the start of the evaluation (e.g. define what input data is required to initiate the task). & Evaluators & (iv) \\
PlanE4 & Justify the use of existing publicly available datasets, given that the datasets, and published analyses of them, may already be part of a GenAI tool's training data. & Evaluators & (iii) \\
PlanE5 & Plan an appropriate data analysis, including meta-analysis if the evaluation aims to summarize results from multiple SLRs. & Evaluators & (iv) \\
PlanR1 & Consider using GenAI tools as an additional research assistant to assist in the specification of research questions and eligibility criteria. & Reviewers & (i) \\
PlanR2 & Specify all the SLR processes and tasks to be supported by GenAI tools. & Reviewers & (i) \\
PlanR3 & Adopt a case-study-based approach to using GenAI tools to provide prospective evaluations of tool performance. & Reviewers & (i) \\
\addlinespace[1pt]
\rowcolor{gray!20}\multicolumn{4}{l}{\emph{Conduct}} \\ \midrule
Con1 & Ensure that any deviations from the plan are identified and reported. & Both & (iv) \\
Con2 & Ensure data collection processes are used correctly. & Both & (iv) \\
Con3 & For GenAI case studies within SLRs, ensure that all human researcher effort required to support GenAI tool use is reported. & Both & (i) \\
Con4 & Ensure human researchers review all GenAI tools' outcomes and take full responsibility for how the outcomes are used. & Both & (i) \\
\addlinespace[1pt]
\rowcolor{gray!20}\multicolumn{4}{l}{\emph{Reporting}} \\ \midrule
Rep1 & Report GenAI tool identification details (tool(s), platform(s), version(s) and date used) & Both & (iii) \\
Rep2 & Report the prompt strategy, the prompt development process, the prompts used to deliver the results reported in the study, and the date when the final prompt was executed. & Both & (iii) \\
Rep3 & Report all GenAI tool settings and parameters. & Both & (iii) \\
Rep4 & Accurately assess and report all costs of GenAI tool use. & Both & (iii) \\
Rep5 & Report the full data collection process including data collection, necessary labelling or annotation, collation, storage and quality assurance. & Both & (iii) \\
Rep6 & Report the results and conclusions of the evaluation results with process reproducibility and future meta-analysis in mind. In particular, support open access to data, analysis algorithms and results. & Both & (iv) \\
Rep7 & Report any study limitations and constraints, including any outstanding concerns regarding the reliability of GenAI tools' results. & Both & (iii) \\

RepE1 & When making recommendations aimed at conducting SLRs, explain how the GenAI tool(s) can be incorporated into current SLR processes. & Evaluators & (i) \\
RepE2 & When making recommendations aimed at conducting SLRs, explain how the GenAI tool use complies with requirements for human oversight. & Evaluators & (iii) \\
RepE3 & Report any additional use of GenAI tools to plan, conduct or report the tool evaluation process. & Evaluators & (iii) \\
RepE4 & Report any financial interest or any other special issues, and how bias was avoided. & Evaluators & (iv) \\
RepR1 & Follow appropriate SLR reporting guidelines (i.e., SEGRESS \citep{kitchenham-2023} or PRISMA 2020 \citep{page-2021}). & Reviewers & (ii) \\
RepR2 & Report any additional use of GenAI tools to report the SLR or assist with managing SLR data. & Reviewers & (iii) \\
\bottomrule
\end{tabularx}
\end{table*}

\subsubsection{Recommendations Related to SLR Processes}
Based on analyses of specific SLR processes.
\Cref{tab:GUESTProcess} lists the recommendations for the individual SLR processes: literature screening (title, abstract, keyword (TAK) and full text), data extraction, qualitative synthesis, and risk of bias (RoB) and strength of evidence (SoE) assessment.
For literature screening, our companion study develops these principles into ten operational recommendations (R1--R10), differentiated by benchmarking and deployment study types \citep{Madeyski26LLM4SCREENLIT}.
We are not yet aware of SE-specific empirical evidence on GenAI tool performance for RoB and SoE assessment, so those recommendations are preliminary (see \Cref{sec:OpenQuestions} for the main open research questions).

\begin{table*}[!t]
\caption{GUEST SLR-process recommendations. ID prefixes: Scr = literature screening, Ext = data extraction, Syn = qualitative synthesis, A = risk-of-bias and strength-of-evidence assessment. ScrRev and RevA mark recommendations for reviewers only.}
\label{tab:GUESTProcess}
\fontsize{8.0}{9.6}\selectfont%
\begin{tabularx}{\textwidth}{@{}lXll@{}}
\toprule
\textbf{Id} & \textbf{Recommendation} & \textbf{Role} & \textbf{Kind} \\
\midrule
\rowcolor{gray!20}\multicolumn{4}{l}{\emph{Literature screening}} \\  \midrule
Scr1 & Use Matthews correlation coefficient (MCC) to ensure results are better than chance.\textsuperscript{a} & Both & (i) \\
Scr2 & Use WMCC to properly account for cost differences associated with False Positives (FPs) and False Negatives (FNs). & Both & (i) \\
Scr3 & Report full confusion matrix counts to support meta-analysis and re-analysis of results.\textsuperscript{a} & Both & (i) \\
\rowcolor{gray!20}\multicolumn{4}{l}{\emph{Reviewer-based validation studies undertaken during the conduct of an SLR}} \\
 \midrule 
ScrRev1 & Planning: Acceptable performance levels for GenAI tools (or the process to set them) should be identified and justified prior to the validation exercise. & Reviewers & (i) \\
ScrRev2 & Planning: Have appropriate contingency plans to address the validation results. & Reviewers & (i) \\
ScrRev3 & Conduct and Reporting: Provide confidence intervals for the selected performance metrics based on resampling without replacement. & Reviewers & (i) \\
\rowcolor{gray!20}\multicolumn{4}{l}{\emph{Data extraction}} \\  \midrule
Ext1 & Plan for data extraction to include descriptive textual information such as study type, study goals, study setting, publication date, first author, and publication type. & Both & (ii) \\
Ext2 & For mapping studies requiring classifying primary studies against pre-defined categories, use MCC (with an extension if multiple outcomes or multiple categories are required). & Both & (i) \\
Ext3 & For mapping studies requiring classifying primary studies against pre-defined categories, avoid WMCC unless there is an explicit justification for assuming differential costs for data extraction errors. & Both & (i) \\
Ext4 & For quantitative data extraction, correctness metrics should use the values explicitly reported in the primary study text as the gold standard whether the extraction was performed by human researchers, or by a tool. & Both & (iv) \\
Ext5 & For more complex data extraction, such as that supporting qualitative analysis, define which sub-tasks in the data extraction process are being automated.\textsuperscript{b} & Both & (i) \\
Ext6 & For more complex data extraction, such as that supporting qualitative analysis, define the starting point of the automated process in terms of knowledge about primary studies.\textsuperscript{b} & Both & (i) \\
Ext7 & For more complex data extraction, such as that supporting qualitative analysis, specify how comparisons between human researchers and GenAI tools will be conducted and assessed. & Both & (i) \\
\rowcolor{gray!20}\multicolumn{4}{l}{\emph{Qualitative synthesis}} \\  \midrule
Syn1 & Specify and report the qualitative synthesis method being adopted and explain why it is suitable given the SLR research questions and the SLR primary studies. & Both & (ii) \\
Syn2 & Specify and report the criteria used to assess GenAI tool performance, given the qualitative synthesis method being adopted. & Both & (i) \\
Syn3 & Specify and report how the synthesis of GenAI tools will be compared with the synthesis provided by human researchers, in particular, how relative performance will be measured and assessed. & Both & (i) \\
\rowcolor{gray!20}\multicolumn{4}{l}{\emph{Risk of bias and strength of evidence}} \\  \midrule
A1 & Define which sub-tasks are being automated.\textsuperscript{b} & Both & (i) \\
A2 & Define the information required as input to each sub-task and how it is obtained.\textsuperscript{b} & Both & (i) \\
A3 & Specify how the assessments made by GenAI tools for each sub-task will be compared with those made by human researchers and how the relative performance will be measured. %
& Both & (i) \\  %
\rowcolor{gray!20}\multicolumn{4}{l}{\emph{Qualitative synthesis, risk of bias, and strength of evidence --- reviewers}} \\  \midrule
RevA1 & Restrict GenAI tool use to providing additional process validation of tasks performed by human researchers. & Reviewers & (i) \\
\bottomrule
\multicolumn{4}{p{0.92\textwidth}}{\footnotesize \textsuperscript{a}\;Although MCC and full confusion-matrix reporting are long established in general classifier evaluation, we classify Scr1 and Scr3 as SLR-specific.
Applying them to SLR screening evaluations is not established practice: conventional metrics such as accuracy and $F_1$ mislead under the severe class imbalance of screening (\Cref{sec:FailureModes}), and our companion study develops the metric guidance for this setting in detail \citep{Madeyski26LLM4SCREENLIT}.} \\
\multicolumn{4}{p{0.92\textwidth}}{\footnotesize \textsuperscript{b}\;We classify Ext5, Ext6, A1, and A2 as SLR-specific rather than general practice.
They presuppose automation, so they do not restate established secondary-study practice, and the decomposition they require is of the specific SLR task: which parts of it are delegated to a GenAI tool, and what knowledge of the primary studies the automated process starts from.}
\end{tabularx}
\end{table*}

The general SLR process changes for reviewers that are implied by these recommendations are summarized in \Cref{fig:SLRProcessChanges} (Appendix~\ref{sec:TestingANdEvaluatingGenAITools}).

\subsection{Empirical Grounding: Failure Modes}\label{sec:FailureModes}
A recurring concern with guidelines of this kind is that they can rest on expert judgement rather than reported evidence.
As an initial exercise to address this concern, we synthesized the failure modes reported in our two empirical companion studies when GenAI tools are applied to SLR tasks. We then mapped each task to the recommendation that responds to it (\Cref{tab:FailureModes}).
The clearest evidence concerns two tasks.
For literature screening, our companion study \citep{Madeyski26LLM4SCREENLIT} reanalyses several software engineering screening evaluations and shows how conventional performance metrics mislead under class imbalance and how readily relevant studies are lost. In one such 9,695-article evaluation, the metric used to choose the best-performing LLM largely determined how much evidence was lost: the LLM ranked best by accuracy missed 63.3\% of the relevant studies and the best by plain MCC missed 43.9\%, whereas the best by the cost-sensitive Weighted Matthews Correlation Coefficient (WMCC) proposed in that study \citep{Madeyski26LLM4SCREENLIT} missed only 5.8\%. We summarize these findings here and refer the reader to that study for the underlying metric analysis and data.
For qualitative synthesis, \citet{Pizard26} report, from autoethnographic trials, failure modes such as plausible-but-incorrect output, loss of interpretation, sensitivity to prompts, and a verification burden as large as the manual task.
For data extraction, risk-of-bias assessment, and strength-of-evidence assessment, direct empirical evidence specific to these tasks is currently limited, so we treat the related guidance as open research questions (\Cref{sec:OpenQuestions}).

\begin{table*}[!t]
\caption{Failure modes reported in empirical studies, mapped to SLR tasks and to the recommendations that address them.}
\label{tab:FailureModes}
\footnotesize
\begin{tabular}{@{}>{\raggedright\arraybackslash}p{1.4cm}p{8.5cm}>{\raggedright\arraybackslash}p{6.6cm}@{}}
\toprule
\textbf{SLR task} & \textbf{Reported failure mode (source)} & \textbf{Recommendation(s) that address it}\\
\midrule
Literature screening
& Conventional metrics (accuracy, specificity, $F_1$) mislead under the severe class imbalance of screening: an LLM can score over 96\% on accuracy yet detect none of the relevant studies. False negatives cause lost evidence (the proportion of relevant studies missed, $1-\mathrm{Recall}$); incomplete reporting of confusion-matrix counts prevents re-analysis and meta-analysis; and workload-saving claims that count the cost of false positives but ignore the much higher cost of false negatives (falsely rejecting relevant studies loses evidence, potentially irretrievably) can hide a serious threat to the validity of the review \citep{Madeyski26LLM4SCREENLIT}.
& Above all, use WMCC so the far higher cost of false negatives (lost evidence) than false positives is reflected in the score (Scr2); plain MCC (Scr1) corrects for chance and class imbalance but not for this cost asymmetry. Report full confusion-matrix counts (Scr3), so recall and lost evidence ($1-\mathrm{Recall}$) can be computed and monitored; address the risk of data contamination (Plan4).\\
\midrule
Data extraction
& Direct empirical evidence on GenAI performance for SE data extraction is currently limited.
& Data-extraction recommendations (Ext1--Ext7); treated as an open research question pending further evidence (\Cref{sec:OpenQuestions}).\\
\midrule
Qualitative synthesis
& LLMs produce plausible but incorrect syntheses and deliver plausibility rather than interpretation, so output can look useful while distorting the primary studies; small prompt changes, or even the same prompt, yield different output, undermining traceability and reproducibility; models carry training biases; and verifying that a synthesis is sound can take as much effort as doing it manually \citep{Pizard26}.
& Qualitative-synthesis recommendations Syn1--Syn3 (specify the method; specify the performance criteria; specify how human and tool syntheses are compared); for reviewers, restrict tool use to validating human work (RevA1) and review all output (Con4); planning recommendation Plan4 (built-in biases).\\
\midrule
Risk-of-bias assessment
& RoB assessment depends on methodological knowledge not contained in any single primary study; direct empirical evidence on GenAI performance is currently limited.
& RoB/strength-of-evidence recommendations (A1--A3); framed as open research questions (\Cref{sec:OpenQuestions}).\\
\midrule
Strength-of-evidence assessment
& SoE assessment requires judging the reliability of findings across multiple studies; direct empirical evidence on GenAI performance is currently limited.
& RoB/strength-of-evidence recommendations (A1--A3); framed as open research questions (\Cref{sec:OpenQuestions}).\\
\bottomrule
\end{tabular}
\end{table*}

\section{A First Validation}\label{sec:Validation}

We validated our recommendations to check that they were complete and consistent with related guidance, and that our thought experiments had not missed important issues. To this end, we compared our set of recommendations with those constructed by \citet{baltes-2025}, which we did not use for developing our recommendations, and those reported by \citet{thomas-2026-RAISE1}, which we used while developing our recommendations.  The Baltes et al. study, which refines and extends the authors' earlier workshop position paper \citep{wagner2025} and is therefore a closely related line of work rather than a fully independent one, provided a recent SE baseline against which to cross-check our recommendations, while the study by Thomas et al. provided a check that the different methods we trialled to report our results had not led to any important issues being lost. This is a face validation, i.e., an assessment of whether our recommendations appear, on inspection, to address what they claim to address. The item-by-item comparison and the rationale for each change are reported in the supplementary material (Appendix~\ref{sec:FirstValidationDetails}).

Six of the eight guidelines of Baltes et al. are covered by our recommendations, once we added a recommendation to report GenAI tool identification details (Rep1). The remaining two, ``Report System and Prompt Design'' and ``Report Session Traces'', are only partly covered. Their uncovered aspects mainly concern the architecture and runtime behaviour of GenAI tools. These are issues for tool developers, a role outside our scope, as we note in our threats to validity (\Cref{sec:ThreatsToValidity}).

For Thomas et al., we considered only the recommendations for evidence synthesists (reviewers) and evidence synthesis methodologists (evaluators). We excluded those addressed to the broader evidence synthesis ecosystem. Most of the reviewer and evaluator recommendations were already covered by our recommendations, explicitly or through our stated assumptions. The comparison led to the following changes. We added five recommendations: Rep1 (report GenAI tool identification details), Plan2 (justify the choice of study design), Rep5 (report data collection and management), RepE4 (evaluators should report any financial or other interest and how bias was avoided), and Con4 (reviewers and evaluators remain responsible for the use of GenAI and its outputs). We extended Plan1 to cover the critical assessment of candidate tools, Plan5 to cover data collection, labelling, storage, and quality assurance, and Rep6 to cover open access. We also clarified in \Cref{sec:Context} our assumptions about reviewer and evaluator competence and about good evaluation practice. We made no change for three recommendations. Two of them, on research papers written by AI and on reporting exclusion criteria, had an unclear intention. The third, registering review-related studies, is not current practice in SE.

The validation also highlighted two general points about recommendations of this kind. First, it is important to make the underlying assumptions explicit. For example, reporting evaluation studies is expected of research-based reviewers and evaluators, but may conflict with the commercial interests of industry-based tool builders. Second, each recommendation should make clear how it can be addressed. We could not act on some of Thomas et al.'s recommendations because we could not determine what they required. Relating each of our recommendations to specific steps of the evaluation and SLR processes is intended to reduce this risk for our readers.

\section{Threats to Validity}\label{sec:ThreatsToValidity}
Given our research characteristics and the use of several methods, we did not adopt a particular framework to analyse the threats to validity.  Instead, as suggested in \citet{Lago24}, we considered the dimensions regarded as being most significant for our research objectives. In particular, we studied the construct, internal, and external validity, as well as the reliability of our process.

\subsection{Construct Validity}
Construct validity concerns whether our recommendations, and the risks they are intended to address, correspond to the real risks that the use of GenAI tools pose to the validity of an SLR.
The main threat is that recommendations derived from thought experiments could target the wrong issues, or miss important ones.
We addressed this threat in three ways.
First, we grounded the recommendations in empirical evidence: the failure modes they address are drawn from empirical studies of GenAI use in SLR tasks, including our companion studies on screening \citep{Madeyski26LLM4SCREENLIT} and qualitative synthesis \citep{Pizard26}, and are mapped to the tasks they affect (\Cref{sec:FailureModes}).
Second, we informed the recommendations with a rapid review of existing guidelines for using GenAI in research (\Cref{sec:DevelopingLLMEvaluationANdUseGuidlines}).
Third, we cross-checked our set against the independent guidelines of \citet{baltes-2025}, which we deliberately did not use while developing our recommendations, and found that they agreed on most points (\Cref{sec:Validation}).

\subsection{Internal Validity}
Internal validity concerns whether our derivation process could have produced biased or incorrect recommendations.
The main threat is that a thought experiment draws on the researchers' own experience, so unstated or unrecognized assumptions can shape its results.
We saw direct evidence of this: comparing our evaluator recommendations with those of \citet{thomas-2026-RAISE1}, we found several items that they stated explicitly but that we had treated as restatements of standard good practice.
We therefore added those generally accepted good-practice items for completeness, even though we had assumed the role of experienced evaluators.
Even so, we cannot guarantee that we have articulated every assumption.
Two features of our process reduce this threat.
The recommendations were developed and reviewed by several authors with extensive experience of conducting and reviewing SLRs, and they were corroborated by an independent guideline set (\Cref{sec:Validation}).
A further threat affects the rapid review that informed our work: a single researcher selected the candidate papers, which could have introduced selection bias (Appendix~\ref{sec:DetailsOfRR}).

\subsection{External Validity}
External validity concerns how far our recommendations generalize beyond the setting in which we produced them.
First, our thought experiments and briefing notes rest on our experience as software engineering researchers, so some recommendations may be less relevant to other disciplines, and researchers in those fields will need to judge which ones transfer.
Because our recommendations state general good practice, researchers must also interpret them in the context of their own study, with its particular research questions and hypotheses.
Second, we address only the evaluator and reviewer roles, not the GenAI tool developer role.
That role is important in software engineering and would benefit from its own recommendations, and is taken up by related RAISE guidance on building tools \citep{thomas-2025-RAISE2}.
Third, GenAI tools are changing quickly and their performance varies across disciplines and study types, so recommendations that hold today, or in one setting, may need revisiting in another.
We return to both points in \Cref{sec:Discussion}.

\subsection{Reliability}
Reliability concerns whether other researchers, repeating our process, would arrive at similar recommendations.
A thought experiment and a rapid review are both judgement-based, so an exact replication is unlikely.
We have sought to reduce this threat by documenting the process so that others can follow and scrutinize it: we report the two researcher roles and the three process stages that framed the thought experiments (\Cref{sec:DevelopingRecommendations}), the briefing notes for each SLR task (Appendix~\ref{sec:SLRProcessDetails}), and the research question, eligibility criteria, search strings and known-paper validation for the rapid review (Appendix~\ref{sec:DetailsOfRR}).
This documentation will not make the process deterministic, but it makes our reasoning transparent and open to challenge.

\section{Discussion}\label{sec:Discussion}

Our recommendations support the viewpoint that the use of GenAI tools in the context of academic research must be based on human oversight. Although GenAI tools are already capable of providing accurate summaries of the research literature, such summaries cannot be equated to \emph{systematic} literature reviews because currently they do not deliver associated documentation, allowing their summaries to be replicated by other human researchers using a trustworthy systematic literature review process. The requirement for human supervision will not disappear even if we have GenAI tools that are theoretically capable of delivering a complete SLR, together with associated documentation. Scientific progress has been undermined by expert opinions in the past, and results based on one (or more) GenAI tool cannot be regarded as inherently more reliable than results based on expert opinion. To use such output, we still require human researchers to take responsibility for validating all the SLR processes and their outcomes. 

Furthermore, if we allow GenAI tools to take over some SLR tasks, human researchers may lose the ability to properly assess the reliability of  GenAI tool outcomes. In the context of screening literature for inclusion in an SLR, \citet{spillias-2024} point out that, for PhD students the goal of undertaking an SLR is not just to contribute to the literature, but also to learn more about their chosen topic. We would add that it is also important to learn about the SLR process. It is unlikely that researchers who do not know how to undertake an SLR manually will understand how best to prompt a GenAI tool to perform an SLR task, or how to evaluate the outcome of its use.\footnote{See \citet{binz-2025} for a broader discussion of the implications of GenAI tools for scientific research.}

Currently, one of the most problematic issues for evaluations seems to be the danger of contaminated test data. This issue is not unique to the evaluation of GenAI tool performance. However, in the case of GenAI tools, it is \emph{not} caused by poorly understood or badly compiled data; it is due to GenAI tool processes that are invisible to the researcher conducting the study. This means, even if we suspect there are data problems (for example, the evaluation results are too good to believe), it is extremely difficult to detect and confirm data contamination. This means that standard data mining approaches based on re-analysing publicly available datasets, which have served empirical software engineering research well in the past (when data sets are used with care), are unreliable for assessing GenAI tool performance. \citet{Dennstadt2024} point out that ``To objectively assess how well an LLM-based solution can evaluate scientific publications for new research questions, large cultivated [sic] and independent prospective data sets on many different topics would be needed''. It is important that researchers compiling such data sets provide the results of their own GenAI tool analyses, so other researchers can use their results without re-running the GenAI tool analyses. \citet{Huotala25} provide this kind of benchmark for SE researchers. Their benchmark is based on past completed SLRs, which may be part of the GenAI tools' training data. The re-analysis by \citet{Madeyski26LLM4SCREENLIT} used the reported results, so it did not add the contamination risk of re-running the original data analyses.

In addition, we agree with \citet{thomas-2026-RAISE1} and \citet{Devane-2022} that one important option for gathering evidence about GenAI tool performance is to incorporate evaluation of GenAI tools into our basic SLR process. For example, using GenAI tools to perform SLR tasks in parallel with human reviewers would mean that each SLR would provide a \emph{prospective} case study of GenAI tool support for SLR tasks. Furthermore, the technology is changing rapidly, so it would be foolish to assume that any process recommendations will not need to be adapted as the technology evolves.\footnote{See, for example, Jovanovi{\'c} and Campbell's discussion of large reasoning models and large concept models \citep{jovanovic2025reasoning}.} Thus, if we make the evaluation of tools to support SLRs part of our SLR process, we will be in a better position to keep our evaluation methods in line with technology changes.

Using GenAI tools in an SLR is also an iterative and incremental activity rather than a single linear pass.
Planning, prompt development, tool testing and validation, and conduct form feedback loops, which we describe in the summary of process changes (Appendix~\ref{sec:TestingANdEvaluatingGenAITools}).

Our discussion in \Cref{sec:TheNatureOfLLMs} and comments in \citet{woelfle-2022} suggest that it is challenging to ensure GenAI tools produce trustworthy results, because the underlying design of such tools does not readily support reproducibility and transparency. Bearing this in mind, we should not forget that if our goal is to improve the way we conduct SLRs, tools based on more deterministic algorithms should also be investigated. For example, \citet{suster2024} report an experiment used to compare FlanT5XL with a supervised tool called RobotReviewer that showed RobotReviewer outperforming FlanT5XL. In addition, if we want simply to improve the validity of SLR reports, using GenAI tools as an additional research assistant provides prospective assessment of tool performance and may reveal errors or omissions made by human researchers.

Our recommendations do not assume that only a single GenAI tool is used. Where it improves robustness, a review team can run several tools, or the same tool with several prompts, and combine the outputs --- an ensemble --- with human judgement. Tools such as AISysRev \citep{Huotala26AISysRevArXiv} support multi-model title--abstract screening. Combining tools can reduce idiosyncratic errors, but it does not remove the risk of data contamination, which can affect several tools trained on overlapping data, and it increases cost, so the trade-off should be reported.

Our recommendations were developed for GenAI tools applied to individual SLR tasks, but agentic AI platforms --- end-to-end systems in which LLM agents plan and execute multiple review stages automatically \citep{thomas-2025-RAISE2} --- fall within their scope, because each embedded stage raises the same process issues we discuss.
However, two of those issues become more severe.
First, our validation process operates on a task-by-task basis, and end-to-end platforms do not always expose the intermediate outputs of each stage.
\citet{thomas-2025-RAISE2} identify two evaluation strategies for such platforms: evaluating each stage in isolation when the platform permits, or evaluating only the final report.
The second strategy inherits the problems discussed in~\Cref{sec:TheNatureOfLLMs}, since a report whose provenance cannot be inspected is at best equivalent to expert opinion, and the evaluation is invalid if the reviews used to judge the report were part of the tool's training data \citep{thomas-2025-RAISE2}.
Second, the behaviour of an agentic run is steered not only by the initial prompts but also by the context that accumulated during the run, such as the plans the agents generate and the outputs of the tools they call.
Reporting the initial prompts is therefore not enough to make an agentic study reproducible. 
We agree with \citet{thomas-2025-RAISE2} that reports generated by end-to-end platforms should not be treated as a rigorous evaluation of such platforms, and we regard the rigorous evaluation of such platforms as an open research question.

We hope that our recommendations will help SE researchers to improve the evaluation of GenAI tools, because improved evaluation studies will support more reliable meta-analyses. Single evaluation exercises cannot provide all the evidence we need to properly assess the risks and benefits of GenAI tool use. However, we urge meta-analysts to expect discipline-related study heterogeneity. For example, \citet{spillias-2024} point out that studies from the medical domain may deliver better performance when screening papers for inclusion in SLRs because they have a better-defined and more stable vocabulary than other disciplines. Another issue is that medical studies often rely on RCTs, which can be assessed using tools such as the Cochrane RoB-2 method. When aggregating medical primary studies based on RCTs, meta-analysts are better able to identify poor quality studies than when aggregating primary studies based on a wider range of experimental methods.

\section{Open Research Questions}\label{sec:OpenQuestions}
For literature screening and qualitative synthesis we can draw on empirical evidence (\Cref{sec:FailureModes}).
For three further tasks --- data extraction, risk-of-bias (RoB) assessment, and strength-of-evidence (SoE) assessment --- there is little or no empirical evidence on GenAI tool performance for the heterogeneous study types found in SE, and what evidence exists (largely from medical randomized controlled trials) is contradictory.
Our recommendations for these tasks are therefore preliminary and process-oriented, and we set out the main questions we believe must be answered before firmer, evidence-based guidance is possible.

For RoB assessment:
\begin{itemize}[noitemsep]
\item Which evaluation instruments are appropriate for the heterogeneous study types found in SE SLRs? The SE community has not agreed on such instruments (Appendix~\ref{sec:RiskOfBiasAssessment}), and without agreement GenAI tool performance cannot be assessed consistently.
\item Can GenAI tools apply RoB instruments reliably? Current evidence is contradictory: applying RoB questionnaires to randomized controlled trials, \citet{suster2024} found LLMs ``seldom surpassing trivial baselines'', whereas \citet{lai-2024} reported ``substantial accuracy and consistency''.
\item Can GenAI tools help to construct or refine RoB instruments for SE methods, rather than only apply existing ones?
\end{itemize}

For SoE assessment:
\begin{itemize}[noitemsep]
\item Can GenAI tools assess the strength of evidence for a finding, given that this requires reasoning about confounders and bias across a set of studies using knowledge not contained in any single primary study?
\item How should a GenAI tool's SoE judgements be validated against human assessment, when that human assessment is itself subjective?
\end{itemize}

For data extraction, the question is narrower: how reliably can GenAI tools extract the heterogeneous quantitative and qualitative data items found in SE primary studies, and how should that reliability be measured?

Finally, several questions cut across all of these tasks.
Can a GenAI tool serve as a judge or oracle for an SLR task, given that the human researchers used as its comparison can themselves misjudge, and what should serve as the reference ``correct'' answer when no infallible human baseline exists?
And why are two human researchers required for these subjective tasks?  Is this an irreducible safeguard, or could a validated GenAI tool justifiably replace one of them, and what level of agreement with human researchers would justify such a replacement?
We raise these as open questions rather than resolving them, because answering them will require the kind of empirical evidence the field is only beginning to gather.

\section{Conclusions}\label{sec:Conclusions}
GenAI tools are a hot topic. They are widely adopted by industry and widely discussed in academia. They represent a new technology that few researchers understand but that has great potential to change practices in industry and academia. Furthermore, the nature of GenAI tools means we can no longer rely on our usual processes for evaluating the accuracy of software tools. Thus, we need to rethink our evaluation methods when we investigate the performance of GenAI tools.

In this paper, we have presented GUEST, a preliminary set of recommendations that extend existing research processes to address some of the problems faced by researchers using, testing, and evaluating GenAI tools in the context of SLR conduct. We argue that GenAI tools are unable to produce SLRs, but that the ability of such tools to produce summaries of research papers provides a strong incentive for further research. These recommendations will help researchers to develop more rigorous evaluation methods that properly address both the limitations and the potential of GenAI tools. They will help journal editors and reviewers avoid accepting papers simply because the topic is important, rather than because the authors have made a compelling case that their evaluation methods, and hence their results, are trustworthy.

\appendix
\counterwithin{table}{section}
\counterwithin{figure}{section}

\section{Details of our Rapid Review}\label{sec:DetailsOfRR}
At the start of our review we were aware of two relevant reports:
\begin{enumerate}
    \item A preliminary version of \citet{thomas-2026-RAISE1}, Responsible use of AI in Evidence SynthEsis (RAISE 2026) 1: Recommendations for practice.
    \item \citet{wagner2025}, Towards Evaluation Guidelines for Empirical Studies involving LLMs.
\end{enumerate}

 \subsection{Goal and Research Questions}

At the beginning of our research, we sought to identify papers that would inform our development of guidelines for evaluating and using GenAI tools in the context of SLRs. Specifically, we searched for papers that already provided guidelines to help researchers assess the performance of GenAI tools in SLRs, as well as any related work that could help us to formulate useful guidelines.

This led to the research question: \emph{RQ Which papers report guidelines or recommendations for evaluating and using of GenAI tools in the context of conducting SLRs?}

\subsection{Research Method Strategy}
Our choice of methodology was affected by the following issues:
\begin{itemize}
\item We did not want to limit our search to Software Engineering and Computer Science sources, as guidelines for conducting SLRs have often first been developed within the medical research community.
\item We expected to find large numbers of irrelevant papers as a consequence of requiring cross-domain searches. We decided that our search goal should be to obtain coverage of themes (i.e. issues that were considered important by any existing guideline papers) rather than to attempt to identify all existing papers. We expected any attempt at achieving full coverage would fail because of the large numbers of papers being published. Furthermore, we had no intention of relying solely on existing research, we wanted to adopt a risk assessment approach using our own analysis of the issues facing use of GenAI tools, the processes used to reduce threats to validity for SLRs, and our experiences of performing tool evaluations. We wanted to make use of our own experiences in order to identify potential problems before they become embedded in \emph{de facto} standards in the same way that MMRE did.
\item We wanted a reproducible baseline for identifying relevant articles, so we intended to include only articles available in journals or conference proceedings, not on self-publication sites. However, we did not intend to exclude editorials or letters.
\item Finally, we wanted to deliver the results of such a review in as expeditious a manner as possible to practitioners and researchers given the incredible growth in research into, and use of, GenAI based tools. A key feature of a Rapid Review (RR) ---as the name implies--- is its timeliness. We also intended to use the evidence ourselves as input for developing our guidelines. Therefore, we followed the recommendation that RRs be used primarily in collaborations with a clearly defined requester or knowledge user, especially in low-resource settings \citep{pizard-2025}.
\end{itemize}

These issues led us to adopt a Rapid Review \citep{cartaxo-2018,pizard-2025} rather than a full SLR. We decided to use SCOPUS as our only digital library because:
\begin{enumerate}
\item Using a single source reduces the issue of including multiple versions of the same report in the set of candidate primary studies.
\item SCOPUS indexes a wide range of important journals and conferences in many different domains, including the medical and computer science domains.
\item Its search processes conform with the rules of Boolean algebra meaning that we could use more complex search strings without inflating the number of irrelevant studies found.
\item Scopus indexes articles published in journals and conferences, including editorials and letters.
\end{enumerate}

\subsection{Eligibility Criteria}\label{sec:EligibilityCriteria}
We were seeking articles that:
\begin{itemize}
\item I1: Presented guidelines or recommendations for the evaluation of GenAI tools. 
\item I2: Presented guidelines or recommendations for using GenAI tools to support academic research.
\item I3: Had the full text of the article available in English.
\item I4: Had a context that involved the use of systematic literature reviews or tasks associated with SLRs.
\end{itemize}
We did not want our searches to be restricted to a specific GenAI tool or LLM. Neither did we want to restrict our search discipline. These criteria were the basis of our keyword-based search process.
We also excluded studies based on the following criteria:
\begin{itemize}
    \item E1: Articles that would not address our research questions, including those that used the results of one or more SLRs to generate recommendations for practice, such as health care guidelines. We planned to exclude any such articles during the literature screening process.
    \item E2: Articles that were self-published and available only on a site such as arXiv.
\end{itemize}

\subsection{Search Process}
Our search process was developed to identify two types of paper:
\begin{enumerate}
\item Studies that evaluated the use of GenAI tools to support various types of SLR (hereafter, Search 1).
\item Studies that evaluated the use of GenAI tools to support specific SLR tasks rather than SLR types (hereafter, Search 2).
\end{enumerate}

Our search strings were intended to address inclusion items I1, I2, and I4 and exclusion item E2 reported in~\ref{sec:EligibilityCriteria}. The search strings were initially designed by Pizard and reviewed and refined by Madeyski.

Each search string was built from groups of synonyms for the main concepts in our research question, combined with Boolean operators.
The string for Search~1 combined a group of SLR-type terms (drawn from standard SLR methodology vocabulary, for example, ``systematic review'', ``mapping study'' and ``scoping review'') with a group of technology terms (general terms such as ``large language model'', ``LLM'' and ``generative AI'', plus the names of specific GenAI front-end systems). String validation was based on assessing the effectiveness of the string (without any restriction related to guidelines papers) in finding the 13 papers listed without asterisks in Table~\ref{tab_knownpapers}. In addition, \Cref{tab_knownpapersnot} reports other known papers that were not expected to be found by our searches.
For Search~2 we evolved the string in two ways: we broadened the SLR-type group to name individual SLR tasks (for example ``screening'', ``data extraction'', ``risk of bias'', ``qualitative synthesis'' and ``strength of evidence''), so that we would also find papers addressing a single task, and we added a group of guideline terms (``guideline'', ``guidance'' and ``recommendation'') together with further GenAI model names. String validation was based on assessing the effectiveness of the string (without any restriction related to guideline papers) at finding all of the 18 papers in Table~\ref{tab_knownpapers}. The sources of the known papers were the authors of this paper, some citations from identified papers, and a set of 20 papers found by an SLR that evaluated the use of GenAI to support qualitative analysis methods \citep{leca-2025}. The papers related to qualitative analysis are indicated by an asterisk and these were only used to validate String 2.

Search 1 was based on a stepwise search strategy and conducted on 13/03/2025. First, we performed a general search for studies on LLMs and SLRs (see first row of \Cref{tab:search1}). Second, since authors sometimes used proprietary technology names instead of general terminology, we used the names of specific frontend systems that we were familiar with, and those mentioned in \citet{Pearson2024} and \citet{Gibney2025}. Third, the results of the individual searches were merged (obtaining 2462 papers) and validated against the known papers, finding the 13 known papers that we expected to find. As we expected, none of the papers listed in~\Cref{tab_knownpapersnot} were found by the search, although some were subsequently published formally. Finally, from this list of papers, we selected those studies that included ``guidance'' or ``guideline'' in their titles, resulting in 13 candidate primary studies. 

Madeyski and Pizard continued refining the search process because they were concerned that we would miss papers that referred to papers addressing issues related to specific SLR tasks rather than SLRs in general. For Search~2 Madeyski changed which document fields were matched. Search~1 had used the combined \texttt{TITLE-ABS-KEY} field, whereas in Search~2, the \texttt{TITLE} and \texttt{AUTHKEY} fields were searched separately.
In Scopus, \texttt{AUTHKEY} matches only the keywords an author assigns to a document, whereas the combined keyword field (\texttt{KEY}) used within \texttt{TITLE-ABS-KEY} also matches Scopus-assigned index terms, trade names and chemical names.\footnote{See the Scopus search field definitions: \url{https://schema.elsevier.com/dtds/document/bkapi/search/SCOPUSSearchTips.htm}.}
Restricting the keyword match to \texttt{AUTHKEY} therefore returned results that were less polluted.

\Cref{tab:search2} shows the SCOPUS string used for Search 2. It was conducted on 25/03/2025 and returned 873 papers. These included the 18 known papers: 13 were the same relevant papers found by Search 1, and 5 were related to qualitative analysis. The search string was then restricted to select papers related to guidelines (see~\Cref{tab:search2}). This search string was applied on 29/03/2025 and delivered 20 candidate primary studies.

\begin{table*}[!t]
\caption{Relevant Known papers (The asterisk identifies papers included in this set after expanding the search strings to consider the stages of the SLR. They were not used to assess Search 1.)}
\label{tab_knownpapers}
\footnotesize
\begin{tabular}{lp{8.6cm}p{3cm}p{3cm}}
\toprule
Id   & Title     & Reference & Source            \\ \midrule
KN1  & Automated Paper Screening for Clinical Reviews Using Large Language Models: Data Analysis Study                                                                                                                                     &   \citet{Guo2024}                                                           &  \citet{Huotala2024}  \\
KN2  & Can ChatGPT write a good boolean query for systematic review literature search?                                                                                                                                                     &     \citet{Wang2023}                                                         & \citet{Petersen2025}      \\
KN3  & Can large language models replace humans in systematic reviews? Evaluating GPT-4's efficacy in screening and extracting data from peer-reviewed and grey literature in multiple languages                                           &     \citet{Khraisha2024}                                                         & \citet{Huotala2024}  \\
KN4*  & CollabCoder: A Lower-barrier, Rigorous Workflow for Inductive Collaborative Qualitative Analysis with Large Language Models                                                                                                          & \citet{gao-2024}             & \citet{leca-2025}       \\

KN5*  & Exploring the Use of Artificial Intelligence for Qualitative Data Analysis: The Case of ChatGPT                                                                                                                                     & \citet{morgan-2023}              & \citet{leca-2025}      \\
KN6  & Harnessing the power of ChatGPT for automating systematic review process: methodology, case study, limitations, and future directions                                                                                               &   \citet{Alshami2023}                                                           & \citet{felizardo-2024} \\
KN7  & Leveraging Large Language Models for Literature Review Tasks - A Case Study Using ChatGPT                                                                                                                                           &    \citet{Zimmermann2024}                                                          & Shepperd            \\
KN8  & Leveraging the potential of generative AI to accelerate systematic literature reviews: an example in the area of educational technology                                                                                             &   \citet{Castillo-Segura2023}                                                          & Shepperd            \\
KN9 & On the road to interactive LLM-based systematic mapping studies                                                                                                                                                                     &      \citet{Petersen2025}                                                        & Madeyski             \\
KN10* & Performing an Inductive Thematic Analysis of Semi-Structured Interviews With a Large Language Model: An Exploration and Provocation on the Limits of the Approach                                                                   & \citet{dePaoli-2024}           & \citet{leca-2025}       \\
KN11 & PRISMA Systematic Literature Review, including with Meta-Analysis vs. Chatbot/GPT (AI) regarding Current Scientific Data on the Main Effects of the Calf Blood Deproteinized Hemoderivative Medicine (Actovegin) in Ischemic Stroke &  \citet{Anghelescu2023}                                                            & \citet{felizardo-2024} \\
KN12* & Prompts, Pearls, Imperfections: Comparing ChatGPT and a Human Researcher in Qualitative Data Analysis                                                                                                                              & \citet{wachinger-2024} & \citet{leca-2025}       \\
KN13 & Screening articles for systematic reviews with ChatGPT                                                                                                                                                                              &      \citet{Syriani2024}                                                        & Shepperd           \\
KN14 & Should we collaborate with AI to conduct literature reviews? Changing epistemic values in a flattening world                                                                                                                        &     \citet{Ngwenyama2024}                                                         & Shepperd           \\
KN15 & The Promise and Challenges of Using LLMs to Accelerate the Screening Process of Systematic Reviews                                                                                                                                  &        \citet{Huotala2024}                                                      & \citet{leca-2025}\\
KN16* & The use of Generative AI in qualitative analysis: Inductive thematic analysis with ChatGPT                                                                                                                                          & \citet{perkins-2024}        & \citet{leca-2025}       \\
KN17 & Transforming literature screening: The emerging role of large language models in systematic reviews                                                                                                                                 &   \citet{Delgado-Chaves2025}                                                        & Kitchenham          \\

KN18 & Zero-shot generative large language models for systematic review screening automation                                                                                                                                               &      \citet{Wang2024}                                                        & Shepperd         \\ \bottomrule  
\end{tabular}
\footnotesize
\end{table*}

\begin{table*}[!t]
\caption{Other known papers not expected to be found by SCOPUS}
\label{tab_knownpapersnot}
\footnotesize
\begin{tabular}{p{7cm}p{2.6cm}p{2.6cm}p{3cm}}
\toprule
Title  & Reference & Source  & Reason   \\ 
\midrule
Assessing the ability of chatgpt to screen articles for systematic reviews & \citet{Syriani2024} & Shepperd & Published version available after search \\
Automated Title and Abstract Screening for Scoping Reviews Using the GPT-4 Large Language Model & \citet{wilkins2023} & \citet{Huotala2024} & Not peer reviewed \\
Automating research synthesis with domain-specific large language model fine-tuning & \citet{Susnjak_2025} &  Shepperd & Not in SCOPUS \\
Bio-SIEVE: Exploring Instruction Tuning Large Language Models for Systematic & \citet{robinson2023} & \citet{Huotala2024} & Not peer reviewed \\
ChatGPT application in
Systematic Literature Reviews in Software Engineering: an evaluation of its accuracy to support the selection activity & \citet{felizardo-2024} & - & Indexed by SCOPUS after search.\\
Exploring the Use of AI in Qualitative Analysis: A Comparative Study of Guaranteed Income Data    & \citet{hamilton2023exploring}  & \citet{leca-2025}   & Not mention any variant of generative AI or LLMs in the title or author keywords.    \\
Prompting is all you need: LLMs for systematic review screening &  \citet{Cao2024}  & Kitchenham  & Published version available after search \\
Towards Evaluation Guidelines for Empirical Studies involving LLMs &  \citet{wagner2025} & Shepperd  & Published version available after search \\
Unsupervised title and abstract screening for systematic review: a retrospective case-study using topic modelling methodology                                                                                                       &      \citet{Natukunda2023}                                                        & \citet{felizardo-2024} & Not mention any variant of generative AI or LLMs in the title or author keywords.    \\
Utilizing Large Language Models to Update Systematic Literature Reviews: A Case Study on Time Pressure and Well-Being in Software Engineering &  \citet{Guo2024b} & Shepperd          & Published version available after search \\ 
\bottomrule
\end{tabular}
\end{table*}

\begin{table*}[!t]
\caption{Search strings for Search 1}
\label{tab:search1}
\footnotesize
\begin{tabular}{p{3cm}p{13cm}}
\toprule
General Search  & TITLE-ABS-KEY ( ( "systematic review*" OR "systematic literature review*" OR "mapping stud*" OR "mapping review*" OR "systematic map*" OR "literature review*" OR "scoping review*" OR "scoping stud*" OR "rapid review*" ) AND ( "large language model*" OR llm* OR "generative AI" ) ) AND PUBYEAR \textgreater 2022 AND PUBYEAR \textless 2026 \\
Specific searches                                                             & TITLE-ABS-KEY ( ( "systematic review*" OR "systematic literature review*" OR "mapping stud*" OR "mapping review*" OR "systematic map*" OR "literature review*" OR "scoping review*" OR "scoping stud*" OR "rapid review*" ) AND ( \#SPECIFIC FRONTEND SYSTEMS\#) ) AND PUBYEAR \textgreater 2022 AND PUBYEAR \textless 2026                       \\
Specific Frontend Systems related to LLMs and generative IA (separated by semicolon). & chatgpt OR o3-mini; Scopus AI; Perplexity; Research Kick; Scite; PaperQA2; Undermind; Elicit; DeepSeek; Llama; Claude; Olmo \\ \bottomrule                                   
\end{tabular}
\end{table*}

\begin{table*}[!t] 
\caption{Search string for Search 2}
\label{tab:search2}
\footnotesize
\begin{tabular}{p{16.5cm}}
\toprule
( TITLE ( ( "systematic review*" OR "systematic literature review*" OR "mapping stud*" OR "mapping review*" OR "systematic map*" OR "literature review*" OR "scoping review*" OR "scoping stud*" OR "rapid review*" OR "screening" OR "eligibility assessment" OR "data extraction" OR "data collection" OR "quality assessment" OR "risk of bias" OR "peer evaluation" OR "peer review" OR "qualitative synthesis" OR "meta-synthesis" OR "meta synthesis" OR "thematic analysis" OR "qualitative analysis" OR "qualitative data analysis" OR "strength of evidence" OR "certainty of evidence" ) AND ("guideline*" OR "guidance" OR "recommendation*" ) AND ( "generative artificial intelligence" OR "generative AI" OR "large language model*" OR llm* OR "GPT*" OR o1 OR "o1-mini" OR "o3-mini" OR claude OR grok OR llama OR gemini OR bard OR titan OR nova OR phi OR orca OR mai OR qwen OR deepseek OR chatgpt OR "Perplexity" OR "Scopus AI" OR claude OR "Research Kick" OR scite OR paperqa2 OR undermind OR elicit ) ) OR AUTHKEY ( ( "systematic review*" OR "systematic literature review*" OR "mapping stud*" OR "mapping review*" OR "systematic map*" OR "literature review*" OR "scoping review*" OR "scoping stud*" OR "rapid review*" OR "screening" OR "eligibility assessment" OR "data extraction" OR "data collection" OR "quality assessment" OR "risk of bias" OR "peer evaluation" OR "peer review" OR "qualitative synthesis" OR "meta-synthesis" OR "meta synthesis" OR "thematic analysis" OR "qualitative analysis" OR "qualitative data analysis" OR "strength of evidence" OR "certainty of evidence" ) AND ( "guideline*" OR "guidance" OR "recommendation*" ) AND ( "generative artificial intelligence" OR "generative AI" OR "large language model*" OR llm* OR "GPT*" OR o1 OR "o1-mini" OR "o3-mini" OR claude OR grok OR llama OR gemini OR bard OR titan OR nova OR phi OR orca OR mai OR qwen OR deepseek OR chatgpt OR "Perplexity" OR "Scopus AI" OR claude OR "Research Kick" OR scite OR paperqa2 OR undermind OR elicit ) ) ) AND PUBYEAR \textgreater 2022 AND PUBYEAR \textless 2026 \\ 
\bottomrule
\end{tabular}
\end{table*}

\subsection{Final Primary Study selection}

Search 1 delivered 13 articles that used the terms guidelines or guidance in their titles. Search 2 identified 20 articles that used the terms  guidelines or guidance or recommendations. The article abstracts, title and keywords were then assessed by Kitchenham using the inclusion and exclusion criteria reported in ~\ref{sec:EligibilityCriteria}. However, particular attention was paid to inclusion criterion I3 and exclusion criterion E1 because neither of these criteria were addressed by the search strings.

Search 1 and Search 2 were processed separately. Based on their abstracts, Kitchenham identified three relevant papers from Search 1 and five from Search 2, which are shown in \Cref{tab:PapersIncludedBasedOnAbstracts}. The most common reason for exclusion was due to criterion E1 i.e., the paper presented recommendations for practice based on using GenAI tools to summarize SLR results.  Subsequently, two of the  five  papers found by Search 2 \citep{chen-2024, Luo-2024} were excluded because the full text was not in English (i.e., they failed inclusion criteria I3). 

Search 2 found another paper of interest, i.e., \citet{woelfle-2022}. Although not a paper aimed at producing recommendations, it reported  a rigorous large-scale study of the evaluation of using GenAI tools to support quality assessment of systematic literature reviews. In addition, one of its co-authors was John Ioannidis who is recognized as one of the world's leading systematic review and meta-analysis researchers. This paper seemed potentially very useful for identifying good practices for evaluating GenAI performance on SLR tasks requiring assessment of primary study methodology.

\begin{table*}[!t]
\caption{Guidelines Papers included in the Rapid Review based on Abstract and Title}
\label{tab:PapersIncludedBasedOnAbstracts}
\footnotesize
\begin{tabular}{p{1.4cm}p{8cm}p{2.5cm}p{1.1cm}p{2.3cm}}
\toprule
Authors & Title & Reference & Found By  & Final Decision      \\
\midrule
Inam et al & A review of top cardiology and cardiovascular medicine journal guidelines regarding the use of generative artificial intelligence tools in scientific writing & \citealp{inam-2024} & Search 1 & Include \\
Kim et al. & ChatGPT and large language model (LLM) chatbots: The current state of acceptability and a proposal for guidelines on utilization in academic medicine & \citealp{kim-2023} & Search 1 & Include \\
Lubowitz & Guidelines for the Use of Generative Artificial Intelligence Tools for Biomedical Journal Authors and Reviewers & \citealp{lubowitz-2024} & Search 1 & Include \\
Veiga & Ethical guidelines for the use of generative artificial intelligence and artificial intelligence-assisted tools in scholarly publishing: a thematic analysis & \citealp{Veiga-2025} & Search 2 & Include \\
Uribe et al. & Integrating Generative AI in Dental Education: A Scoping Review of Current Practices and Recommendations & \citealp{uribe-2025}& Search 2 & Include \\
Perkins \& Roe & Academic publisher guidelines on AI usage: A ChatGPT supported thematic analysis & \citealp{perkins-2024guidelines} & Search 2 & Include \\ \hline
Chen & Utilizing ChatGPT in Systematic Reviews and Meta-Analyses & \citealp{chen-2024} & Search 2 & Exclude (failed I3)  \\
Luo & The application of large language models in the field of evidence-based medicine & \citealp{Luo-2024} & Search 2  & Exclude (failed I3)\\
\bottomrule
\end{tabular}
\end{table*}

The two searches were processed separately. Investigating the two sets of results we found that:
\begin{itemize}
\item None of the papers would have been found by a search restricted to the computing domain.
\item None of the six papers referenced any of the other five papers. 
\item Superficial analysis suggested that none of the papers found by Search 1 was found by Search 2 and vice versa. However, Search 2 found all three papers found by Search 1, but they were not identified when the search results were restricted to guideline papers.  Search 1 found one of the papers identified by Search 2 \citep[i.e.,][]{uribe-2025}. However, it was not included in the papers selected from Search 1 because it used the term  "recommendations", but not "guidelines" or "guidance" in its title.
\end{itemize}

    \subsection{Data Extraction and Synthesis}
Our raw data comprised the recommendations obtained from each paper. Santos extracted the guidelines specified by six papers, i.e., each of the three papers found by Search 2, the two papers that we were aware of when we started our study (i.e., the preliminary version of \citet{thomas-2026-RAISE1} and \citet{wagner2025}) and an initial version of a paper that Madeyski, Kitchenham and Shepperd authored discussing evaluation of LLMs in the context of literature screening process. Kitchenham extracted the guidelines specified by the three papers identified by Search 1.

Data synthesis was intended to integrate the recommendations from different papers using thematic analysis. In practice, this was an iterative process in which we tried various methods of organizing a thematic framework.

The first synthesis was undertaken by Santos based on the raw data he extracted from the six articles. He extracted 41 recommendations from the articles and then identified 7 higher-level themes. He mapped the 41 recommendations to these themes and finally identified what the high-level themes meant in the context of SLRs. This approach had a number of issues:
\begin{itemize}
    \item The level of abstraction was appropriate for guidelines related to the use of GenAI tools to support academic research. 
    \item Thomas et al.'s recommendations relating to the use of GenAI tools to support SLR conduct arose from different usage scenarios, which had overlapping recommendations, and it was unclear which specific recommendations we should include.
    \item Wagner et al.'s recommendation related to general tool evaluation studies, and although useful, the recommendations were not focused on support for SLR conduct.
    \item Madeyski et al.'s recommendations were aimed at assessing the performance of GenAI tools in assessing the eligibility of candidate primary studies. They covered details related to that task, and could not be easily incorporated with recommendations from studies reporting issues related to other SLR tasks. Nonetheless, it seemed apparent that recommendations about specific SLR tasks were of potential value.
\end{itemize}

The second synthesis was undertaken by Kitchenham. Initially, it was intended to analyse the papers found by Search 1 to validate the high-level recommendations extracted from the synthesis of papers by Search 2. The second synthesis did not find any issues relevant to SLRs that had not already been found by Santos. However, it was unable to integrate recommendations from sources which were at different levels of abstraction, and gave us no confidence that the current recommendations had covered all the major GenAI tool evaluation issues. 

In order to address the problems found during the second synthesis, we decided to produce three different types of guidelines using Thomas et al.'s approach of defining guidelines for different research roles:
\begin{enumerate}
    \item To support all researchers using AI tools for academic research, Kitchenham integrated the recommendations from the six relevant papers to produce \Cref{tab:Guidelines}. For reasons of completeness, we included an additional theme (i.e., G6) which was related to peer review. This issue was not relevant to guidelines intended for use in the context of SLR conduct, but is relevant to academic research. 
    \item To support researchers intending to assess how well GenAI tools can support SLR tasks, we decided to undertake a thought experiment working through the processes that need to be taken to plan, conduct and report evaluations of tools to support SLRs. 
    \item To support researchers intending to use GenAI tools to support the conduct of a specific SLR, we decided to undertake a thought experiment that would involve working through the processes used to plan, conduct and report an SLR, and assessing how the use of GenAI tools would affect those processes.
\end{enumerate}

\subsection{Limitations}

A major limitation of our search was that it was limited to a single source (i.e., SCOPUS) and hence there is a high probability that it will have missed some relevant papers. However, SCOPUS covers an extremely large range of journals and conferences and, in particular, indexes all the major computing and software journals and conferences. Our searches also recovered all of a pre-specified set of known relevant papers (Table~\ref{tab_knownpapers}), providing direct evidence that the single-source search was sufficient for our goal of theme coverage.

Another limitation arises because the selection of relevant papers from the candidate primary studies was done by a single researcher, which might also have led to incorrectly rejecting some relevant papers found by the searches. The possibility of missing papers during search and selection means that we may have failed to identify important issues concerning the use of LLMs to support SLRs. This was the goal of our RR, and we found only two papers that directly addressed that goal \citep[i.e.,][]{chen-2024,Luo-2024}. These papers were omitted from our study because they were not available in English. However, the title and abstracts of those two papers suggested that our search should have found other similar papers if any such papers were indexed by SCOPUS, and so it was probable that we would have found them if they existed.  %

In terms of identifying guidelines for using GenAI tools to support academic research, our RR seems to have been more successful. The various themes reported in \Cref{tab:Guidelines} were repeated by many of the individual papers, and the individual papers did not cite each other. In addition, with one caveat, our set of guidelines matches recommendations reported by \citet{nahar-2025}, who analysed the recommendations currently being used by the organizers of computer conferences. Thus, it is likely that we have achieved completeness with respect to the most important issues.

A further limitation is the search horizon: the searches were completed on 29/03/2025 and were not re-run before submission.
The RR's goal was theme coverage to inform the development of our recommendations, not an exhaustive or continuously updated survey.
That development is complete, and the recommendations are grounded in the failure-mode synthesis (\Cref{sec:FailureModes}) and our companion studies \citep{Madeyski26LLM4SCREENLIT,Pizard26}, so re-running the searches would not, by itself, change them.
We therefore addressed post-search developments narratively instead: guidance published after the searches --- \citet{baltes-2025}, \citet{farotimi-2025}, and \citet{nahar-2025} --- is incorporated and positioned in \Cref{sec:GuidelineBackground} and \Cref{sec:DevelopingLLMEvaluationANdUseGuidlines}.
We also verified that successive versions of the RAISE recommendations left unchanged the list that we had used \citep{thomas-2026-RAISE1}.

\section{Details of the Thought Experiment Results}\label{sec:ThoughExperimentResults}
This section reports the detailed outcomes of the thought experiments related to evaluators of GenAI tools of SLRs and to reviewers using GenAI tools to support their review.

\subsection{Evaluating GenAI tools}\label{sec:EvaluatingGenAITools}
This section addresses the processes needed to plan, conduct and report GenAI tool evaluations.

\subsubsection{Planning GenAI tool Evaluations}\label{sec:PlanningGenAIToolEvaluations}

Evaluators need to specify their goals and hypotheses clearly. In particular, they should clarify whether the GenAI tools being evaluated are intended to replace human researchers (i.e., to fully automate a specific process) or to support human researchers (i.e., act as an assistant, replacing one of the human researchers). If investigating full automation, evaluators should discuss how their proposals can support current ethical requirements for human oversight.

Another critical planning issue is identifying the data needed to address their research questions and hypotheses. There are two basic evaluation approaches:
\begin{enumerate}
    \item Retrospective analysis. This involves identifying and extracting appropriate data from previously completed SLRs. Keeping in mind the issue of data leakage if data sets are reused \citep{woelfle-2022}, evaluators planning to use data extracted from previously published SLRs must ensure that the data they require has not already been widely used in previous evaluation studies.
    \item Prospective analysis. This involves identifying and extracting appropriate data during the conduct of an SLR. Evaluators can use GenAI tools to perform tasks in parallel with human researchers conducting an SLR. In addition, evaluators can also undertake some of the SLR tasks themselves to better assess the performance levels achieved by human researchers.
\end{enumerate}

From a scientific viewpoint, a prospective analysis is preferable because it avoids the possibility that evaluators design their research questions to suit the available data. In addition, prospective analysis provides some protection against the risk of data leakage. It should be noted that SLRs are retrospective studies, and that the complexity of the SLR process arises from the need to reduce the risk of bias inherent in such studies.

The risk of data leakage is a particular problem for evaluating GenAI tool performance. This can occur if retrospective datasets are used, particularly well-known curated datasets that have been used to evaluate other non-generative SLR tools. However, it can also occur if evaluators test and retest multiple LLMs using multiple prompt strategies on the same datasets or dataset partitions without putting in place procedures to prevent models from learning or retaining the data from previous evaluation sessions.

Many issues are similar for researchers evaluating GenAI tool performance in the context of SLRs, and also for researchers training and assessing GenAI tools for use on a specific SLR. However, from the perspective of evaluators, the planning process should:
\begin{itemize}
    \item Identify an open GenAI tool as a baseline and explain the choice of any other GenAI tools being evaluated \citep{thomas-2026-RAISE1,wagner2025,woelfle-2022}. Use of an open tool can be pinned to a specific, archived version that other researchers can rerun, which supports reproducibility. Proprietary, hosted tools also change, but it is often hard to tell exactly what has changed, an earlier version usually cannot be re-accessed, and the tool may become unavailable or prohibitively expensive.
    \item Produce a data-management plan for:
        \begin{itemize}
            \item Collecting and storing all performance evaluation data and applying any necessary quality control processes.
            \item Storing prompts, outputs, and model versions \citep{thomas-2026-RAISE1,wagner2025,woelfle-2022}. 
        \end{itemize}
    \item Put in place procedures needed to ensure data privacy if GenAI tool use requires uploading of PDFs of published papers, sensitive data, or patent-level information without appropriate permissions and safeguards.
    \item Specify the basic prompt strategy, which must include:
    \begin{itemize}  
        \item Ensuring that the tool is prompted to deliver the information needed to investigate the nature of disagreements between tools and humans --- for example, its answer to each eligibility criterion or evaluation question, and the supporting text extracts or page references on which its decision rests.
        \item Ensuring that final prompts are duplicated to assess GenAI tool reliability \citep{woelfle-2022}.
        \item Identifying the process to be used to measure the power and environmental costs of GenAI use. For hosted tools the provider controls the infrastructure, so precise energy use is hard to measure. In practice, we recommend reporting feasible proxies such as the number of API calls or tokens, and the monetary cost.
    \end{itemize}
    \item If required, given the study goals and hypotheses, compare GenAI tool results with a baseline derived from two or more human researchers and report agreement metrics \citep{thomas-2026-RAISE1,wagner2025,woelfle-2022}.
    \item Include an evaluation of environmental and power costs in any assessment of GenAI tool efficiency.
\end{itemize}

In addition, evaluators need to consider the impact of obtaining data from multiple SLRs, in particular:
\begin{itemize}
    \item The data management plan must address the collection and collation of data from multiple SLRs.
    \item The data analysis plan must consider how to aggregate the results from several different SLRs.
\end{itemize}
Also, for prospective evaluation studies, evaluators must specify the procedures that will be followed when collaborating with the SLR review teams.

Evaluators must take into account both the nature of specific SLR tasks and the limitations of GenAI tools when planning their experimental process and performance measures \citep{thomas-2026-RAISE1,woelfle-2022,Madeyski26LLM4SCREENLIT}. Individual SLR processes and their requirement are discussed in Appendix~\ref{sec:SLRProcessDetails}.  %

\subsubsection{Conducting GenAI tool Evaluations}

Evaluators should follow the processes and procedures defined during planning and keep a record of any deviations from the planned experimental procedures. Evaluation of GenAI tools does not change this.  

However, if needing to obtain fair comparisons between human researchers and GenAI tools, human researchers would need to be rigorous with respect to both time keeping for any SLR tasks they conduct, and to ensuring that they also record any additional information that evaluators require to assess the validity of the GenAI tool.%

\subsubsection{Reporting GenAI tool Evaluations}

As is usual for empirical studies, evaluators must report their experimental goals and design and the conduct of their study, including any deviations from their planned process. This section identifies any additional requirements that apply to the evaluation of GenAI tools. 

Evaluators should report:
\begin{enumerate}
    \item The rationale for evaluating the specific GenAI tools \citep{woelfle-2022}.
    \item A description of the SLR(s) that provided the evaluation data.
    \item If data were obtained from multiple SLRs, the contribution of the different SLRs to generalizability should be discussed.
    \item The role of any human participants contributing to the assessment of human performance should be described \citep{woelfle-2022}.
    \item The prompt engineering process, including prompt templates and the methods used to validate prompt outcomes. This information may be included in the final report or in accessible supplementary materials \citep{thomas-2026-RAISE1,wagner2025,woelfle-2022}. 
    \item Model versioning and configuration details, including all parameter values \citep{thomas-2026-RAISE1,wagner2025,woelfle-2022}. 
    \item Any metrics used to assess tool performance \citep{wagner2025,woelfle-2022}.
    \item GenAI tool intra-tool reliability, which can be assessed by running final prompts several times \citep{woelfle-2022}. 
    \item The results of the evaluation in terms of the effectiveness (validity) and efficiency (cost-benefit) \citep{Madeyski26LLM4SCREENLIT}.
    \item Any concerns regarding possible GenAI tool bias \citep{thomas-2026-RAISE1}.
    \item Any financial or other personal interests in the evaluation results and, if necessary, how any risk of evaluator bias due to reported interests was mitigated \citep{thomas-2026-RAISE1}.
\end{enumerate}

In addition, evaluators should discuss any limitations or constraints associated with their study, for example:
\begin{itemize}
    \item Are the results topic or discipline dependent? For example, GenAI performance in the medical domain may depend on the use of the well-defined, unambiguous terminology used in many medical research papers.
    \item Given the current requirements for human oversight in academic research, are the results compatible with this requirement? If they are, are there any process issues users of research need to be aware of? 
\end{itemize}

These recommendations assume that the evaluators have not used GenAI tools to support their evaluation process. However, if the evaluators have used GenAI tools not just as the objects of the evaluation but also as tools to assist the evaluation (e.g., for data collection and collation, or for report writing), they should ensure that their report adheres to the general reporting requirements for GenAI tool use described in the recommendations for reviewers and \Cref{tab:Guidelines}. For example, \citet{spillias-2024}, who undertook a case study to evaluate GenAi tool performance on literature screening, included an explicit declaration of their use of generative AI and AI-assisted technologies in the writing process.

\subsection{Using GenAI tools}\label{sec:TestingANdEvaluatingGenAITools}
In this section, we identify the extensions to the SLR process and reporting guidelines that are needed to address the issues raised by using GenAI tools. We consider the three main phases of the SLR process (Planning, Conduct, and Reporting), together with a new phase we refer to as Tool Testing and Validation. This new phase takes place in parallel with the planning and conducting phases.

\subsubsection{Tool Testing and Validation}\label{sec:ToolTestingAndValidation}
There are several different scenarios to consider:
    \begin{itemize}
        \item Reviewers may be concerned mainly with the conduct of a specific SLR. In such circumstances, they are likely only to be interested in using tools that reduce effort or that can be used to improve SLR validity. However, if review teams have only limited expertise in GenAI tools, they may need to include a GenAI researcher in their research team.
        \item Reviewers may be collaborating with evaluators to conduct a case study of GenAI tool performance. In this case, the nature of the collaboration needs to be defined. At one extreme, the collaboration can be fully independent, where the reviewers conduct their SLR without the use of any GenAI tools, but make all their process results available to a separate evaluation team that conducts the same processes using GenAI tools. A problem with this approach is that by using a GenAI tool an independent evaluation team might identify a flaw in the work of the reviewers. It would violate independence if the evaluators alerted the reviewers to the flaw, but it would endanger the validity of the SLR if they did not and would damage the prospects for further collaboration. Alternatively, the evaluators may be part of the SLR team. This would mean that it would not be possible to investigate automation of the full SLR process, because the requirements of the SLR would mean that human researchers maintained oversight of each process and GenAI tool results could not be included in the SLR without human validation.
    \end{itemize}

Regardless of how the testing and validation are organized, it is critical that all stakeholders understand and accept the goals of the activity, are fully aware of how it will be managed and know what their personal roles and responsibilities will be. 

The first testing and validation task is to identify all SLR processes that are to be GenAI-assisted and to identify appropriate candidate tools. Candidate tools can be identified based on previous tool evaluation studies, which need to be critically evaluated in order to assess tool capability, risks and limitations \citep{thomas-2026-RAISE1}. All reviewers who are required to use GenAI tools should be given any necessary training.

During SLR planning, researchers might consider using GenAI tools to support:
\begin{enumerate}
\item Research question development.
\item Boolean query development.
\item Specification of eligibility criteria.
\end{enumerate}

In the case of query development, \citet{Wang2023} conducted an experiment investigating whether ChatGPT could write an effective Boolean query. They reported that improvements in precision sometimes came with decreased recall, although this could be mitigated by better MeSH terms (MeSH is a standardized, controlled vocabulary thesaurus created by the U.S. National Library of Medicine). A useful element of their study is that they provide a good description of how they developed their prompts to support unguided and guided query refinement. In particular, the guided process was based on a current automated query formulation method. However, in our opinion, their study had several serious threats to validity:
\begin{itemize}
\item They evaluated results with respect to precision, F1, and recall, without considering the different costs of false negatives and false positives. 
\item Although they reported F1 values, they did not discuss the fact that all their tests report low F1 values (the highest F1 values were close to 0.6, and most were less than 0.5) indicating a high proportion of false positives and false negatives.
\item They used a well-known public data source and refined their prompts by multiple iterations of the same task without considering the possibility of data leakage.
\item Dependence on MeSH means their results do not generalize to other topic areas.
\end{itemize}

Assessing the use of GenAI tools for any of these three planning tasks would not necessarily require a separate evaluation process.
It could be restricted to using the tool and checking the outcomes against the opinions of the SLR research team. The procedures adopted for prompting the tool and checking the outcomes should be reported. In particular, the prompt inputs, tool output, and the final version of the output agreed by the research team should be reported in the protocol and the final SLR report or supplementary materials.

Using GenAI tools to support subsequent SLR processes is more complex because subsequent processes are extremely varied. For processes that involve performing the same task on a set of different studies, such as screening candidate primary studies based on the inclusion and exclusion criteria (i.e.,  eligibility criteria) based either on abstracts, or full text, and extracting information about the type of a primary study in the context of mapping study data extraction, reviewers could undertake a GenAI tool evaluation to assess whether using a tool (or tools) could allow tasks usually requiring two human researchers to be performed by one human researcher and one (or more) GenAI tools.

For these processes, the basic test and validation process would be the same as an \emph{independent sample-based prospective evaluation} which would involve:
\begin{enumerate}
\item Selecting a random subset of individual items (e.g., 5\% to 20\% of the required inputs). 
\item Assigning the items to at least two human researchers. We need to have the validation sample performance results from two human researchers because this  represents current best practice for SLRs. Disagreements between the human researchers should be discussed and resolved to produce an agreed decision for each item.
\item Assigning the same items to one or more GenAI tools.
\item Specifying appropriate agreement metrics to measure GenAI performance.
\item Identifying the decision criteria that will be used to assess GenAI tool performance, if decisions about GenAI tool adoption are required.
\item Measuring performance data for processing the items to assess effectiveness and effort, elapsed time and all related costs to assess efficiency (for GenAI tools, this includes human time and effort for prompt engineering and all other costs related to GenAI tool use).
\item Assessing the reliability of GenAI tools, for example, running the agreed prompts at least twice \citep{woelfle-2022}.\footnote{When a large number of items are being assessed (e.g., the candidate primary studies in a screening validation sample), each item contributes a pair of repeated prompt executions, so two repetitions of each prompt are sufficient to assess reliability. If only a few items are involved, researchers might need to consider more repetitions.} 
\item Inconsistent results should be identified as ``Requiring further assessment'' --- they should not be excluded from performance evaluation.
\item Analysing agreement metrics to calculate:
\begin{itemize}
\item Agreement between two human researchers.
\item Agreement between each human researcher and the GenAI tools.
\item Agreement between the GenAI tools and the decision agreed between the human researchers.
\item Sample-based confidence intervals for GenAI performance metrics and GenAI tool reliability.
\end{itemize}
\item Deciding whether, in the context of the SLR goals, the performance of the tool(s), in terms of both cost saving and threats to validity, is good enough to allow the remainder of the tasks to be performed by one human researcher and the tool(s).
\item If the reviewers intend to adopt the use of the tools for their current SLR, the test process and results must be included in the SLR report or supplementary materials.
\end{enumerate}

Additionally, it is important to ensure that the evaluation appropriately identifies the full costs of using a GenAI tool, or tools, to replace a human researcher. In particular,
\begin{itemize} 
\item What changes, if any, would need to be made to the standard human researcher-based SLR process?
\item How would disagreements between the tool(s) and the human researcher be resolved?
\item How would full oversight of GenAI tool output be ensured?
\end{itemize}

Appendix~\ref{sec:SLRProcessDetails} discusses more details related to eligibility screening and data extraction together with process-oriented discussions of the more complex SLR tasks including qualitative synthesis, risk of bias assessment and strength of evidence assessment. We agree with \citet{Thomas-2026-RAISE3} that current GenAI tools cannot be relied on to automate all complex SLR tasks fully. Nonetheless, in  Appendix~\ref{sec:SLRProcessDetails} we discuss the possibility that evaluators could collaborate with reviewers to assess GenAI performance on all or part of these processes. %

\subsubsection{SLR Planning}
SLR planning procedures need to be extended to cover the use of GenAI tools. Specifically, the protocol should include the following information for any GenAI-assisted SLR task, including:
\begin{enumerate}
    \item The basic prompt strategy, which should include ensuring that the tool is prompted to produce additional information to support assessment of the validity of its outputs \citep{thomas-2026-RAISE1, wagner2025,woelfle-2022}.
    \item A data-management plan \citep{thomas-2026-RAISE1} for:
        \begin{enumerate}
            \item Storing prompts, outputs, and model versions.
            \item Labelling, collecting and storing all human and tool performance measures.
            \item Any quality assurance processes applied to collected data.
        \end{enumerate}
    \item The procedures needed to ensure data privacy if GenAI tool use requires uploading PDFs of published papers,  sensitive data, or patent-level information without appropriate permissions and safeguards.
    \item Specifying how any GenAI tool testing and validation will be integrated with SLR conduct. 
    \item The processes used to monitor tool output for errors and bias. Process methods must include: 
     \begin{itemize}
         \item Running the finally agreed prompt twice and checking for consistency \citep{woelfle-2022}. 
        \item Comparing the GenAI tool output with that of at least two human researchers \citep{thomas-2026-RAISE1, wagner2025,woelfle-2022}.
        \item For any SLR process that involves running multiple equivalent tasks, reporting agreement matrices for tasks (excluding tasks that were part of the GenAI tool test) \citep{Madeyski26LLM4SCREENLIT}.  
        \item Checking outputs for possible GenAI bias, such as ignoring or downgrading studies with female authors or authors with non-European names, or that express minority opinions or uncommon results \citep{thomas-2026-RAISE1, Pizard26}.
    \end{itemize} 
    \item Compiling an environmental and computing cost statement for use of the GenAI tool.
    \item Defining the process to be adopted if GenAI errors or biases are detected, which would usually involve referring the task back to another human researcher \citep{Madeyski26LLM4SCREENLIT,woelfle-2022,Pizard26}.
    \item Contingency plans to deal with inadequate GenAI tool performance, which could be:
    \begin{itemize}
        \item Referring the task(s) back to human reviewers not previously involved in the specific task(s).
        \item Revising the prompt strategy.%
    \end{itemize}
\end{enumerate}

\subsubsection{SLR Conduct}
SLR guidelines have always specified that the conduct of an SLR should follow the processes and procedures defined in the protocol and report any deviations from the protocol. The use of GenAI tools does not change this.   

However, human participants in GenAI tool training and validation may need to be more rigorous with respect to:
\begin{enumerate}
    \item Time keeping. Usually, members of the SLR review team are not required to provide accurate reports of the time they spend on particular SLR tasks. However, for validation purposes, human researchers should keep accurate records, including an allowance for break periods, while performing tasks which are part of GenAI tool validation, without which they are likely to become tired and error-prone, but excluding time lost due to external interruptions.
    \item Providing any extra information needed to support comparisons between AI tool results and results obtained by human researchers. For example, if the GenAI tool is required to provide information to support or explain its response, human researchers will need to do the same, even if such details would not normally be recorded.
\end{enumerate}

\subsubsection{SLR Reporting}
Current SLR standards (PRISMA-2020 \citep{page-2021} and SEGRESS \citep{kitchenham-2023}) already require full reporting of the use of GenAI tools, including training and validation details \citep{thomas-2026-RAISE1}. However, certain issues need to be emphasized:
\begin{enumerate}
    \item Explain and justify the use of GenAI tools \citep{thomas-2026-RAISE1,woelfle-2022}. 
    \item Ensure humans review all GenAI-assisted results to reduce the risk of fabricated conclusions, data, graphics, quotations, or references \citep{Pizard26}. 
    \item Report the prompt engineering process, including prompt templates, and the methods used to validate prompts, either in the final report or in accessible supplementary materials \citep{thomas-2026-RAISE1,wagner2025, woelfle-2022, Pizard26}.
    \item Report model versioning and configuration details, including all parameter values in the methods section \citep{Thomas-2025,wagner2025, woelfle-2022, Pizard26}.
    \item Report any concerns regarding possible GenAI tool bias \citep{thomas-2026-RAISE1, Pizard26}.
    \item Report the full environmental impact and computing-cost statement, including both training and use of the GenAI tool(s).
    \item Adhere to legal and ethical issues relating to the use of GenAI tools reported in \Cref{tab:Guidelines}, and in addition:
    \begin{itemize}
        \item Do not upload copyrighted or sensitive data to commercial APIs.
        \item Ensure that GenAI-assisted results do not breach plagiarism regulations. 
        \item Report any financial interests or other special interests associated with the adoption of GenAI tools.
    \end{itemize}
\end{enumerate}

\subsubsection{Summary of Process Changes}
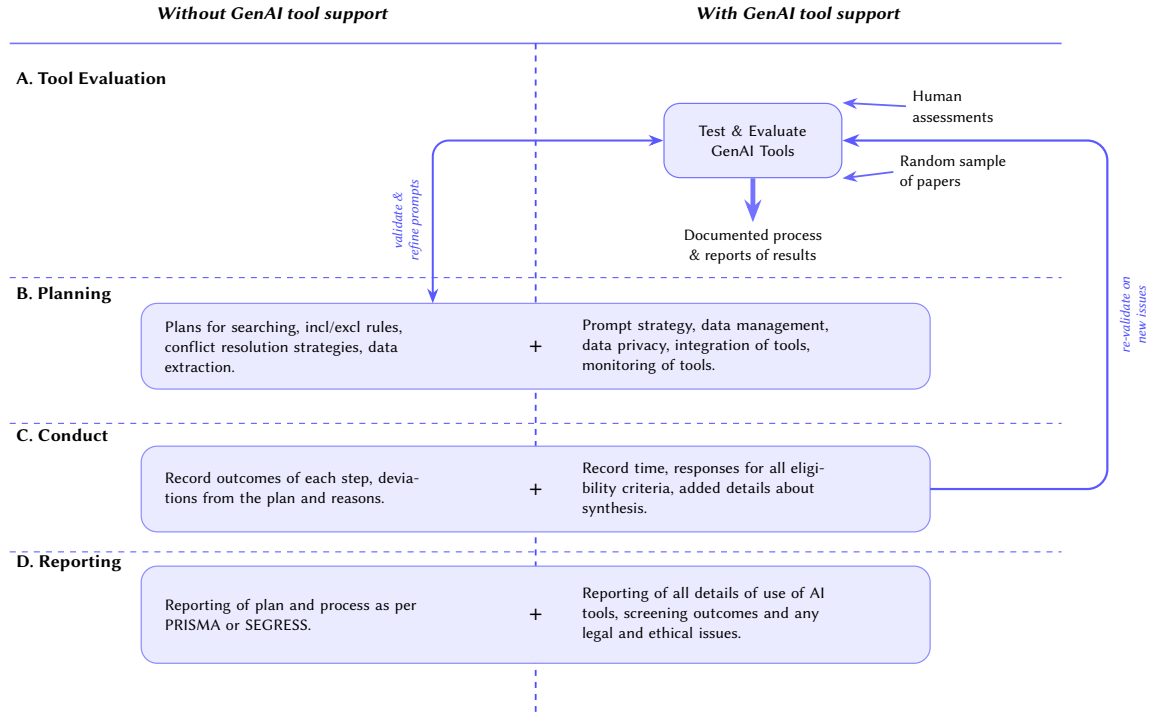
\begin{figure*}[!t]
\centering
{\sffamily\resizebox{\textwidth}{!}{%
\begin{tikzpicture}[
  font=\small,
  band/.style={font=\bfseries},
  bigbox/.style={rounded corners=8pt, draw=blue!45, fill=blue!8, line width=0.6pt},
  evalbox/.style={rounded corners=8pt, draw=blue!55, fill=blue!8, line width=0.6pt, align=center, text width=2.9cm, minimum height=1.3cm},
  txt/.style={align=left, text width=4.4cm, font=\small},
  fb/.style={-{Stealth[length=3mm]}, draw=blue!65, line width=1.1pt},
  fb2/.style={{Stealth[length=2.6mm]}-{Stealth[length=2.6mm]}, draw=blue!65, line width=1.0pt},
  inarrow/.style={-{Stealth[length=2.6mm]}, draw=blue!60, line width=0.9pt},
  down/.style={-{Stealth[length=4mm]}, draw=blue!55, line width=2.4pt},
]
\def\xLft{-9.2}
\def\xRgt{9.2}
\def\xC{0}
\def\xL{-4.6}
\def\xR{4.6}
\def\xFB{10.0}
\def\ybot{0.4}
\node[font=\bfseries\itshape] at (\xL,12.6) {Without GenAI tool support};
\node[font=\bfseries\itshape] at (\xR,12.6) {With GenAI tool support};
\draw[blue!55, line width=0.8pt] (\xLft,12.1) -- (\xRgt,12.1);
\draw[blue!70, dashed, line width=0.9pt] (\xC,12.1) -- (\xC,\ybot);
\node[band, anchor=west] at (\xLft,11.5) {A. Tool Evaluation};
\node[band, anchor=west] at (\xLft,7.7)  {B. Planning};
\node[band, anchor=west] at (\xLft,5.25) {C. Conduct};
\node[band, anchor=west] at (\xLft,3.0)  {D. Reporting};
\draw[blue!70, dashed] (\xLft,8.0)  -- (\xRgt,8.0);
\draw[blue!70, dashed] (\xLft,5.45) -- (\xRgt,5.45);
\draw[blue!70, dashed] (\xLft,3.2)  -- (\xRgt,3.2);
\node[evalbox] (eval) at (\xR-0.8,10.4) {Test \& Evaluate\\ GenAI Tools};
\node[align=left] (ha) at (\xRgt-1.9,11.0) {Human\\ assessments};
\node[align=left] (rs) at (\xRgt-1.9,9.85) {Random sample\\ of papers};
\draw[inarrow] (ha.west) -- (eval.north east);
\draw[inarrow] (rs.west) -- (eval.south east);
\node[align=center] (docout) at (\xR-0.8,8.55) {Documented process\\ \& reports of results};
\draw[down] (eval.south) -- (docout.north);
\draw[bigbox] (-6.9,6.05) rectangle (6.9,7.55);
\node[txt, anchor=west] at (-6.6,6.8) {Plans for searching, incl/excl rules, conflict resolution strategies, data extraction.};
\node[font=\Large\bfseries] at (\xC,6.8) {$+$};
\node[txt, anchor=west] at (0.7,6.8) {Prompt strategy, data management, data privacy, integration of tools, monitoring of tools.};
\draw[bigbox] (-6.9,3.55) rectangle (6.9,5.05);
\node[txt, anchor=west] at (-6.6,4.3) {Record outcomes of each step, deviations from the plan and reasons.};
\node[font=\Large\bfseries] at (\xC,4.3) {$+$};
\node[txt, anchor=west] at (0.7,4.3) {Record time, responses for all eligibility criteria, added details about synthesis.};
\draw[bigbox] (-6.9,1.25) rectangle (6.9,2.95);
\node[txt, anchor=west] at (-6.6,2.1) {Reporting of plan and process as per PRISMA or SEGRESS.};
\node[font=\Large\bfseries] at (\xC,2.1) {$+$};
\node[txt, anchor=west] at (0.7,2.1) {Reporting of all details of use of AI tools, screening outcomes and any legal and ethical issues.};
\draw[fb, rounded corners=10pt] (6.9,4.3) -- (\xFB,4.3) -- (\xFB,10.4) -- (eval.east);
\node[align=center, font=\footnotesize\itshape, text=blue!70, rotate=90] at (\xFB+0.45,7.35) {re-validate on\\ new issues};
\draw[fb2, rounded corners=8pt] (-1.8,7.55) -- (-1.8,10.4) -- (eval.west);
\node[align=center, font=\footnotesize\itshape, text=blue!70, rotate=90] at (-2.25,9.0) {validate \&\\ refine prompts};
\end{tikzpicture}%
}}
\caption{Summary of SLR Process Extensions}
\label{fig:SLRProcessChanges}
\end{figure*}

\Cref{fig:SLRProcessChanges} summarizes the general process changes required for using GenAI tools to conduct an SLR. It does not consider the detailed process change discussed in Appendix~\ref{sec:SLRProcessDetails}. We only report the extensions to the general process, not the savings from GenAI tool use, since the human researchers will still need to perform a part of all the SLR tasks manually, even with the support of GenAI tools.

The process is iterative and incremental rather than strictly linear.
Planning fixes the acceptable performance levels and the initial prompts.
Tool testing and validation then check the tool against those levels.
If the tool falls short, the team refines the prompts or revises the plan --- for example, by invoking the contingency plans agreed during planning --- and validates again before the tool is used during conduct.
Issues noticed during conduct, such as unexpected outputs or borderline cases, may likewise send the team back to refine prompts or re-validate.
\Cref{fig:SLRProcessChanges} shows the main extensions.
In practice, the steps it lists are revisited as needed.

\section{Details of SLR Process and GenAI Considerations}\label{sec:SLRProcessDetails}
This section discusses the detailed tasks needed to conduct the following SLR processes, and considerations for incorporating GenAI in these.
\begin{enumerate}[noitemsep]
    \item Eligibility of candidate primary studies based on title, abstract and keyword (TAK) information.
    \item Eligibility based on full text assessment.
    \item Data extraction.
    \item Risk of Bias (RoB) assessment.
    \item Synthesis of qualitative primary studies.
    \item Strength of Evidence (SoE) assessment.
\end{enumerate}
The goal of this section is to provide a framework to help develop preliminary task-specific SLR recommendations.
\subsection{Literature Screening}

In this section, we summarize the most important issues from our revision of \citet{Madeyski26LLM4SCREENLIT}. We consider both abstract screening and full-text screening.

Human-based TAK and full-text eligibility assessment require the following process steps:
\begin{enumerate}
    \item Defining the eligibility criteria. These should be based on the SLR research questions (which would also have driven the candidate's primary study search process). Note, eligibility criteria may need to be revised when proceeding from TAK-based screening to full-text-based screening.
    \item Trialling the eligibility criteria on a random selection of candidate primary studies, to ensure that the criteria are properly understood and used correctly by human researchers.
    \item Assessing each of the candidate primary studies identified by the search process against the eligibility criteria and retaining any study that cannot be rejected from the list of candidate primary studies. For a study claiming to be a formal SLR, this process should be conducted by two human researchers working independently, and disagreements would need to be identified and resolved either by discussion or by independent arbitration conducted by a third researcher. Mapping studies, scoping studies and rapid reviews often weaken this requirement.
\end{enumerate}

\citet{Madeyski26LLM4SCREENLIT} point out that for studies related to screening candidate primary studies, evaluators should plan for highly imbalanced data. Evaluators must also recognize that the cost of False Negatives is lost evidence, whereas the cost of False Positives is additional human effort, so False Negatives are more of a problem than False Positives in the context of SLR validity. Therefore, we recommend:
\begin{enumerate}
    \item Use performance metrics that are reliable when classification groups are imbalanced. The Matthews Correlation Coefficient (MCC) is a good option because, like all correlation coefficients, values close to zero suggest that our predictions are not much different from chance.
    \item False negatives are more important than false positives, so either use WMCC to support comparisons of performance, or conduct a cost-benefit analysis based on the cost ratio of FP to FN.
    \item Full confusion matrix counts are essential to support meta-analysis.
\end{enumerate}
These recommendations are consolidated as Scr1--Scr3, and the validation-study items below as ScrRev1--ScrRev3, in \Cref{sec:Recommendations}.

For researchers conducting an SLR who need to decide whether or not to deploy a GenAI tool, deployment decisions should be based on the results of a validation study where:
\begin{itemize}
    \item Researchers will need to decide which GenAI tools to consider, which prompt strategies to use and which deployment process(es) to evaluate.
    \item Acceptable rates of false negatives and their relative costs relative to false positives should be set in advance, based on the type of SLR and the requirements of stakeholders.
    \item To support decision making, validation results should include estimates of performance metric variability (e.g., confidence intervals based on resampling).
    \item If the validation results support GenAI tool use, mechanisms to maintain human oversight need to be specified.
 \end{itemize}   

 Screening full text is similar to TAK screening, however:
\begin{enumerate}
    \item We would not expect the extreme imbalance observed in TAK screening. However, our recommendations for appropriate metrics are unchanged.
    \item It may be necessary to amend/refine the eligibility criteria and/or prompt design. \citet{Cao2024} changed their prompt design for full text screening. However, \citet{Khraisha2024} used the same prompt process for full text screening.
    \item There may not be enough primary studies to allow a realistic validation study. In such circumstances, reviewers may decide to use GenAI tools in the role of an additional research assistant while still involving two human researchers in each eligibility assessment. This would provide both additional quality assurance on the screening task and the opportunity to assess GenAI tool capability.
\end{enumerate}

\subsection{Data Extraction}\label{sec:DataExtraction}
 Data Extraction involves the following subtasks:
 \begin{enumerate}[noitemsep]
     \item Creating a data extraction form or set of forms,
     \item Validating data extraction forms,
     \item Using the data extraction forms,
     \item Converting extracted data into usable formats.
 \end{enumerate}

\subsubsection{Creating Data Extraction Forms} 
This involves defining what data items to extract. General issues related to data form definition include:
\begin{itemize}
    \item Defining the scope of the data collection exercise, which depends on research questions and planned synthesis method.
    \item Addressing the need for both structured fields and free-text fields.
\end{itemize}

The  main difficulty in the context of SE, is that data items tend to be  very heterogeneous:
\begin{enumerate}
    \item For quantitative SLRs, data includes effect sizes, sample sizes, p-values, and confidence intervals. In addition, outside the framework of families of experiments \citep{basili-2002}, unlike medical researchers, SE researchers do not have well-defined vocabularies to describe their metrics and experimental conditions. For these reasons, it may be necessary to specify the unit and/or metric definition used in each primary study, as well as the value of each required data item.
    \item For qualitative SLRs, data includes: study context, methodology descriptions, participant characteristics, and findings. 
    \item For mapping studies, data usually involves multi-class categorization of primary studies along characteristics such as research method, SE domain, contribution type, and venue type.
\end{enumerate}

For researchers conducting an SLR, GenAI could assist in form creation (e.g., suggesting relevant data items for the given research questions), but the form(s) would need to be validated by human researchers. 

\subsubsection{Validating Data Extraction Forms} 
This involves piloting the agreed data collection form on a subset of primary studies. The human-based validation process involves:
\begin{enumerate}
    \item Two or more researchers independently extracting data from a few randomly selected studies.
   \item Discussing any disagreements in order to refine data form definitions and instructions.
    \item Iterative refinement until consistent use of the form is achieved.
\end{enumerate}

Researchers considering the use of GenAI tools to support a specific SLR could incorporate a GenAI tool into the data form validation process. GenAI tools could replace one human extractor or provide additional support for human data extraction, but this would require evaluation of: 
\begin{enumerate}
    \item The accuracy achieved for extracting numerical data, which requires an exact match. However, if the data item is used in a specific primary study, the gold standard ``true'' value exists in that primary study and applies whether the extraction is performed by human researchers or by a GenAI tool. The existence of a gold standard value offers an exception to the general approach of relying on the results obtained by two human researchers to assess GenAI tool performance\footnote{We note that, since primary study analyses and reporting could be flawed, the gold standard value could still prove to be incorrect when judged against a re-analysis of the raw data. While human researchers might be able to recognize implausible or impossible values during data extraction, a GenAI tool targeting data extraction might not be able to make such assessments.}.
    \item For mapping studies, classifications of primary study type require an exact match. Nonetheless, in the short term, this is likely to be the most viable context for incorporating GenAI tool data extraction into the SLR process because classifying a primary study against a set of defined categories is similar to checking candidate primary studies against eligibility criteria. For example, \citet{Khraisha2024} treated each extraction field as a binary detection problem and used the same performance metrics they used for data screening.
    \item The accuracy of qualitative/textual data. For this, semantic similarity may be sufficient, but it is not necessarily easy to assess. It is likely to require additional effort to evaluate the equivalence between the textual results obtained by human researchers performing the data extraction task and the outputs of GenAI tools.
    \item Handling of missing data when a primary study simply does not report a data item that is required, e.g., no standard deviation reported, no explicit description of participant selection method.
    \item Managing hallucination risks, including the fabrication of values not present in the source paper, fabrication of textual content, misattribution of textual information (to the wrong study, the wrong RQ, the wrong experimental condition), and plausible confabulation (i.e., output that is internally consistent and sounds reasonable but does not correspond to anything in the source.
\end{enumerate}

Data extraction performance metrics need to address all the above issues. \citet{Madeyski26LLM4SCREENLIT} point out that reliable metrics such as the Matthews correlation coefficient (MCC) can  be extended to address the classification issue (i.e. multiple categories for a nominal scale variable) and multiple outcomes  (i.e., missing correct, incorrect). However, there is currently little research literature addressing this SLR stage. 

\subsubsection{Using the Data Extraction Form} This means applying the data collection form and process across all primary studies (except those used for the validation exercise - unless researchers agree that further cross-checking is needed). 

The human data extraction process involves at least two researchers who extract independently, and disagreements are resolved. When researchers are conducting an SLR, GenAI tools can support this process or might replace a human researcher. However, to maintain human oversight, tools should be prompted to provide source quotes/page references for each extracted value to support traceability and verification.

A key difference between performance metrics used for screening candidate primary studies and performance metrics used for data extraction is that data extraction errors are harder to detect downstream (e.g., a wrong, missing or spurious effect size can silently corrupt meta-analysis). This makes the issue of validating GenAI tool performance critical. In addition, there is no inherent difference between types of error, which implies researchers should rely on MCC as the main performance metric without needing to weight different sorts of errors.

\subsubsection{Collating Extracted Data} This involves converting extracted data into usable formats, and preparation for synthesis. This includes issues such as:
\begin{itemize}
    \item standardizing units, 
    \item converting between effect size measures
    \item handling incomplete reporting,
    \item highlighting anomalous values for additional validation.
\end{itemize}

GenAI tools could assist with unit conversions, anomaly detection and deriving missing statistics, but such computations would need to be verified because GenAI tools are known to make arithmetic errors.

\subsubsection{GenAI Performance Evaluation} For researchers wanting to assess the performance of GenAI tools outside the context of a specific SLR, it is critical to define the specific aims of the investigation (i.e., which subtasks are being addressed) and the starting and end points of the study. For example, is the data form available or is the GenAI tool(s) expected to construct a data form? Are the GenAI tool(s) expected to provide data that is reliable enough to be used without further validation? If the tool(s) is required to replace human researchers as opposed to assisting them, how will human oversight be maintained?

\subsection{Risk of Bias (RoB) Assessment}\label{sec:RiskOfBiasAssessment}
The goal of risk of bias assessment is to assess whether there are any weaknesses in the research methodology adopted by each of the included primary studies. Poor research methodology casts doubts on the trustworthiness of primary study results and findings which, in turn, would undermine any SLR results relying on the primary study. Although risk of bias is considered essential for quantitative and qualitative SLRs intended to provide recommendations for SE practice or academic research, it is not necessary for mapping or scoping studies. It is optional for rapid reviews.

Characteristics of the risk of bias (RoB) assessment task are that:
\begin{enumerate}
    \item RoB assessment relies on having an appropriate set of criteria for assessing the methodological weaknesses of a primary study. Such criteria are usually organized as a set of questions with explanations about how the questions should be answered and are referred to as an evaluation (or assessment) instrument (or tool).
    \item RoB evaluation questions include:
    \begin{itemize}[noitemsep]
        \item Was the study prospective or retrospective?
        \item Was the method appropriate to address the research question(s)?
        \item Were best practices for the specific method used?
        \item Were known weaknesses of the methods properly addressed?
        \item Were data correctly analysed/synthesized?
        \item Were reported findings clearly linked to the analysis/synthesis results?
        \end{itemize}
    \item To provide reliable assessments, RoB evaluation questions are usually extended to provide more details about how to answer them. For example, for formal (laboratory) experiments, good practices include randomized assignment to treatment and methods to avoid participant and experimenter bias, while weaknesses include small sample sizes, reliance on student participants and small-scale or over-simplified tasks and task materials.
    \item Evaluation assessments require a means of aggregating the answers to each question into an overall assessment of the RoB associated with a primary study. Such an assessment is usually based on a subjective qualitative ordinal scale of the type: Critical RoB, High RoB, Moderate RoB, Low RoB
    \item Quantitative and qualitative primary studies usually require different evaluation instruments because they ask different types of questions using different research methods. 
    \item Even in the case of quantitative primary studies, an evaluation instrument usually involves some subjective judgements. For this reason, RoB evaluation is usually performed by two or more human evaluators and disagreements need to be discussed and resolved.
\end{enumerate}
In the next sections, we discuss the four main tasks involved in RoB evaluation: 
\begin{enumerate}[noitemsep]
    \item Creating Evaluation Instruments.
    \item Validating Evaluation Instruments for a specific SLR.
    \item Using an Evaluation Instrument
    \item Converting the evaluation results into a RoB assessment
\end{enumerate}

\subsubsection{Creating Evaluation Instruments}
Currently, research has concentrated on evaluating whether GenAI tools can be trained to use the RoB or the updated RoB2 instrument \citep[for a comparison of the two instruments see][]{nejadghaderi-2024}. This is only applicable to quantitative experiments that conducted randomized controlled trials (referred to as RCTs), and is, therefore, of little value in the context of SE studies. However, there are major disagreements about the performance of GenAI tools even for these well-defined instruments that assess a well-defined experimental method. For example, \citet{suster2024} assessed LLM performance by applying the RoB2 questionnaire to 213 papers describing randomized controlled trials. They used four LLMs (FlanT5XL, ChatGPT (gpt-3.5-turbo), Meditron-70B, and Med42) and reported that ``The results fall short of expectations, with LLMs seldom surpassing trivial baselines''. In contrast, \citet{lai-2024} assessed the RoB questionnaire on 30 papers, and, using ChatGPT and Claude they concluded that the LLMs ``demonstrated substantial accuracy and consistency in evaluating RCTs, suggesting their potential as supportive tools in systematic review processes''. These studies used different risk of bias questionnaires, different primary studies, different GenAI tools, and different prompting strategies. So, they illustrate why it is essential that we develop appropriate procedures for evaluating GenAI tools. Without agreement on reliable research methods, it will not be possible to evaluate the RoB of current research, which will handicap attempts to meta-analyse results and to assess the reasons for differences. 

Thus, a critical issue for SE researchers intending to evaluate GenAI performance on the RoB tasks is to identify an appropriate evaluation instrument. Furthermore, the SE community needs to come to some agreement about such instruments or we will be unable to aggregate results from different experiments.  

Studies relating to using GenAI tool to construct evaluation instruments would need to consider:
\begin{enumerate}
    \item The scope of any generated evaluation instrument i.e., what type of empirical studies should be considered. The scope of a specific evaluation instrument would depend on the type of empirical studies expected in a specific SLR.  %
    \item How to construct RoB evaluation questions. Evaluation questions need to be based on the known process requirements for using a specific method, any known weaknesses of the specific method, and the extent to which the study's design has correctly conducted the required process and addressed any weaknesses. To generate evaluation instruments, it is necessary to find source material from which to generate and/or refine appropriate evaluation questions. 
    \item How to assess the validity of any constructed evaluation instrument.
\end{enumerate}

This process is difficult because: 
\begin{enumerate}
    \item The different study types used in SE include data mining studies, case studies, quasi-experiments, formal randomized experiments, opinion surveys and a variety of different qualitative methods. Furthermore, a specific SR may include numerous different study types, all of which would need to be assessed with respect to an evaluation instrument calibrated to the likely methodological weaknesses associated with each of the adopted research methods. It should also be noted that a single primary study may use several different methods.
    \item SE researchers do not always correctly label their study types, nor do they always analyse their data correctly \citep[see, for example,][]{Vegas-2016,wohlin-2021,kitchenham-2020,kitchenham-2019}. Thus, evaluators cannot rely on selecting an appropriate evaluation instrument based solely on the method claimed by study authors. Furthermore, reference to a specific analysis method does not guarantee that the method was applied correctly. It is also clear that publication in a reputable journal does not, by itself, guarantee low RoB.
    \item \citet{woelfle-2022} report that GenAI tool accuracy using an evaluation instrument depends on the level of subjectivity of the instrument. They assessed GenAI tool performance on three different evaluation instruments and confirmed that the GenAI tools performed best on the instrument which involved the least amount of subjective assessment, and worst on the instrument that involved extensive subjective assessments.
\end{enumerate}

Given that SE researchers are seldom able to use randomized clinical trials (which are more rigorous than standard laboratory experiments), and SE methodologists have reported problems with the use of empirical methods, we do not believe that it is currently feasible to use GenAI tools to support the generation of evaluation instruments for use in a specific SLR. Nonetheless, research aimed at investigating the construction of RoB instruments would be extremely useful to improve current empirical SE practice.

\subsubsection{Validating an Evaluation Instrument}
In order to use an agreed evaluation instrument in the context of a specific SLR, it is essential both to validate the instrument and to train researchers to use the instrument: 
\begin{enumerate}
    \item Initial validation requires the researchers responsible for constructing the instrument (or tailoring an existing instrument to the types of primary studies included in a specific SLR) to apply the instrument independently to a few randomly selected primary studies and compare their results. Disagreements need to be discussed with a view to determining whether the instrument needs further refinement. 
    \item Subsequently, if other researchers need to use the evaluation instrument, they should also apply the instrument to the selected primary studies. Then, the research group need to discuss any remaining disagreements in terms of possible changes to the instrument, and/or gaining a better common understanding of how to use the instrument.
\end{enumerate}

For researchers conducting an SLR, this task would provide an opportunity to investigate whether GenAI tools could be used to support the RoB evaluation of the remaining primary studies

\subsubsection{Using the Evaluation Instrument}
The evaluation instrument(s) must then be used to assess each primary study. There are different approaches to conducting the human researcher-based process. Issues include deciding:
\begin{enumerate}
    \item How many researchers evaluate each primary study, assuming that at least two researchers should assess each primary study. 
    \item Whether each researcher initially applies the evaluation instrument independently and they discuss disagreements at a subsequent meeting, or they both read the primary study independently but work together to complete the evaluation instrument.
    \item Whether the process is done in tranches, i.e., an initial set of primary studies is evaluated, followed by a discussion of any problems and issues arising from the first set of reviews. This is useful if a fairly large and diverse set of primary studies is being evaluated.
    \item How the evaluation results are converted into a RoB assessment (see the following section).
\end{enumerate}

During the conduct of an SLR, it would be possible to incorporate a prospective study aimed at assessing the performance of GenAI tools alongside the research process conducted by humans. A prospective study would investigate
\begin{enumerate}
    \item Whether the performance of one or more of the GenAI tools is reliable enough to replace one or both human researchers. In the case of a single GenAI tool, the tool should be applied several times to the same primary study using the same prompt. In the case of multiple GenAI tools, the results from the different tools can be compared. In either case, it is necessary to identify the rate of disagreements. 
    \item Whether the performance of one or more of the GenAI tools working with a single human researcher is reliable enough to replace one human researcher. 
\end{enumerate}
In both cases, a high rate of disagreement between human researchers and GenAI tools would suggest that the value of the GenAI tools was limited. 

It is also worth noting that GenAI tools might be useful for highlighting text related to specific evaluation instrument questions, even if human researchers remain responsible for answering the question. 

\subsubsection{Converting the evaluation results into a RoB assessment}
The final issue with the use of evaluation instruments is converting the set of evaluation criteria into a statement of risk of bias.  It is important to note that RoB is not additive. If one criterion has a high risk of bias, the study as a whole will be classified as having a high risk of bias, even if other criteria have a low risk of bias. Formalizing the aggregation process for each evaluation instrument is necessary for human researchers and would also support investigations of the performance of GenAI tools.

\subsection{Qualitative Synthesis}\label{sec:QualitativeSynthesis}
For Qualitative synthesis (QS), findings are extracted from individual studies, and are then integrated and interpreted to build a collective understanding, identify common patterns, and generate new interpretations or hypotheses.
Qualitative synthesis methods are difficult to understand, use and report, because of the following aspects, which we summarize from \citet{Pizard26}:
\begin{enumerate}
\item There is a wide range of methods that can be used to conduct QS (many adapted from qualitative analysis), which have different approaches and processes. 
\item One important aspect of qualitative synthesis methods is the form of data transformation, which exists along a continuum. At one end are aggregative approaches, which summarize and integrate findings across studies with minimal reinterpretation. At the other end are interpretive (or configurative) approaches, which develop new conceptual understandings by reanalysing and reinterpreting evidence.
\item  Another important aspect is the researchers' epistemological stance. This can be described as a spectrum ranging from an idealist position, in which reality is constructed through multiple, alternative human interpretations, to a realist position, in which reality exists independently of thought and can be directly observed and known. In general, idealist stances align with constructivist perspectives, whereas realist stances align with positivist perspectives.
\item Researcher subjectivity plays a significant role in the reliability of QS and their outcomes. and it must be considered in a way that aligns with the epistemological stance. In realist approaches, such as content analysis, subjectivity is viewed as a potential source of bias that needs to be measured and controlled to ensure coding reliability. Conversely, idealist approaches recognize subjectivity as an inherent and valuable aspect of research. These approaches emphasize the importance of critically reflecting on the researcher’s positionality and influence on the study.
\item Several other concepts are important in QS. These include analytical orientation (inductive or deductive elaboration), the use of coding as an intermediate instrument, conceptual innovation (the use of concepts not present in primary studies\footnote{It may be difficult to distinguish conceptual innovation by human researchers from plausible confabulation by GenAI tools}.), data heterogeneity, and the intended use of the synthesis outcomes.
\end{enumerate}

\subsubsection{The Qualitative Synthesis Process} The QS process involves three main steps:
\begin{enumerate}
    \item \textbf{Plan the synthesis}: Define the characteristics of the synthesis to be conducted and decide on the method(s) to be used, the role of the researchers, and how the process and its results will be validated. This involves considering the research questions, the intended use of the synthesis results, the types of data found in the primary studies, and the resources of the review team. In software engineering, it is quite common to use different methods for the RQs of an SR. For example, some questions that aim to characterize primary studies can be answered with categorization, while other open questions may be addressed with more interpretative approaches that allow for an inductive orientation (e.g., thematic analysis).
    \item \textbf{Conduct the synthesis}: Carry out the synthesis of the primary studies. Generally, this involves following the plan developed earlier and executing the steps of the selected methods. However, it can also include incorporating activities based on the results obtained. For instance, if categorizing studies, new dimensions to explore may be found\footnote{This is also common in mapping studies.}. It is also important to maintain a record of the process, decisions made, and intermediate results to allow for auditing of the process later.
    \item \textbf{Validate the synthesis}: In this stage, evaluate whether the process was conducted and the results are adequate. We recommend using criteria predefined in the planning stage. 
  \end{enumerate} 
\citet{Pizard26} propose the following list of characteristics for a sound QS, as the basis of evaluation criteria:
\begin{enumerate}
\item The findings should be grounded in data from the primary studies and account for the observed patterns \citep{sandelowski-2007,lewin-2018,flemming-2021}. They should provide a full and fair representation of the perspectives and understandings of the primary study authors \citep{sandelowski-2007,krefting-1991}. Any contradictory evidence from the primary studies should be reported and, if possible, explained. 
\item The findings should have practical relevance to the research question, and provide information to support their transferability to other contexts  \citep{sandelowski-2007,lewin-2018,krefting-1991}. 
\item The findings should be appropriately weighted and considered for use, which may require identifying and reporting the methodological limitations of the primary studies that underpin them \citep{lewin-2018,krefting-1991}. 
\item The choice of synthesis method should be appropriate and well justified \citep{sandelowski-2007,lewin-2018,krefting-1991}, and should be consistent with the research question, the chosen epistemological approach, and the available primary study data.  
\item The conduct of the method should be consistent with good practice, its reporting should be transparent, and there should be sufficient information to enable auditing \citep{sandelowski-2007,lewin-2018,krefting-1991}.  
\item The researchers should address their subjectivity appropriately and in alignment with the epistemological approach they have chosen \citep{krefting-1991}. In realist approaches, addressing subjectivity typically involves independent coding by multiple researchers and the calculation of agreement metrics. In contrast, more idealist (non-positivist) approaches require authors to demonstrate reflexivity about the interpretive process. This includes critically examining potential influences, such as their background, positions, and interests (especially on the research topic), as well as biases introduced by the software tools used \citep{Weitzmann-2003}. 
\end{enumerate}

\subsubsection{Considerations for Using GenAI tools to Support Qualitative Synthesis}
Given the heterogeneity of methods and possible adaptations to use in the QS process and the few studies on the subject, it is difficult to consider an appropriate and safe way to use GenAI in QS.

It is critical to ensure that any QS is sound using criteria such as those discussed above. The mechanisms to achieve this should be considered and defined during the planning stage. We suggest using predefined evaluation criteria, always involving humans in the process, and proactively identifying and addressing the risks associated with incorporating GenAI into the QS process.

The current state of GenAI and the results reported by Pizard et al. indicate that when conducting an SLR:
\begin{enumerate}
\item GenAI tools could be used for realistic aggregative synthesis (e.g., classifying studies into predefined categories with little interpretation), taking on the role of an independent reviewer. For validation, similar evaluations to those proposed for study selection involving reviewer agreement could be conducted.
\item For more idealist or interpretative syntheses, many risks arise. It is important to identify and report biases introduced by the researchers or the tools used. However, GenAI generally presents many inherent biases that cannot be as easily studied or identified. In addition, in many idealist approaches, cross-checking between independent reviewers is not suitable as a form of validation, so incorporating GenAI as an additional researcher does not add value and could even introduce additional problems. In these cases, we agree with the views of other qualitative analysis researchers who suggest that GenAI tools can be used as a secondary analysis/synthesis method to validate the work done and to obtain new ideas and perspectives complementary to the work previously carried out by researchers \citep{binz-2025}. 
\end{enumerate}
In either case, outcomes or findings from any GenAI tool would need to be scrutinized by human researchers before being considered for inclusion in the final SLR report.

For researchers evaluating GenAI tool performance, prospective studies could act as trials for GenAI tools (even if there is no intention of incorporating the GenAI tool outcomes into the ongoing SLR). In addition, and subject to considerations of data leakage, retrospective studies could be based on existing SLRs by applying GenAI tools to the defined set of primary studies. The critical issues to address would be to define the following:
\begin{enumerate}
    \item Appropriate criteria (and/or performance metrics) to assess GenAI tool performance. 
    \item A method to assess whether agreement between GenAI tools and human researchers was better than chance.
    \item A procedure to assess the importance of disagreements between the GenAI tool and human researchers, particularly in the case of interpretative synthesis.
\end{enumerate}

The briefing notes in this section are based on research by \citet{Pizard26}. A limitation of that study is that it did not include an independent data extraction stage. Instead, the complete papers were provided to the GenAI tool as input to conduct the steps of the synthesis methods. More research is needed to evaluate data extraction and qualitative synthesis separately.

\subsection{Strength of Evidence Assessment}\label{sec:StrengthOfEvidenceAssessment}
Assessing the strength of evidence (also called the certainty of evidence) of individual review findings aims to assess whether the individual findings of an SLR are trustworthy. It is an important part of any quantitative or qualitative systematic literature review \citep{dyba-2008}. It can also be of value in the context of Rapid Reviews. However, it is not usually relevant for Mapping or Scoping reviews. 

The human-based assessment process involves five steps:
\begin{enumerate}
\item Identifying an evaluation instrument either by creating a new instrument or by selecting (and amending if necessary) an existing instrument.
\item Collating the information required from each primary study for each finding. Some information should be available from the RoB evaluation, but other information may be available in tables of data extracted during the synthesis process or only by extracting further data from each primary study.
\item Trialling the instrument and training the human evaluators on at least one finding. 
\item Applying the instrument to each finding and resolving any disagreements between evaluators.
\item Agreeing and reporting the overall strength of evidence for each finding. Note, Lewin et al (2018) recommend reporting the completed strength of evidence tables for each finding.
\end{enumerate}
Like risk of bias assessment, the strength of evidence is usually based on a subjective assessment of a number of criteria relating to the trustworthiness of a specific finding. This is usually conducted by two or more researchers using an evaluation instrument, including criteria such as: 
\begin{enumerate}
    \item Methodological limitations: The extent to which there are concerns about the design or conduct of the primary studies that contributed evidence to an individual review finding, based on the risk of bias assessment of the relevant primary studies \citep{dyba-2008,lewin-2018,Pizard26}. 
    \item Synthesis/Analysis Method: The extent to which the choice of synthesis/analysis method used to construct a finding is appropriate and justified \citep{Pizard26}. For qualitative analysis, consider:
        \begin{itemize}
        \item How the researchers addressed their own subjectivity.
        \item How the researchers aligned their synthesis method with their own epistemological approach.    
        \end{itemize}
    \item Synthesis/Analysis Conduct: The extent to which the conduct of the synthesis/analysis method is adequate, its reporting is transparent, and sufficient information is provided to enable auditing (i.e., reporting is based on specific SLR reporting guidelines) \citep{Pizard26}.
    \item Coherence: The extent to which the review finding is grounded in data from the relevant primary studies and accounts for the observed patterns \citep{lewin-2018}. 
    \item Adequacy of data: An overall determination of the degree of richness and quantity of data supporting a review finding \citep{dyba-2008,lewin-2018}. 
    \item Practical Relevance: The extent to which the body of evidence from the primary studies supporting a review finding is applicable to the context (perspective or population, phenomenon of interest, setting) specified in the review question \citep{dyba-2008,lewin-2018,Pizard26}. 
    \item Risk of publication bias: based on whether SLR findings could be biased due to characteristics of the set of relevant primary studies  \citep{dyba-2008,Pizard26}. Examples of sources of bias include:
    \begin{itemize}
        \item Only having small-scale positive studies, particularly concept validation studies, rather than rigorous evaluations.
        \item Only having studies that agreed with one another, with no reports of any minority opinions.
        \item Only having English language studies with male authors of European/North American ancestry.
        \item Having studies that were mainly produced by a single research group.
    \end{itemize}
\end{enumerate}
Individual assessments for each criterion need to be scored on a qualitative scale, such as No Serious Problem, Minor Problem, or Serious Problem. Individual assessments are then combined (after disagreements are resolved) into an overall assessment on a qualitative Strength of Evidence scale, such as: Strong (no serious problems), Medium (not more than one serious problem), Low (two or three serious problems), Very Low (four or more serious problems).

Researchers conducting SLRs, who want to use GenAI tools to support their research, will need to be sure that any gains (e.g., reduced human effort or more consistent assessments) of using such a tool outweigh the potential risks. An important issue is that there is seldom a large number of findings from a single SLR, so it remains unclear whether the effort involved, both in training a GenAI tool to perform subjective assessments and in assessing whether the tool is performing adequately, outweighs any potential gains from using such a tool. 

However, there are two areas where GenAI tools are particularly likely to be valuable, even if the main strength of evidence assessments is conducted by two human researchers:
\begin{enumerate}
    \item Using one or more GenAI tools to help organize the information from primary studies related to each individual finding, and extracting any additional contextual information needed to support the strength of evidence assessment. 
    \item Using one or more GenAI tools as additional independent researchers. Independent assessments reduce the risk of bias arising when the human researchers producing the strength of evidence assessment were also responsible for the conduct of the SLR.
\end{enumerate}

Researchers evaluating GenAI capabilities should start by investigating the GenAI tool's capability to support human researchers. Such investigations can be organized as case studies in collaboration with researchers conducting SLRs. However, for researchers looking to fully automate systematic reviews, the strength of evidence is a critical task without which a systematic review is formally incomplete.  Case studies undertaken in collaboration with researchers conducting SLRs would seem to be a sound approach to collecting the data needed to evaluate all aspects of fully automating strength of evidence assessment.

\subsection{General Issues Raised by the Briefing Notes}
Although addressing different SLR processes, we can see some common issues arising from the qualitative data synthesis, risk of bias analysis, and strength of evidence assessments briefing notes:
\begin{enumerate}
\item These tasks are more complex than the literature screening task because they rely on the understanding and experience of researchers outside the context of a specific SLR. For example, SLR researchers need knowledge of best practices for a variety of different quantitative and qualitative research methods.
\item They involve more different sub-tasks than literature screening, and those sub-tasks vary in complexity. Some of the tasks would be supported by flexible text searching tools with copy and paste facilities. Other tasks require an assessment of the extent to which a study provides reliable findings, which are hard to define at a level of detail to ensure consistent, justifiable assessments. 
\item It may be difficult for human researchers to agree on what constitutes a correct assessment, making any assessment of the validity of a GenAI tool assessment problematic.
\item A particular problem with strength of evidence assessment is that it is based on issues that cannot be assessed by reviewing a specific primary study. It can only be assessed by considering the weaknesses associated with a set of primary studies.
\end{enumerate}

For these three processes, researchers conducting SLRs would have the option of using one or more GenAI tools as an extra set of eyes. GenAI outcomes could be incorporated into an SLR report, if they were scrutinized by two human researchers who performed the same tasks, and the results from the GenAI tool(s) were directed to identify all supporting evidence. The purpose of using GenAI tools would not be to reduce the effort to conduct the SLR, but to:
\begin{enumerate}
    \item Increase the depth and breadth of SLR process validation and, hence, improve the trustworthiness of the SLR findings.
    \item Deliver prospective case studies of GenAI performance for these processes and their individual sub-tasks.
\end{enumerate}

Data extraction depends very much on the goal of the SLR and the methods adopted by individual primary studies. For mapping and scoping studies, the closer the data extraction processes is to a Yes/No classification process, the more it resembles the literature screening tasks and can use the same basic prompt approaches and performance metrics used for literature screening. If the classification process is extended to cover multiple categories and multiple output values, the performance metrics recommended for use in literature screening can be adapted \citep{Madeyski26LLM4SCREENLIT}.

However, it is more difficult to specify a process to support data extraction for qualitative data synthesis or quantitative data analysis. Again, SLR researchers have the option of using GenAI tools, not to save time and effort, but for process validation and prospective case studies of the task.

\section{Details of First Validation}\label{sec:FirstValidationDetails}

This appendix reports the detailed comparison of our recommendations with the SE-oriented guidelines of \citet{baltes-2025} and the evidence synthesis recommendations of \citet{thomas-2026-RAISE1}. The comparison is a face validation. For each external guideline, we assessed whether our recommendations address it, and, if not, whether we should change our recommendations. Where we made a change, we report the change and its rationale.

\subsection{Comparison with most recent SE Guidelines}

The Baltes et al. guidelines refine and extend the authors' earlier workshop position paper \citep{wagner2025}. They are therefore a closely related line of work rather than a fully independent one. We did not use them while developing our recommendations. Table~\ref{tab_comparisonSE} reports the results of the comparison.

\textbf{Summary of Comparison with SE Guidelines.} Although our recommendations addressed five of the guidelines proposed by Baltes et al., they consider far more viewpoint than do our guidelines, so some discrepancies were due to this difference in scope. In particular, “Report system traces” is primarily of importance for evaluations performed as part of AI tool development, and “Report System and Prompt Design” was covered with response to prompts but not system design which again is mainly of concern for AI tool developers.  However, “Report Model Version, Configuration and Customisations” was relevant but was omitted from our guidelines. This error was corrected by the inclusion of Rep1.

\subsection{Comparison with most recent Systematic Review Guidelines}

Thomas et al. provide recommendations for the whole ``evidence synthesis ecosystem''. The ecosystem comprises the two researcher roles we consider, evidence synthesists (reviewers) and evidence synthesis methodologists (evaluators). It also comprises six other roles: AI tool building teams; organizations that produce evidence synthesis; funders and commissioners of evidence synthesis and AI; publishers of evidence synthesis; users of evidence synthesis; and trainers of evidence synthesis methods. Our thought experiments considered only reviewers and evaluators, and the relationship between them. This relationship concerns the overlap between the roles and the opportunity to conduct prospective evaluations while conducting SLRs. We therefore compared our recommendations only with the Thomas et al. recommendations for reviewers and evaluators. These recommendations were unchanged across the versions published in 2024, 2025, and 2026. We did not compare our recommendations with those related to developing the ecosystem.

\textbf{Comparison with evidence synthesist guidelines.} Table \ref{tab_comparisonSRsRev} shows the 14 guidelines plus one additional context issue that we extracted from the Thomas et al. guidelines. Three reviewer's related guidelines found by Thomas et al. were not addressed:
\begin{enumerate}
    \item ID 1.2 “Reviewers must take a critical view of available tools” In this case the Plan1 recommendation was extended to cover this issue of critical evaluation of candidate tools.
    \item ID 1.6 “Reviewers need to be aware that research papers may be written by AI”. In this case we were unsure about the intention and implication of the guideline, so we made no change to our guideline.
    \item ID 1.9a (Report) “AI tool identification details”. This was omitted in error. A new recommendation labelled “Rep1” was included to address this omission.
\end{enumerate}

In addition, several guidelines address reviewers' responsibility and accountability for the use of GenAI and its outputs (ID 1.1, ID 1.4, ID 1.5, ID 1.7, ID 1.10). These issues were implicit in our original recommendations. To address them explicitly, we added Con4. For reviewers, ID 1.9c (financial or other interests in the AI tool) and ID 1.9d (tool limitations) are covered by our recommendation to follow current reporting standards (RepR1).

Table~\ref{tab_comparisonSRsEva} shows the result of comparing our recommendations with the recommendations for evidence synthesis methodologists. Only four of the 16 recommendations, and the context issue, were fully covered by our original guidelines. Disagreements are as follows:
\begin{enumerate}
    \item Three recommendations ID 2.1.i  “Adopt best practice”, 2.1.iii “Use the most appropriate study design” and 2.5 “Report results” related to standard evaluation practice which we had not expected to need to report. For ID 2.1.i and ID 2.5, we emphasized their importance in \Cref{sec:Context}. For ID 2.1.iii, we added recommendation Plan2 but changed the wording to “Justify the choice of study design”, because a reader would find it difficult to assess whether the study methodology was appropriate if the rationale for using it had not been explained.
    \item ID 2.3 “Register review-related studies”. Since SE researchers currently do not register their SLRs or review-related studies, we did not include this in our recommendations.
    \item ID 2.5 “Report results” This is implicit in our reporting recommendations and noted in our assumptions (see \Cref{sec:Context}).
    \item ID 2.6 “Ensure full and transparent reporting of evaluations”. This is implicit in our assumptions (see \Cref{sec:Context}), but rather than a single recommendation, the intent of this concern is spelt out in Rep1 (new), Rep2, Rep3, Rep4, Rep5 (new), Rep6 (extended), Rep7.
    \item ID 2.6a “Identify GenAI tools and usage details” as reported in Table 2, this recommendation was omitted and is now included as Rep1.
    \item ID 2.6b “Report data collection process”, ID 2.6d. Data collection and collation, and ID 2.6e. Data quality assurance are now addressed by two recommendations relating to planning and reporting i.e., Plan5 (which was extended) and Rep5.
    \item ID 2.6c “Report any exclusion criteria”. The scope and intention of this recommendation were unclear, so we have not made any change to our recommendations.
    \item ID 2.7 “Support open access”. We have extended Rep6 to explicitly mention this issue.
    \item ID 2.8 “Any financial interest should be declared and bias avoidance explained.” We have introduced RepE4 to address this issue for evaluators.
\end{enumerate}

\textbf{Performance metrics.} In the RAISE-2 document \citep{thomas-2025-RAISE2}, Thomas et al. discuss performance metrics. However, they provide general definitions of metrics and their use, rather than SLR process-focused recommendations such as ours. For this reason, we did not undertake an item-based comparison of our performance metric recommendations with RAISE-2.

Overall Summary:
\begin{enumerate}
    \item We did not understand the scope and intention of two of the Thomas et al. recommendations, so we made no changes to our recommendations.
    \item One recommendation from Thomas et al. related to registering studies which is currently not done by SE reviewers or evaluators, so we made no change to our recommendations
    \item Five discrepancies related to basic good practice. Our response was to extend \Cref{sec:Context} to make the assumptions with respect to reviewer and evaluator experience and competence clearer. We also added a new recommendation Plan2 “Justify the choice of study design”.
    \item Three discrepancies related to explicitly specifying the components of the data management activities (definition, collection, labelling, storage, quality assurance). To address these discrepancies, we extended Plan5 and added Rep5.
    \item We introduced two new recommendations to address issues omitted from our recommendations: Rep1 (Report GenAI tool identification details) and RepE4 (Evaluators - Report any financial interest or any other special issues, and how bias was avoided).
    \item We extended Plan1 to cover the critical evaluation of candidate tools.
    \item We extended Rep6 to explicitly refer to open access.
    \item We made explicit that final responsibility lies with reviewers and evaluators.
\end{enumerate}

Table~\ref{tab_validationChanges} summarizes the recommendations we added or modified as a result of the comparisons with Baltes et al. and Thomas et al., and the guidelines that each change addresses.

\begin{table*}[htbp]
\centering
\caption{Recommendations added or modified as a result of the first validation}
\label{tab_validationChanges}
\footnotesize
\begin{tabular}{ccp{14cm}}
\toprule
ID & Change & Reason \\
\midrule
Rep1 & Added & Report GenAI tool identification details (name, version, configuration, dates of use). Omitted from our original recommendations. Addresses Baltes et al. ID 2 and Thomas et al. ID 1.9a and ID 2.6a. \\
Plan2 & Added & Justify the choice of study design, so that readers can assess whether it was appropriate. Addresses Thomas et al. ID 2.1.iii. \\
Rep5 & Added & Report data collection and management (collection, labelling, storage, quality assurance). Addresses Thomas et al. ID 2.6b, ID 2.6d, and ID 2.6e. \\
RepE4 & Added & Evaluators should report any financial or other interest and how bias was avoided. Addresses Thomas et al. ID 2.8. \\
Con4 & Added & Make explicit that final responsibility for the use of GenAI and its outputs lies with reviewers and evaluators. Addresses Thomas et al. ID 1.1, ID 1.4, ID 1.5, ID 1.7, and ID 1.10. \\
\midrule
Plan1 & Modified & Extended to cover the critical evaluation of candidate tools. Addresses Thomas et al. ID 1.2. \\
Plan5 & Modified & Extended to cover planning data collection and storage, labelling or annotation, and quality control. Addresses Thomas et al. ID 2.6b, ID 2.6d, and ID 2.6e. \\
Rep6 & Modified & Extended to mention open access explicitly. Addresses Thomas et al. ID 2.7. \\
\bottomrule
\end{tabular}
\end{table*}

\textbf{Lessons from the validation.} Two general points emerged. First, the assumptions underlying a set of recommendations should be explicit. For example, reporting evaluation studies may be obvious for research-based reviewers and evaluators. It is more difficult for industry-based tool builders, who may be required to protect the commercial interests of their companies. Second, it should be clear how each recommendation can be addressed. We could not act on some of the Thomas et al. recommendations because we could not determine what they required. Relating each of our recommendations to specific steps of the evaluation and SLR processes is intended to help readers interpret them correctly.

\onecolumn
{\footnotesize
\begin{longtable}{lp{2.5cm}p{9cm}p{3cm}}
\caption{Comparison with SE Guidelines}
\label{tab_comparisonSE} \\
\toprule
ID & Guideline & Comments & Addressed by our recommendations \\ \midrule
\endfirsthead
\toprule
ID & Guideline & Comments & Addressed by our recommendations \\ \midrule
\endhead
\bottomrule
\endfoot
\bottomrule
\endlastfoot
1 & Declare LLM Usage and Role & Researchers \textbf{must} clearly declare that an LLM was used. Researchers \textbf{should} report the exact purpose of using an LLM in a study, the tasks it was used to automate, and the expected benefits in the paper & Yes, see Plan1, Rep1, RepE3 and RepR2 \\
2 & Report Model Version, Configuration, and Customizations & Researchers \textbf{must} report the exact LLM model or tool version, configuration, and experiment date in the paper. For fine-tuned models, they must describe the fine-tuning goal, dataset, and procedure. Researchers \textbf{should} include default parameters, explain model choices, compare base- and fine-tuned model using suitable metrics and benchmarks, and share fine-tuning data and weights (or alternatively justify why they cannot share them). & Partly in Rep3, Rep1 introduced to address reporting full GenAI tool identifier information \\
3 & Report System and Prompt Design & Researchers must describe in the paper the full architecture of LLM-based tools, from standalone uses to agentic systems, including the LLM’s role and its interactions with other components. Hosting and access should be described; for time-sensitive measurements, researchers must clarify whether local infrastructure or cloud services were used. Researchers must publish all prompts as supplementary material, including prompt templates with representative instances and the dynamic generation process where applicable, and must include representative examples in the paper ; sensitive content must be anonymized. If full prompt disclosure is not feasible, summaries or examples should be provided. The prompting strategy and prompt reuse across models and configurations must be specified, and how the prompts were developed should be described; for few-shot prompts, how the examples were selected must be explained in the paper. Researchers must describe the configuration mechanisms used (e.g., context files, skills, subagents) and summarize in the paper which tools were exposed; all configuration artifacts and the complete tool catalog (schemas, definitions, Model Context Protocol (MCP) servers) should be published as supplementary material . For agentic systems, researchers must specify the agents’ roles, reasoning frameworks, and communication flows; where external tools are used, the model’s reasoning, tool calls, and interactions with users or the environment must be reported separately. For retrieval-augmented generation (RAG) and similar methods, researchers must describe how external data was retrieved, stored, and integrated. Where legally possible, the implementation should be open-sourced; nondisclosed proprietary components must be acknowledged as reproducibility limitations. & Partly see Rep2. Some assume tool developers and that role is outside our context \\
4 & Report Session Traces & To address model non-determinism and ensure reproducibility, especially when targeting SaaS-based commercial tools, researchers should include full interaction logs (prompts and responses) as supplementary material if privacy and confidentiality can be ensured. Traces containing sensitive information must be anonymized. For agentic systems, interaction logs should cover human-in-the-loop exchanges with the agent, including feedback and approval decisions. Researchers should report the complete runtime trace as supplementary material, covering both external tool invocations (tool name, arguments, result, ordering) and which configured artifacts (skills, context files, subagents reported under System and Prompt Design) were activated, so that readers can attribute task outcomes to the model, the tools, or their interaction pattern. Where tool-native trace formats are used, the file format and tool version must be described; an open format with a documented schema should be preferred. Developed plans should be reported as supplementary material if available. When full trace disclosure is not feasible, representative or anonymized examples should be provided, and unobservable aspects of commercial-tool runs must be acknowledged as reproducibility limitations. & No. Issues related to tool development are outside our context. \\
5 & Use Suitable Baselines, Benchmarks, and Metrics & Researchers must justify all benchmark and metric choices in the paper, must discuss their reliability and validity (especially construct validity), and should summarize benchmark structure, task types, and limitations. They should operationally define the phenomenon being measured, justify the sampling strategy used to select problems for inclusion in the benchmark, isolate the target capability from confounders where possible, and perform an error analysis. When creating or releasing a benchmark, data sources and collection dates must be disclosed for each release. Where possible, traditional (non-LLM) baselines should be used for comparison. Researchers must explain in the paper why the selected metrics are suitable for the specific study; prior adoption in related work alone does not constitute sufficient justification. They should report established metrics to make study results comparable, but can report additional metrics that they consider appropriate. Due to the inherent non-determinism of LLMs, experiments should be repeated; the result distribution should then be reported using descriptive statistics. If comparing models or tools, researchers should use appropriate inferential statistics rather than relying solely on summary statistics. Researchers should justify the number of experiment repetitions, for example through a power analysis or by monitoring convergence of descriptive statistics. Latency must be reported when it can affect study outcomes (e.g., interactive user studies, latency comparisons). & Yes, see Plan1, Plan2 \& Plan3 \\
6 & Use an Open LLM as a Baseline & When researchers can control which LLM a study uses, they should include an open-weight LLM, ideally an open-source LLM, as a baseline and report inter-model agreement. Researchers should ensure the open-weight baseline is independently reproducible from their supplementary material. & Yes, see PlanE2 \\
7 & Use Human Validation for LLM Outputs & If assessing the quality of generated artifacts is important and no reference datasets or suitable comparison metrics exist, researchers should use human validation for LLM outputs. If they do, they must define the measured construct (e.g., usability, maintainability) and describe the measurement instrument in the paper. Researchers should consider human validation early in the study design, and not as an afterthought. They should build on established reference models for human-LLM comparison. When developing or adapting measurement instruments, researchers must share them. When LLMs replace, rather than augment, humans in research tasks such as annotating software artifacts, coding interview transcripts, or simulating study participants, researchers must explain whether and how the replacement is justified and should ground it with inter-model and model-to-human agreement. For qualitative coding of interpretive data, they should justify why an LLM is methodologically appropriate. When aggregating LLM judgments, methods and rationale should be reported and inter-rater agreement should be assessed. Confounding factors should be controlled for; where applicable, researchers may perform a power analysis to estimate the required sample size. For value-laden or culturally contingent constructs, researchers should describe rater demographics beyond expertise and discuss potential demographic biases. When evaluating agentic tools, user feedback on the agent’s proposed actions should be assessed and acceptance statistics reported. & Yes, Plan2, Plan3, Plan4, Plan5 \& Plan6 \\
8 & Report Limitations and Mitigations & Researchers \textbf{must} transparently report study limitations, including the impact of non-determinism and generalizability constraints. The paper must specify whether generalization across LLMs or across time was assessed, and discuss model and version differences. Authors \textbf{must} describe measurement constructs and methods, disclose any data leakage risks, and avoid leaking evaluation data into LLM improvement pipelines. They \textbf{must} provide model outputs, discuss sensitive data handling, ethics approvals, and justify LLM usage in light of its resource demands & Yes, see Rep7 and RepE4. \\
\end{longtable}

\footnotesize
\begin{longtable}{lp{2.5cm}p{9cm}p{3cm}}
\caption{Comparison with Systematic Review Guidelines for Evidence Synthesists (Reviewers)}
\label{tab_comparisonSRsRev} \\
\toprule
Id & Guideline & Details & Addressed \\ \midrule
\endfirsthead
\toprule
Id & Guideline & Details & Addressed \\ \midrule
\endhead
\bottomrule
\endfoot
\bottomrule
\endlastfoot
Context & Evidence synthesists need appropriate knowledge/expertise & SLR related expertise and tool knowledge (what is available, what has been validated and how to use it). & Yes. \Cref{sec:Context}. \\ \midrule
\multicolumn{4}{l}{Remain ultimately responsible for the evidence synthesis} \\ \midrule
1.1 & Reviewers are accountable for their SLR. & Evidence synthesists are accountable for the content, methods and findings of their evidence synthesis, including the decision to use AI, how it is used, and its impact on the synthesis. & Yes, see Con4. \\
1.2 & Reviewers must take a critical view of available tools & When considering using an AI system or tool, be critical of its evaluations (see building and validating an AI tool and conducting evaluations to build a cumulative evidence base in RAISE 2), to understand whether it does what it states at an adequate level, as well as its limitations, and whether it can be applied to the context of the specific synthesis. & No – Plan1 recommendation extended to cover this issue \\
1.3 & Reviewers should justify AI use & The use of AI should be justified and should involve demonstrating that the tools are methodologically sound, that they do not undermine the trustworthiness or reliability of the synthesis or its conclusions, and that they are appropriate for the specific evidence synthesis. & Yes, see Plan1 \\
1.4 & An AI tool cannot be considered an author & AI should not receive credit for authorship of an evidence synthesis, although it is sometimes employed as a writing aid. Human authors are accountable for the proper use of AI in their evidence synthesis; an AI tool cannot satisfy all authorship criteria & Yes, see Con4. \\
1.5 & AI should not be used to falsify or fabricate results & AI should not be used to fabricate or falsify evidence synthesis data, results or findings & Yes, see Con4. \\
1.6 & Reviewers need to be aware that research papers may be written by AI. & Evidence synthesists should be aware of and account for the growing prevalence of research papers that are partially or fully written by AI. & No – Unsure about the intention and implication of this recommendation. \\
1.7 & Reviewers should not rely on the correctness of GenAI outputs & Evidence synthesists should not treat a chatbot (e.g., ChatGPT) as a knowledge base for evidence synthesis tasks; that is, do not rely on it for factually accurate answers & Yes, see Con4. \\ \midrule
\multicolumn{4}{l}{Report AI use in your evidence synthesis manuscript transparently} \\ \midrule
1.8 & Declare AI use & Authors should declare when they have used AI if it makes or suggests judgements, such as, in relation to the eligibility of a study, critical appraisals (including risk of bias assessments), extraction of bibliographic, numerical or qualitative data about a study or its results, synthesis of data from two or more studies, assessments of the certainty of evidence (including GRADE domains or overall certainty ratings for an outcome or finding), drafting text that summarises the overall strength of evidence, related implications for decision making or research, or plain language summaries & Yes, see Rep1-Rep7. \\
1.9 & Adhere to established reporting standards & Adhere to established reporting standards for evidence synthesis, such as the Preferred Reporting Items for Systematic Reviews and Meta-Analyses (PRISMA) or RepOrting standards for Systematic Evidence Syntheses (ROSES). PRISMA, for example, includes items on reporting automation tools used at different stages of the synthesis process. & Yes, see recommendation RepR1 \\
1.9a & AI tool identification details & The name(s) of the AI system(s)/tool(s)/platform(s), version(s) and date(s) it was used or they were used & No. Introduced recommendation Rep1 to address the issue explicitly \\
1.9b & Purpose of tools use & The purpose of using the AI tool and specify which parts of the evidence synthesis process were impacted. (Note: Authors should cite or reference user guidance or report how AI was used, including any modifications that were applied); the justification for using the AI tool, including: i.) evidence that the tool is methodologically sound and is unlikely to undermine the trustworthiness or reliability of the synthesis or its conclusions (e.g., citing or referencing evaluations of its performance that detail the impact of errors, limitations and generalisability, and), ii.) how it has been validated (and piloted, if applicable) to ensure that it is appropriate for use in the context of the specific evidence synthesis. Wherever possible and practical, make the inputs (e.g., prompt development), outputs, datasets, and code publicly and freely available (for instance, on repositories or as supplementary material), and describe any steps taken to review and verify AI-generated content. & Yes, see recommendation Plan1 \\
1.9c & Any special interest in the AI tool & Any financial or non-financial interests in the AI tool, if applicable, alongside the funding sources of the AI tool & Yes, implicit in the reporting recommendation RepR1 for reviewers. \\
1.9d & Tool limitations & Discuss any limitations of using AI in the review processes, including any potential biases, and comment on the potential impact of limitations, where relevant. & Yes. Implicit in RepR1 for reviewers \\ \midrule
\multicolumn{4}{l}{Ensure ethical, legal and regulatory standards are adhered to when using AI} \\ \midrule
1.10 & Uphold legal and regulatory guidelines & Ensure ethical, legal, and regulatory standards are upheld in the application of AI to your synthesis. For instance, consider issues related to plagiarism, provenance, copyright, intellectual property, jurisdiction, licensing, confidentiality, compliance, and privacy responsibilities, including data protection laws. The need to stay up to date on technologies available and utilize cost-effective technology is important but should not come at the expense of responsible use. & Yes, see Con4. \\
\end{longtable}

\footnotesize
\begin{longtable}{lp{2.5cm}p{9cm}p{3cm}}
\caption{Comparison with Systematic Review Guidelines for Evidence Synthesist Methodologists (Evaluators)}
\label{tab_comparisonSRsEva} \\
\toprule
Id & Guidelines & Discussion & Addressed \\ \midrule
\endfirsthead
\toprule
Id & Guidelines & Discussion & Addressed \\ \midrule
\endhead
\bottomrule
\endfoot
\bottomrule
\endlastfoot
Context & Evidence synthesist methodologists need appropriate knowledge/expertise & & Yes (\Cref{sec:Context}) \\ \midrule
\multicolumn{4}{l}{Adhere to open science practices when researching and evaluating AI systems:} \\ \midrule
2.1.i & Adopt best practice & Adhere to best practice when training or testing the application of AI within an evidence synthesis. & No – changed the discussion of context to say “We assumed that any rigorous evaluation of GenAI tool requires an experimental framework consistent with best practice that …” \\
2.1.ii & Use appropriate comparators & Use appropriate comparators (e.g., current best practice, a human as the gold standard, another AI system that has previously been evaluated against current best practice. Some metrics, such as user experience, may not need a comparator. & Yes see recommendations Plan3, PlanE2 and Plan6 \\
2.1.iii & Use the most appropriate study design & Use the most appropriate study design to address your research question, e.g., compare AI systems to non-AI processes across large numbers of evidence syntheses; conduct a single study comparing AI systems to non-AI processes; or conduct a feasibility evaluation to test whether an AI system is applicable to the context of a specific evidence synthesis, etc. & No – General good practice, however, now included for reviewers and evaluators (see recommendation Plan2) \\
2.2 & Address GenAI tool biases & When validating AI tools, methodologists should report any equity issues or biases within the AI systems, including potential biases in the systems’ training data and strategies for users to help mitigate possible bias-related issues. & Yes, see recommendation Plan4 \\
2.3 & Register review-related studies & If you conduct a Study Within a Review, register a study within  review and use template protocols from the study within a Review Network, if possible. This allows the findings from your single Study Within a Review to be combined with others, thereby building a cumulative evidence base. & No – not included in our recommendations because SE researchers do not currently register their SLR protocols, nor do they register review-related studies within a Review Network. \\
2.4 & Use appropriate evaluation methods & Evaluations should be conducted using appropriate methods including using key performance metrics adhering to protocols, judiciously interpreting data, and drawing conclusions from the findings with a clear explanation of the basis of the conclusions. & Yes, see recommendations Plan2 \& Plan3 \\
2.5 & Report results & Evaluations should be made publicly available, ideally published in peer-reviewed journals (although publishing as a preprint can also be beneficial, given the rapid developments in the field). They may also be uploaded to a repository or included as supplementary material for an evidence synthesis manuscript. & No specific recommendation – but implicit in reporting recommendations and noted in our assumptions about the viewpoints (see \Cref{sec:Context}) \\
2.6 & Ensure full and transparent reporting of evaluations & Ensure full and transparent reporting of evaluations, including full methods and publishing of null results & Partly implicit in our assumptions (see \Cref{sec:Context}) and spelled out in reporting recommendations Rep1 to Rep7 \\
2.6a & Identify GenAI tools and usage details & Report the name of the AI system(s)/tool(s)/platform(s), version(s) and date(s) it was used or they were used. & No, Rep1 included to address this \\
2.6b & Report data collection process & Report how and where the data were collected (as the scope or source of the data may result in bias in the evaluation). & No – Implicit in reporting standards but now included explicitly in recommendation Rep5 and Plan5 applying to reviewers and evaluators. \\
2.6c & Any exclusion criteria & Any exclusion criteria that were applied (as this may result in bias in the evaluation). & No- scope and intention of this is unclear. \\
2.6d & Data collection and collation & How data were labelled or annotated as part of building the model. & No – new recommendations address this see Plan5 and Rep5 \\
2.6e. & Data quality assurance & Any quality assurance/control processes used to clean the data, including handling missing data and outliers. & No – new recommendations address this see Plan5 and Rep5 \\
2.6f & The prompt engineering process & Prompt engineering and the documentation of how prompts were developed and validated (for AI that require prompts, such as LLMs). & Yes, see Rep2 \\
2.7 & Support open access & Whenever possible, openly share datasets, source code, and other materials used to develop and run the evaluation(s), or clarify how they can be accessed, including any restrictions on their use or licensing considerations. & No, Rep6 extended to include this issue which we think is important for reviewers conducting evaluations as part of an SLR \\ \midrule
\multicolumn{4}{l}{Commit to objective and impartial evaluations and validation of AI systems:} \\ \midrule
2.8 & Any financial interest should be declared and bias avoidance explained. & If methodologists have a financial interest in the tool being evaluated, this should be declared. While making every effort to ensure that the research is objective, the evaluation or validation should not be presented as independent. It is also good practice to declare any non-financial interests & No, included RepE4 to address this issue \\
\end{longtable}
}
\twocolumn

\printcredits

\section*{Funding}
\noindent This research did not receive any specific grant from funding agencies in the public, commercial, or not-for-profit sectors.

\section*{Acknowledgements}
\noindent We thank Shari Lawrence Pfleeger for reviewing this paper and suggesting changes that improved its clarity and readability.

\bibliographystyle{cas-model2-names}
\bibliography{refs}

\bio{}
Barbara Kitchenham retired in 2019 after working in the field of software engineering for over 50 years both in industry and academia. She has published over 200 software engineering journal and conference papers. Her main research interests are the application of evidence-based practice to software engineering, the use of statistical methods in software engineering including meta-analysis and robust methods, and open and reproducible science. In 2019, she was awarded the IEEE Technical Committee Distinguished Women in Science \& Engineering (WISE) Leadership award.
\endbio

\bio{}
Sebastián Pizard is Assistant Professor at the Computer Science Institute of the Faculty of Engineering, Universidad de la República, Uruguay. He is a researcher in Uruguay's National System of Researchers (SNI) and in the Informatics Area of PEDECIBA, and earned his PhD in Informatics from PEDECIBA in 2024. His research centers on Evidence-Based Software Engineering, the adoption of scientific evidence by industry and government, empirical software engineering, and qualitative methods applied to computing. Before academia, he spent a decade in the software industry, and he has over fifteen years advising state agencies and companies on behalf of his Faculty.
\endbio

\bio{}
Lech Madeyski is a Deputy Head of the Department of Applied Informatics at Wrocław University of Science and Technology. He works at the intersection of AI, data science, and software engineering. With Barbara Kitchenham and David Budgen, he developed SEGRESS, a unified reporting standard for software engineering secondary studies, extending the medical PRISMA 2020 guideline. He is the founder and Editor-in-Chief of the e-Informatica Software Engineering Journal and has served on programme committees of numerous conferences, including the ICORE A*-ranked ICSE'2027, ICSE'2026, FSE'2025 and ASE'2023. He is listed among the world's top 2\% of scientists by Stanford University and Elsevier.
\endbio

\bio{}
Ronnie de Souza Santos is an Assistant Professor in Software Engineering and Director of the MEng Program in Software Engineering at the University of Calgary. He works at the intersection of software engineering, human aspects, and responsible AI, with research interests including software quality, fairness, generative AI, and software engineering education. He leads PLURISE, the Plural Software Engineering for a Plural Society research group. Before joining the University of Calgary, he held academic and research positions at Cape Breton University, Dalhousie University, and Carleton University. He received his PhD in Computer Science from the Federal University of Pernambuco, Brazil.
\endbio

\bio{}
Martin Shepperd is Professor of Software Technology and Modelling at Brunel University London. His research interests lie at the intersection of software engineering, machine learning and empirical research methods. He has published three books and more than 180 refereed research articles. He was awarded the 2022 Tage Erlander research prize and professorship at the University of Gothenburg, Sweden. Before entering academia, he worked as a software developer in the financial services sector. He is a Fellow of the British Computer Society.  
\endbio

\bio{}
David Budgen is an Emeritus Professor of Software Engineering in the Department of Computer Science at Durham University in the U.K. His research interests include software design, design environments, healthcare computing and evidence-based software engineering, with his book {\em Software Design} now in its third edition (2021). He was awarded a BSc in Physics and a PhD in Theoretical Physics from Durham University, following which he worked as a research scientist for the Admiralty and then held academic positions at Stirling University and Keele University before moving to Durham University. He is a member of the ACM and the Institution of Engineering \& Technology (IET), and is a Chartered Engineer (CEng).
\endbio

\end{document}